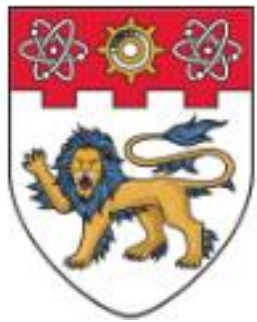

# MEASUREMENT SETUP CONSIDERATION AND IMPLEMENTATION FOR INDUCTIVELY COUPLED ONLINE IMPEDANCE EXTRACTION

**Zhao Zhenyu**

**School of Electrical and Electronic Engineering**

**2021**

# Measurement Setup Consideration and Implementation for Inductively Coupled Online Impedance Extraction

by

# Zhao Zhenyu

School of Electrical and Electronic Engineering

A thesis submitted to the Nanyang Technological University
in partial fulfillment of the requirement for the degree of
**Doctor of Philosophy**

**2021**

## Statement of Originality

I hereby certify that the work embodied in this thesis is the result of original research, is free of plagiarised materials, and has not been submitted for a higher degree to any other University or Institution.

March 8, 2021                         *Zhao Zhenyu*
. . . . . . . . . . . . . . . . .          . . . . . . . . . . . . . . . . . . . . . . . .
       Date                                      Zhao Zhenyu

# Supervisor Declaration Statement

I have reviewed the content and presentation style of this thesis and declare it is free of plagiarism and of sufficient grammatical clarity to be examined. To the best of my knowledge, the research and writing are those of the candidate except as acknowledged in the Author Attribution Statement. I confirm that the investigations were conducted in accord with the ethics policies and integrity standards of Nanyang Technological University and that the research data are presented honestly and without prejudice.

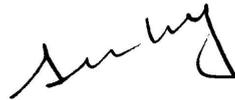

March 8, 2021
................  ............................
Date                                       See Kye Yak

# Authorship Attribution Statement

This thesis contains materials from 4 papers published in the following peer-reviewed journals in which I was the first author.

Chapter 3 is published in:

Z. Zhao, K. Y. See, E. K. Chua, A. S. Narayanan, W. Chen, and A. Weerasinghe, "Time-variant in-circuit impedance monitoring based on the inductive coupling method," *IEEE Trans. Instrum. Meas.,* vol. 68, no. 1, pp. 169-176, Jan. 2019.

The contributions of the co-authors are as follows:
- Prof See provided the initial project direction and edited the manuscript drafts.
- Dr. Chua, Dr. Narayanan, Mr. Chen, and Mr. Weerasinghe assisted in the revision of this manuscript.
- I carried out the work and prepared the manuscript drafts.

Chapter 4 is published in:

Z. Zhao, A. Weerasinghe, W. Wang, E. K. Chua, and K. Y. See, "Eliminating the effect of probe-to-probe coupling in inductive coupling method for in-circuit impedance measurement," *IEEE Trans. Instrum. Meas.*, vol. 70, 2021, Art no. 1000908.

The contributions of the co-authors are as follows:
- Prof See provided the initial project direction and edited the manuscript drafts.
- Mr. Weerasinghe assisted in the collection of the measurement data and the revision of this manuscript.
- Dr. Wang and Dr. Chua assisted in the revision of this manuscript.
- I carried out the work and wrote the drafts of the manuscript.

Chapters 5 and 6 are published in:

Z. Zhao, F. Fan, W. Wang, Y. Liu, and K. Y. See, "Detection of stator inter-turn short-circuit faults in inverter-fed induction motors by online common-mode impedance monitoring," *IEEE Trans. Instrum. Meas.*, accepted.

The contributions of the co-authors are as follows:

- Prof See provided the initial project direction and edited the manuscript drafts.
- Dr. Fan assisted in the collection of the measurement data and the revision of this manuscript.
- Dr. Wang and Dr. Liu assisted in the revision of this manuscript.
- I carried out the work and prepared the manuscript drafts.

Chapter 7 is published in:

Z. Zhao, K. Y. See, W. Wang, E. K. Chua, A. Weerasinghe, Z. Yang, and W. Chen, "Voltage-dependent capacitance extraction of SiC power MOSFETs using inductively coupled in-circuit impedance measurement technique," *IEEE Trans. Electromagn. Compat.*, vol. 61, no. 4, pp. 1322-1328, Aug. 2019.

The contributions of the co-authors are as follows:
- Prof See provided the initial project direction and edited the manuscript drafts.
- Dr. Wang and Mr. Yang assisted in the collection of the measurement data and the revision of this manuscript.
- Dr. Chua, Mr. Weerasinghe, and Mr. Chen assisted in the revision of this manuscript.
- I carried out the work and prepared the manuscript drafts.

|                |                      |
|----------------|----------------------|
| March 8, 2021  | *Zhao Zhenyu*        |
| ................ | ........................ |
| Date           | Zhao Zhenyu          |

# Acknowledgements

I would like to sincerely thank my supervisor, Professor See Kye Yak, for his unfailing guidance and assistance during my PhD study. He is a passionate mentor who always shows positive encouragement in our interactions related to my research direction. He is also a knowledgeable scholar and I have learned a lot from him, which undoubtedly set an example in my future research career.

I wish to express my thanks to Nanyang Technological University, National Research Foundation Singapore, and SMRT Corporation for funding my education and research. I would also like to show my gratitude to my fellow colleagues (Dr. Arun Shankar Narayanan, Dr. Chua Eng Kee, Dr. Fan Fei, Dr. Li Kangrong, Dr. Li Tianliang, Dr. Li Yuhua, Dr. Liao Xinqin, Dr. Liu Yong, Dr. Qu Zilian, Dr. Tengiz Svimonishvili, Dr. Wang Liang, Dr. Wang Wensong, Dr. Zhang Junwu, Dr. Zhao Bo, Mr. Arjuna Weerasinghe, Mr. Sin Kok Kee, Mr. Sun Xin, and Mr. Yang Zhenning) for their kind support and friendship during my PhD study in SMRT-NTU Smart Urban Rail Corporate Laboratory.

Finally, I am most grateful to my family, especially my dear wife Linkai, for their endless support, encouragement, and understanding. Without their devoted love to me, my thesis would never have been completed.

# Contents

# Abstract

The online impedance serves as a key parameter for evaluating the operating status and health condition of many critical electrical systems. For its non-contact nature and ease of on-site implementation, the inductive coupling approach is a superior method to extract the online impedance of the electrical systems. However, all earlier reported works of this approach have assumed that the online impedance of an electrical system is time-invariant for a specific time interval. Therefore, little work has been explored to extract the time-variant online impedance of any electrical systems, especially those systems with frequent impedance changes. Besides, the effect of the probe-to-probe coupling between the inductive probes used in this approach on the accuracy of the extracted online impedance has not been evaluated, especially when the inductive probes are placed very close to each other due to space constraints in some special circumstances. Moreover, in the harsh industrial environment where significant electrical noise and power surges are present, the existing measurement setup of this approach lacks the necessary signal-to-noise ratio (SNR) and front-end protection to the measurement instruments, and consequently, the application of this approach becomes challenging in such an environment.

To overcome the above-mentioned limitations and challenges, this thesis proposes an improved measurement setup of the inductive coupling approach and develops a moving window discrete Fourier transform (DFT) algorithm so that it has the ability to extract not only the time-invariant online impedance but also the time-variant online impedance of electrical systems. In addition, based on a three-port network concept, a comprehensive equivalent circuit model of the proposed measurement setup is described, in which the effect of the probe-to-probe coupling can be taken into account in the model. With the three-port equivalent circuit model, a three-term calibration technique is proposed to deembed the effect of the probe-to-probe coupling with the objective to improve the accuracy of online impedance extraction. Furthermore, by incorporating signal amplification and surge protection modules into the proposed measurement setup, the SNR can be enhanced and the damage to the measurement instruments caused by the power surges can be avoided. With these improvements in the measurement setup of the inductive coupling approach, it opens the door for the application of this approach in many electrical systems, especially those with significant electrical noise and power surges.

Based on the proposed measurement setup and the associated theories, the inductive coupling approach has been validated experimentally to show its ability to detect the incipient stator faults online in the inverter-fed induction motor, which shows its capability for online condition monitoring of the electrical system. Besides, this approach has also been demonstrated with the ability to extract the voltage-dependent capacitances of the silicon carbide (SiC) power metal-oxide-semiconductor field-effect transistor (MOSFET). With the extracted voltage-dependent capacitances of the SiC power MOSFET, it facilitates the evaluation of the switching characteristics of the SiC power MOSFET and the associated electromagnetic interference (EMI) noise caused by the switching of the SiC power MOSFET.

# List of Figures

# List of Tables

# List of Abbreviations

| | |
|---|---|
| 3-D | Three-Dimensional |
| AC | Alternating Current |
| ADC | Analog-to-Digital Converter |
| AFB | Abnormal Friction Bearing |
| AI | Artificial Intelligence |
| AT | Attenuator |
| CH1 | Channel 1 |
| CH2 | Channel 2 |
| CM | Common-Mode |
| DC | Direct Current |
| DFT | Discrete Fourier Transform |
| DM | Differential-Mode |
| DSP | Digital Signal Processing |
| EMI | Electromagnetic Interference |
| FOI | Frequency of Interest |
| IA | Impedance Analyzer |
| IIP | Injecting Inductive Probe |
| IM | Induction Motor |
| ITSC | Inter-Turn Short-Circuit |
| KCL | Kirchhoff's Current Law |
| MDS | Motor Drive System |
| MOSFET | Metal-Oxide-Semiconductor Field-Effect Transistor |
| PA | Power Amplifier |
| PCI | Peripheral Component Interconnect |
| PWM | Pulse Width Modulation |
| RCF | Rotor Cage Fault |
| RF | Radio Frequency |
| RIP | Receiving Inductive Probe |
| SAC | Signal Acquisition Card |
| SG | Signal Generator |
| SGAS | Signal Generation and Acquisition System |
| SGC | Signal Generation Card |
| SiC | Silicon Carbide |
| SID | Signal Injection Device |
| SMPS | Switched-Mode Power Supply |
| SNR | Signal-to-Noise Ratio |
| SOC | State of Charge |
| SOH | State of Health |
| SP | Surge Protector |
| ST | Speed Transient |
| SUT | System Under Test |
| VFD | Variable-Frequency Drive |
| VNA | Vector Network Analyzer |
| WBG | Wide-Bandgap |

# Chapter 1 Introduction

## 1.1. Background

The online impedance is one of the key parameters for evaluating the operation status and health condition of many critical electrical systems. For example, the online impedance of the power grid has a significant impact on the performance of the grid-connected power converters [1]-[3]. It affects not only the converters' inner current control loops [4], [5] but also their voltage control loops [6]. In both cases, inaccurate estimation of the online impedance of the power grid can degrade the performance of the grid-connected power converters and even make them unstable. Moreover, the online impedance of the power grid can also provide useful information to meet certain requirements, such as anti-islanding regulations [7]. In addition to the application in the power grid, its application in the battery energy storage system has also been discussed [8]-[10]. To achieve efficient battery energy management, accurate battery state of charge (SOC) information is needed as a measure of the remaining energy stored in the battery. The online impedance of the battery is an effective way to estimate its SOC [11]. Moreover, the battery impedance would change due to factors such as aging and temperature. Therefore, by comparing its measured impedance with the reference impedance set based on long-term experimental data, it can also be used as a measure of the state of health (SOH) or the battery internal temperature [12]-[14]. In addition to these applications, the online noise source impedance of the switched-mode power supply (SMPS) is essential in the design of power line electromagnetic interference (EMI) filters [15], [16]. The noise source impedance of the SMPS would change with certain parameters such as converter topology, component parasitic parameters, and board layout [17]. Therefore, the impedance of an operating SMPS would be different from that of an offline SMPS, and the online noise source impedance extraction of an SMPS will bring more accurate results for a systematic EMI filter tailored to a specific SMPS [18].

Acknowledging the significance and importance of the online impedance of the electrical systems, developing an effective method to extract their online impedances is necessary. In general, there are mainly three online impedance extraction approaches; namely the capacitive coupling approach [19]-[21], the voltage-current measurement approach [22]-[27], and the inductive coupling approach [28]-[31]. Among them, the measurement setup of the inductive coupling approach does not have any physical electrical connection to the energized system under test (SUT). Also, the clamp-on inductive probes used in the measurement setup of this approach can be easily mounted on or removed from the insulated wire that delivers the power to the SUT and therefore it simplifies the on-site implementation [31].

The inductive coupling approach was firstly reported to extract the power line impedance [28]. It was later extended to extract the noise source impedance of the SMPS for systematic EMI filter design purposes [29]. It was further improved to simplify the online impedance extraction process with the cascaded two-port network concept [30], [31]. However, all earlier reported works of this approach have assumed that the online impedance of an electrical system is time-invariant for a specific time interval and therefore they lack the ability to extract the time-variant online impedance of electrical systems, especially those power conversion systems that adopt switching techniques, where frequent impedance changes take place. Besides, there is a minimum separation distance between the inductive probes to minimize possible measurement errors due to the probe-to-probe coupling between the inductive probes. This minimum separation distance may not be always possible in some circumstances, where the inductive probes have to be closely placed due to space constraints. Moreover, the lack of signal amplification and surge protection modules in the conventional measurement setup of this approach makes it is challenging to be used in the harsh industrial environment where strong electrical noise and power surges are present. Due to the above-mentioned limitations and challenges, the scope of application of the inductive coupling approach is rather limited.

## 1.2. Motivation, Objectives, and Contributions

The motivation of this thesis is to develop an improved and more robust measurement setup to address the aforementioned limitations and challenges of the conventional measurement setup, thereby expanding the scope of application of the inductive coupling approach and improving its measurement accuracy. The objectives of this thesis are listed as follows:

- Propose an improved measurement setup of the inductive coupling approach, which can extract not only the time-invariant online impedance but also the time-variant online impedance of electrical systems.

- Develop a calibration scheme for the proposed measurement setup to deembed the effect of the probe-to-probe coupling between the inductive probes with the objective to improve the accuracy of online impedance extraction.

- Incorporate signal amplification and surge protection modules into the proposed measurement setup to expand the scope of application of the inductive coupling approach into the electrical systems where significant electrical noise and power surges are present.

- Explore expanded applications of the inductive coupling approach based on the proposed measurement setup in online condition monitoring, fault detection, and electrical parameters evaluation of electrical systems.

The main contributions of this thesis are summarized as follows:

- Extension of the inductive coupling approach for time-variant online impedance extraction of electrical systems.

- Development of a three-term calibration technique to deembed the effect of the probe-to-probe coupling so that the measurement error due to such coupling can be compensated to preserve the measurement accuracy.

- Incorporation of signal amplification and surge protection modules into the proposed measurement setup and expansion of its applications into the electrical systems with strong background noise and power surges.

## 1.3. Organization of Thesis

This thesis is organized as follows:

Chapter 1 introduces the background, motivation, objectives, and contributions of this thesis.

Chapter 2 presents a review of existing online impedance extraction approaches.

Chapter 3 proposes the improved measurement setup of the inductive coupling approach and introduces the theory behind time-variant online impedance extraction.

Chapter 4 develops a three-term calibration technique for the proposed measurement setup to deembed the effect of the probe-to-probe coupling between the inductive probes with the objective to improve the accuracy of online impedance extraction.

Chapter 5 discusses the additional measurement setup consideration in industrial applications where significant electrical noise and power surges are present.

Chapter 6 discusses and demonstrates the application of the inductive coupling approach in online detection of the incipient stator faults in the inverter-fed induction motor.

Chapter 7 further extends the application of this approach for non-intrusive extraction of the voltage-dependent capacitances of the silicon carbide (SiC) power metal-oxide-semiconductor field-effect transistor (MOSFET).

Finally, Chapter 8 concludes this thesis and proposes future works that are worth exploring.

# Chapter 2 Literature Review of Online Impedance Extraction Approaches

In view of the significance and importance of the online impedance of the electrical systems, many online impedance extraction methods have been reported in the literature, which are mainly classified into three categories; namely the capacitive coupling approach [19]-[21], the voltage-current measurement approach [22]-[27], and the inductive coupling approach [28]-[31]. In this chapter, the three approaches are carefully reviewed, and their merits and drawbacks are also discussed.

## 2.1. Capacitive Coupling Approach

Fig. 2-1 shows the basic measurement setup of the capacitive coupling approach to extract the online impedance of an electrical system under test (SUT), where the SUT is powered by either a direct current (DC) or a low-frequency (e.g. 50/60 Hz) alternating current (AC) power source. The online impedance of the SUT is denoted by $Z_{SUT}$.

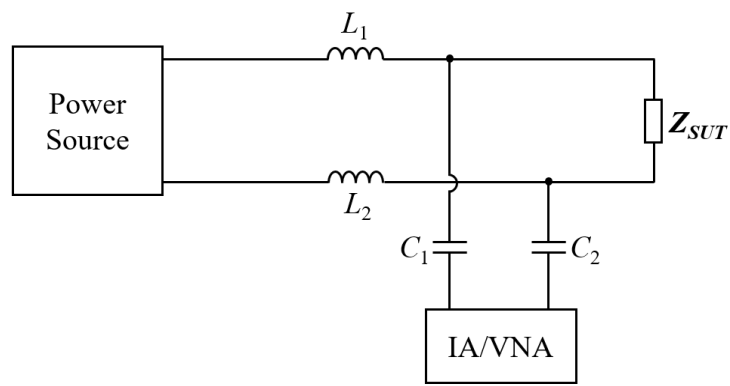

Fig. 2-1. Basic measurement setup of the capacitive coupling approach [19].

To extract $Z_{SUT}$, two coupling capacitors ($C_1$ and $C_2$), two inductors ($L_1$ and $L_2$), and an impedance analyzer (IA) or vector network analyzer (VNA) form the basic measurement setup [19]-[21], in which $C_1$ and $C_2$ are used to block the DC or low-frequency AC power signal from entering into the measurement instrument (IA or VNA) but permit the high-frequency test signal (generated by the IA or VNA) entering into the SUT. In contrast, $L_1$ and $L_2$ are used to block the high-frequency test signal from entering into the power source but they show very low impedance characteristics for the DC or low-frequency AC power signal. Fig. 2-2 shows the simplification of Fig. 2-1 for the high-frequency test signal, where $L_1$ and $L_2$ can be regarded as open circuits, and $\boldsymbol{Z_{C1}}$ and $\boldsymbol{Z_{C2}}$ are the equivalent impedances of $C_1$ and $C_2$, respectively. Similarly, Fig. 2-3 shows the simplification of Fig. 2-1 for the DC or low-frequency AC power signal, where $C_1$ and $C_2$ can be regarded as open circuits.

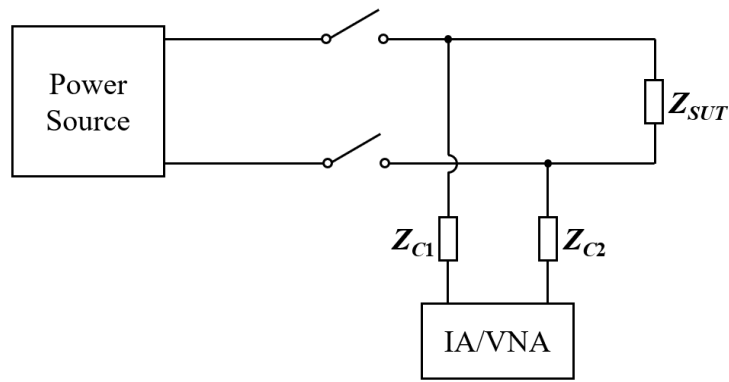

Fig. 2-2. Simplification of Fig. 2-1 for the high-frequency test signal [19].

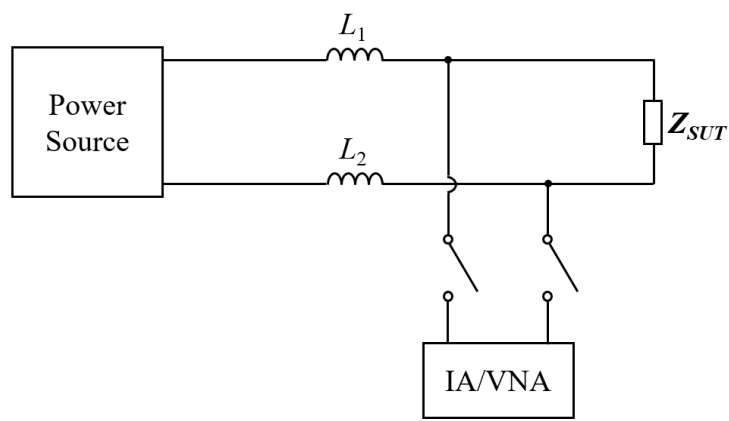

Fig. 2-3. Simplification of Fig. 2-1 for the DC or low-frequency AC power signal [19].

As observed in Fig. 2-2, if an IA is selected as the measurement instrument [19], [20], the resultant impedance of the SUT, $C_1$, and $C_2$ (namely $\boldsymbol{Z_{SUT}} + \boldsymbol{Z_{C1}} + \boldsymbol{Z_{C2}}$) can be directly measured by the IA. After the resultant impedance is obtained, $Z_{SUT}$ can be extracted by deembedding ($Z_{C1} + Z_{C2}$) from the resultant impedance. Alternatively, if a VNA is selected as the measurement instrument [21], Fig. 2-2 can be further equated as a two-port network model, as shown in Fig. 2-4.

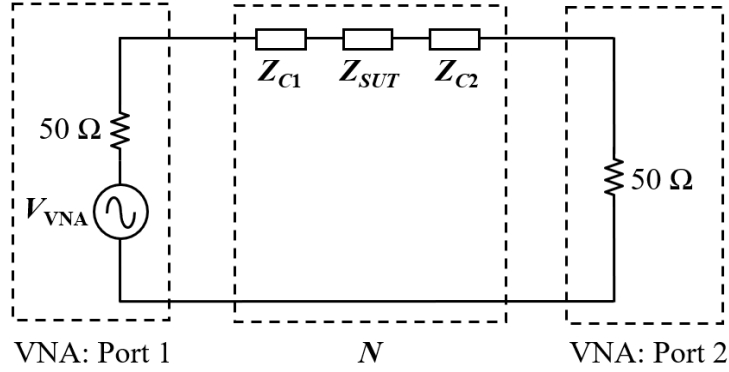

Fig. 2-4. Equivalent circuit model of Fig. 2-2 represented by a two-port network [21].

In Fig. 2-4, the SUT, $C_1$, and $C_2$ can be equated as a two-port network $N$, where $N$ can be expressed in terms of transmission (*ABCD*) parameters.

$$N = \begin{bmatrix} A & B \\ C & D \end{bmatrix} \tag{2-1}$$

Since $B = Z_{SUT} + Z_{C1} + Z_{C2}$ [32], we can directly measure the scattering parameters (*S*-parameters) of $N$ using a VNA and then convert the measured *S*-parameters into $B$ through (2-2) [32].

$$B = 50 \cdot \frac{(1 + S_{11})(1 + S_{22}) - S_{12}S_{21}}{2S_{21}} \tag{2-2}$$

After $B$ is obtained, $Z_{SUT}$ can be extracted by deembedding ($Z_{C1} + Z_{C2}$) from $B$.

## 2.2. Voltage-Current Measurement Approach

The voltage-current measurement approach is a very straightforward method, in which the impedance of the SUT ($Z_{SUT}$) is estimated via extracting the test signal voltage across the SUT and the test signal current passing through it. It should be noted that the test signal can be harmonics already presented in the SUT [24] or a signal injected by a specific device, such as a current transformer [33], a voltage transformer [33], or a SUT-connected converter [24].

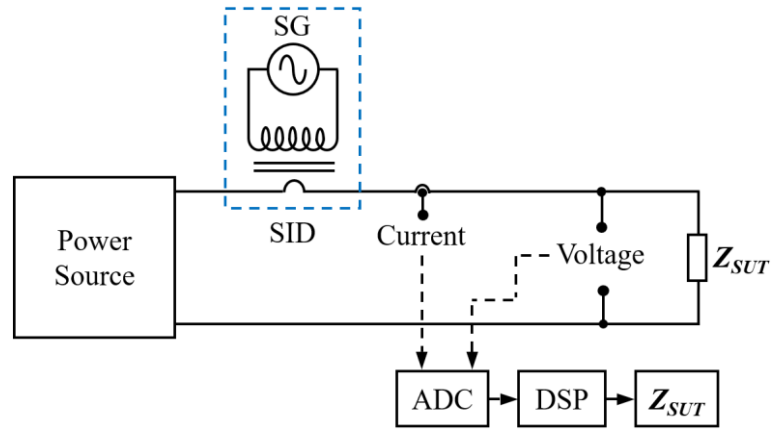

Fig. 2-5. Basic measurement setup of the voltage-current measurement approach using a current transformer as the signal injection device [33].

Fig. 2-5 shows the basic measurement setup of the voltage-current measurement approach using a current transformer as the signal injection device (SID), where a signal generator (SG) is applied to generate a test signal. The test signal is injected into the wire connected between the power source and the SUT via the current transformer. Subsequently, the induced signal voltage across the SUT and the current passing through it are extracted using a current sensor and a voltage sensor, respectively. Finally, an analog-to-digital converter (ADC) is used to sample the analog voltage and current signals, and a digital signal processing (DSP) algorithm is applied to determine $Z_{SUT}$.

## 2.3. Inductive Coupling Approach

The earliest work of the inductive coupling approach was reported in [28]. The work is further improved to simplify the online impedance extraction process with the cascaded two-port network concept [30], [31]. Fig. 2-6 shows its measurement setup for online impedance extraction of a SUT, where the SUT is powered by either a DC or a low-frequency (e.g. 50/60 Hz) AC power source through the wiring connection. The online impedance of the SUT is denoted by $Z_{SUT}$. The measurement setup consists of a VNA and two clamp-on inductive probes. One of the inductive probes serves as the injecting inductive probe (IIP) and the other is used as the receiving inductive probe (RIP). A sweep-frequency excitation (test) signal generated by port 1 of the VNA is injected into the connecting wire between the power source and the SUT via the IIP, and the response signal is received by port 2 of the VNA via the RIP, where the IIP and RIP are clamped onto the connecting wire with the clamping position denoted as $p_1$-$p_2$. $V_S$ and $Z_S$ are the equivalent source voltage and impedance of the power source, respectively. $Z_W$ is the equivalent impedance of the wiring connection except for the part being clamped by the IIP and RIP. Therefore, the resultant impedance of the power source, the wiring connection excluding the part being clamped, and the SUT is $Z_X = Z_S + Z_W + Z_{SUT}$.

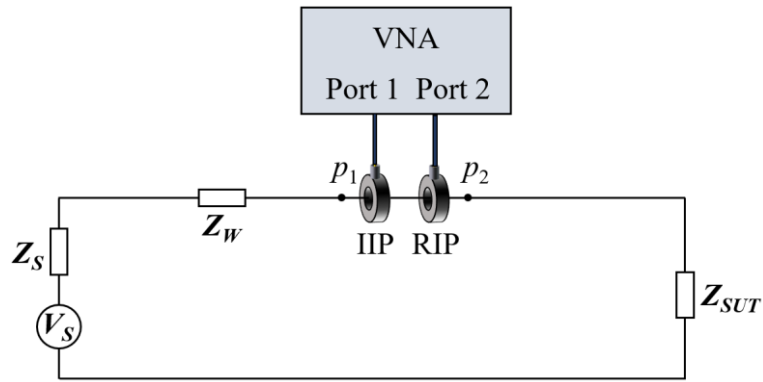

Fig. 2-6. Measurement setup of the inductive coupling approach using a VNA and two clamp-on inductive probes [30].

As shown in Fig. 2-7(a), the IIP with the wire being clamped can be regarded as a two-port network $N_I$, in which $L_{lk1}$ and $C_{p1}$ represent the leakage inductance and the parasitic capacitance between the winding of the IIP and its frame, respectively. Port 1 of $N_I$ denotes the input port of the IIP, and Port 2 of $N_I$ denotes the two ends of the wire being clamped [30]. Similarly, as shown in Fig. 2-7(b), the RIP with the wire being clamped can also be regarded as a two-port network $N_R$, in which $L_{lk2}$ and $C_{p2}$ represent the leakage inductance and the parasitic capacitance between the winding of the RIP and its frame, respectively. Port 1 of $N_R$ denotes the two ends of the wire being clamped, and Port 2 of $N_R$ denotes the output port of the RIP. Thus, based on the cascaded two-port network concept, Fig. 2-8 shows the equivalent circuit model of Fig. 2-6, where $N_X$ is the two-port network for the resultant impedance ($Z_X$) to be measured; $V_{VNA}$ represents the signal source voltage at Port 1 of the VNA.

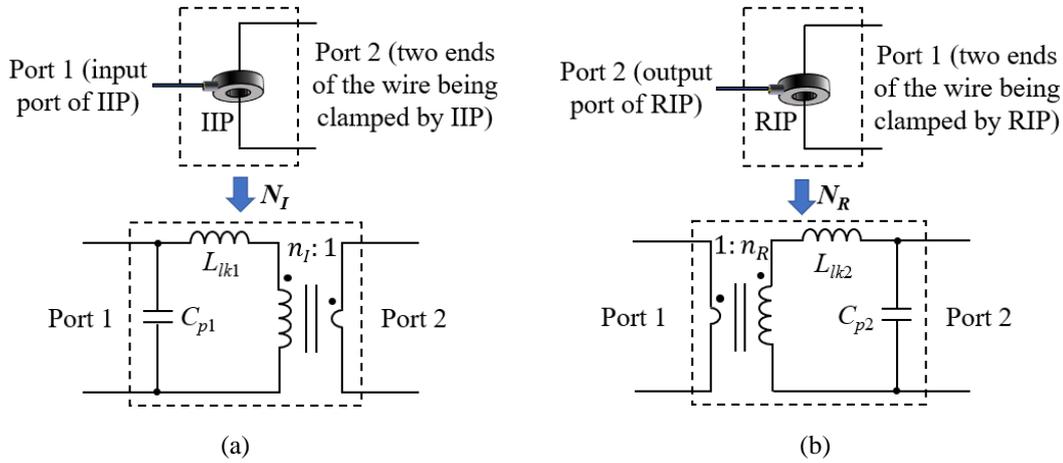

Fig. 2-7. Two-port network of (a) the IIP with the wire being clamped; (b) the RIP with the wire being clamped [30].

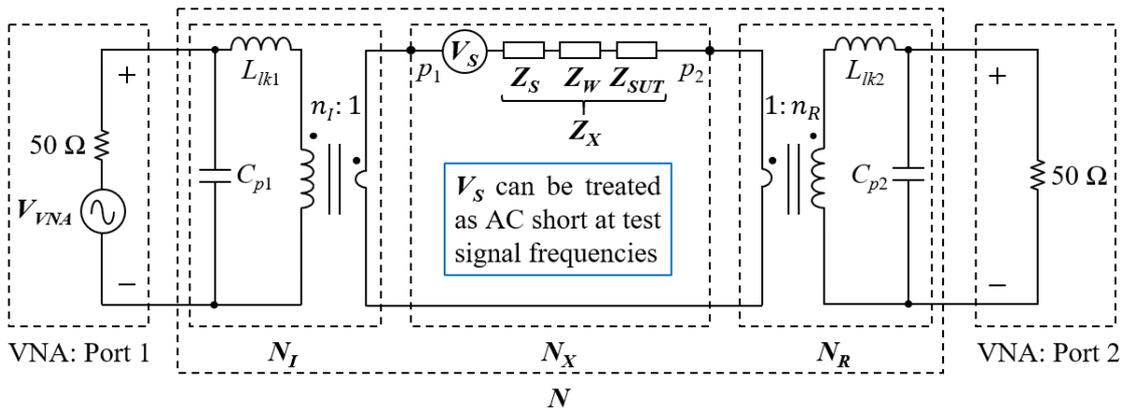

Fig. 2-8. Equivalent circuit model of Fig. 2-6 represented by cascaded two-port networks [30].

Because the frequencies of the test signal generated by Port 1 of the VNA are much higher than the frequency of the power source, $V_S$ can be treated as AC short at test signal frequencies. Besides, since the three two-port networks $N_I$, $N_X$, and $N_R$ are cascaded, the resulting two-port network $N$ can be expressed as

$$N = N_I N_X N_R \tag{2-3}$$

From (2-3), the three two-port networks can be expressed in terms of their respective transmission (*ABCD*) parameters, which can be rewritten as

$$\begin{bmatrix} A & B \\ C & D \end{bmatrix} = \begin{bmatrix} A_I & B_I \\ C_I & D_I \end{bmatrix} \begin{bmatrix} A_X & B_X \\ C_X & D_X \end{bmatrix} \begin{bmatrix} A_R & B_R \\ C_R & D_R \end{bmatrix} \tag{2-4}$$

By solving $N_X$, $Z_X$ can be obtained since $B_X = Z_X$ [32]. From (2-3), $N_X$ can be derived when $N$, $N_I$, and $N_R$ are known. Among them, $N$ can be directly measured via the VNA setup in Fig. 2-6. To obtain $N_I$ and $N_R$ prior to the measurement, a test jig shown in Fig. 2-9 is required. In Fig. 2-9, the inductive probe is clamped onto the inner conductor of the test jig and the outer cylindrical conductor serves as the common reference return path for the VNA. One end of the test jig is terminated with a short so that the inner conductor can be shorted with the outer conductor. The other end of the test jig is connected to one of the VNA's ports. To characterize $N_I$, Port 1 of the VNA connects to the input port of the IIP and Port 2 of the VNA connects to the other end of the test jig, as shown in Fig. 2-9(a). To characterize $N_R$, Port 1 of the VNA connects to the other end of the test fixture and Port 2 of the VNA connects to the output port of the RIP, as shown in Fig. 2-9(b). Thus, based on the test jig in Fig. 2-9, the *S*-parameters of $N_I$ and $N_R$ can be measured by the VNA, respectively. Finally, the *ACBD* parameters of $N_I$ and $N_R$ can be derived from the conversion of the measured *S*-parameters [32]. After $N_I$ and $N_R$ are predetermined, $Z_X$ can be obtained based on the direct measurement of $N$ via the VNA. Thus, $Z_{SUT}$ can be extracted through deembedding ($Z_S + Z_W$) from $Z_X$, where ($Z_S$

+ $Z_W$) can be predetermined by shunting the SUT with a capacitor of suitable value, which provides an AC short circuit to the test signal. In many practical situations, $Z_X$ is dominated by $Z_{SUT}$, as $Z_S$ and $Z_W$ are usually relatively small [30].

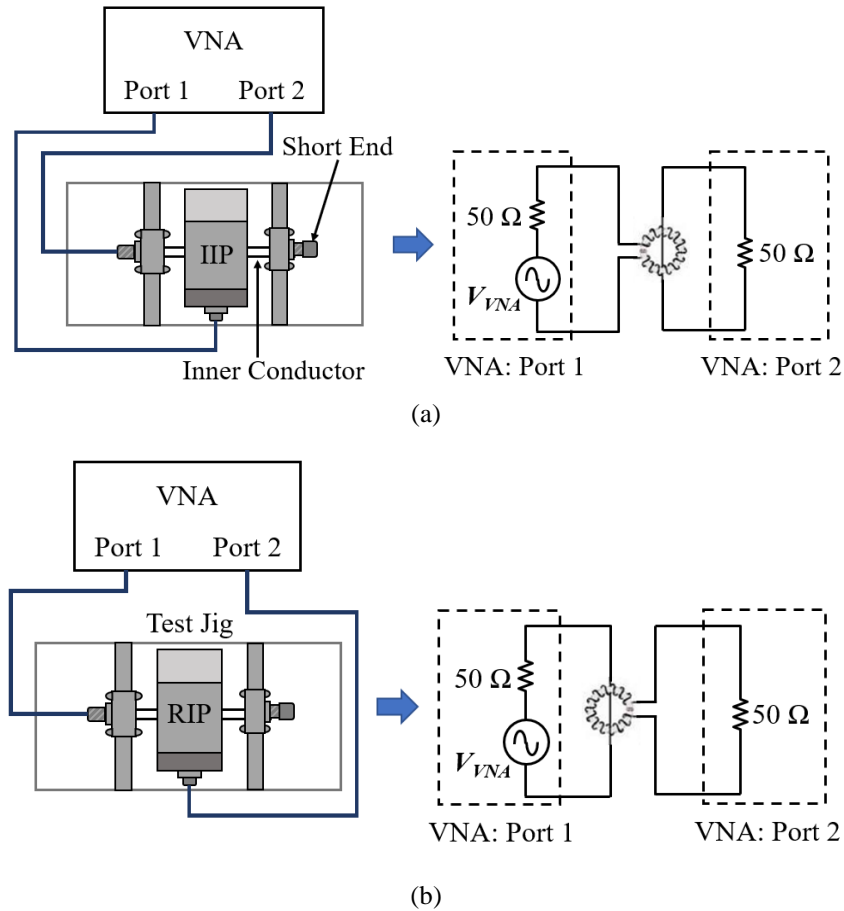

Fig. 2-9. Characterization of (a) $N_I$ and (b) $N_R$ [31].

This chapter reviews and discusses three prevalent online impedance extraction approaches; namely the capacitive coupling approach, the voltage-current measurement approach, and the inductive coupling approach. Among them, the coupling capacitor that forms part of the measurement setup of the capacitive coupling approach and the voltage sensor that forms part of the measurement setup of the voltage-current measurement approach require physical electrical connections to the energized SUT for online impedance extraction. These physical electrical connections pose electrical safety hazards to the service personnel who maintains the instruments on-site. Also, for a SUT that is energized with high operating voltage, the coupling capacitor and the voltage sensor are subject to high dielectric and thermal stresses, which require regular maintenance and replacement; and result in unnecessary downtime of the SUT. In contrast to the capacitive coupling approach and the voltage-current measurement approach, the measurement setup of the inductive coupling approach does not have any physical electrical connection to the energized SUT. Also, the clamp-on inductive probes used in the measurement setup of this approach can be easily mounted on or removed from the insulated wire that delivers the power to the SUT and therefore it simplifies the on-site implementation.

Although the inductive coupling approach has shown its superiority in online impedance extraction, there are still some issues that need to be overcome. This thesis will address these issues and describe how the inductive coupling approach can be further improved and refined. These issues are summarized as follows: the ability to extract the time-variant online impedance of electrical systems, the ability to eliminate the measurement error contributed by the probe-to-probe coupling between the inductive probes, and the ability to extract the online impedance of electrical systems where significant electrical noise and power surges are present.

# Chapter 3 Novel Measurement Setup and Time-Variant Online Impedance Extraction

With the widespread use of active devices such as power electronics devices in most electrical systems [34]-[38], the time-variant electrical parameters of the systems provide valuable inputs for the reliable and efficient operation of the systems. Among the electrical parameters, the time-variant online impedance of an operating system is useful to evaluate the system's operation status and health condition [23]. As described in Chapter 2, the inductive coupling approach has shown its superiority in online impedance extraction because of its non-contact nature and ease of on-site implementation. However, all earlier reported works of this approach have made the simplifying assumption that the online impedance of an electrical system is time-invariant for a specific time interval and its use to extract the time-variant online impedance of electrical systems has not been explored, especially those systems with fast switching rates such as power electronics systems. Therefore, this chapter introduces an extension of this approach and proposes a time-domain based measurement setup. By combining this measurement setup with a moving window discrete Fourier transform (DFT) algorithm, it can extract not only the time-invariant online impedance but also the time-variant online impedance of electrical systems.

The organization of this chapter is as follows. Section 3.1 introduces the proposed measurement setup and the principle behind time-variant online impedance extraction. In Section 3.2, the ability of this measurement setup for time-variant online impedance extraction is verified experimentally using an emulated time-variant electrical system that consists of a time-variant switching circuit. In Section 3.3, the same switching circuit is used as a test case to demonstrate the ability of online impedance monitoring in the abnormality detection of a time-variant electrical system.

## 3.1. Measurement Setup and Principle

Fig. 3-1 shows the proposed measurement setup for online impedance extraction of a SUT, where the SUT is power by either a DC or a low-frequency (e.g. 50/60 Hz) AC power source via the wiring connection. The measurement setup consists of a computer-controlled signal generation and acquisition system (SGAS), an IIP, and a RIP.

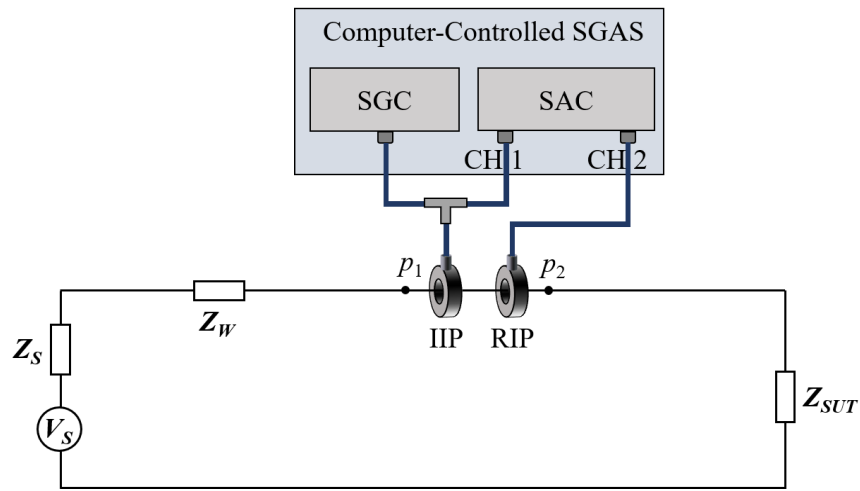

Fig. 3-1. Proposed measurement setup of the inductive coupling approach.

To extract the online impedance of the SUT ($Z_{SUT}$), a signal generation card (SGC) in the SGAS produces a sinusoidal excitation (test) signal of known frequency ($f_{sig}$). The excitation signal is injected into the connecting wire between the power source and the SUT through the IIP, and the RIP monitors the response signal, where the IIP and RIP are clamped onto the connecting wire with the clamping position denoted as $p_1$-$p_2$. Channel 1 (CH1) and channel 2 (CH2) of a signal acquisition card (SAC) in the SGAS sample the excitation signal voltage at the input port of the IIP and the response signal voltage at the output port of the RIP, respectively. $V_S$ and $Z_S$ are the equivalent source voltage and impedance of the power source, respectively. $Z_W$ is the equivalent impedance of the wiring connection except for the part being clamped by the IIP and RIP. Therefore, the resultant impedance of the power source, the wiring connection excluding the part being clamped, and the SUT is $Z_X = Z_S + Z_W + Z_{SUT}$.

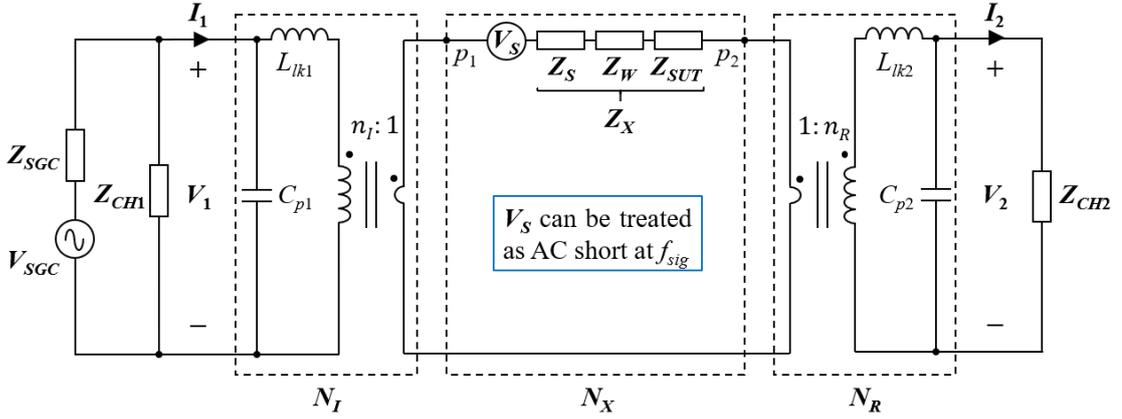

Fig. 3-2. Equivalent circuit model of Fig. 3-1 represented by cascaded two-port networks.

Based on the cascaded two-port network concept, Fig. 3-2 shows the equivalent circuit model of Fig. 3-1. $V_{SGC}$ and $Z_{SGC}$ are the equivalent source voltage and internal impedance of the SGC, respectively. $Z_{CH1}$ and $Z_{CH2}$ are the input impedances of CH1 and CH2 of the SAC, respectively. $V_1$ and $I_1$ are the excitation signal voltage and current at the input port of the IIP, respectively. $V_2$ and $I_2$ are the response signal voltage and current at the output port of the RIP, respectively. $N_I$ is the two-port network of the IIP with the wire being clamped, in which $L_{lk1}$ and $C_{p1}$ represent the leakage inductance and the parasitic capacitance between the winding of the IIP and its frame, respectively. $N_R$ is the two-port network of the RIP with the wire being clamped, in which $L_{lk2}$ and $C_{p2}$ represent the leakage inductance and the parasitic capacitance between the winding of the RIP and its frame, respectively. $N_X$ is the two-port network for the resultant impedance ($Z_X$) to be measured. Since the frequency ($f_{sig}$) of the sinusoidal test signal (generated by the SGC) is much higher than the frequency of the power source, $V_S$ can be treated as AC short at $f_{sig}$. Expressing the three two-port networks ($N_I$, $N_X$, and $N_R$) in terms of their respective *ABCD* parameters, the input port of the IIP and the output port of the RIP are related by

$$\begin{bmatrix} V_1 \\ I_1 \end{bmatrix} = \begin{bmatrix} A_I & B_I \\ C_I & D_I \end{bmatrix} \begin{bmatrix} A_X & B_X \\ C_X & D_X \end{bmatrix} \begin{bmatrix} A_R & B_R \\ C_R & D_R \end{bmatrix} \begin{bmatrix} V_2 \\ I_2 \end{bmatrix} \tag{3-1}$$

Since $I_2 = V_2/Z_{CH2}$ and $N_X$ can be expressed as [32]

$$N_X = \begin{bmatrix} 1 & Z_X \\ 0 & 1 \end{bmatrix} \tag{3-2}$$

Substituting (3-2) into (3-1), $Z_X$ can be determined by

$$Z_X = \frac{1}{A_I(C_R + D_R/Z_{CH2})} \cdot \frac{V_1}{V_2} - \frac{A_R + B_R/Z_{CH2}}{C_R + D_R/Z_{CH2}} - \frac{B_I}{A_I} \tag{3-3}$$

As mentioned earlier, the *ABCD* parameters of $N_I$ and $N_R$ can be precharacterized individually using the test jig in Fig. 2-9. $Z_{CH2}$ can be directly obtained from the datasheet of the SGC [39], [40]. Therefore, $Z_X$ can be obtained from (3-3) through the measurement of $V_1$ and $V_2$. Once $Z_X$ is obtained, $Z_{SUT}$ can be extracted by deembedding ($Z_S + Z_W$) from $Z_X$. To obtain the dynamic voltage equivalents at different time instances, the DFT is performed in a moving window [41]. The moving window size of the DFT is fixed and the details of the DFT in a moving window is given by

$$V_k(m) = \sum_{n=m-N+1}^{m} v(n)e^{-i2\pi k[n-(m-N+1)]/N} \tag{3-4}$$

where *i* is the imaginary unit. *N* is the number of sampling points of the fixed moving window size. *m* is the $(m + 1)^{th}$ discrete sample during the entire measurement process, where *m* equates to the sampling time point *t* multiply by the sampling rate $1/\Delta t$ (namely $m = t/\Delta t$). $V_k(m)$ is the dynamic voltage equivalent at time $t$ and specific frequency point $k$. $v(n)$ is the $(n + 1)^{th}$ discrete sampling voltage value in the time domain. In contrast to the DFT with a fixed window, whose range of $n$ is between 0 and $N - 1$, the range of $n$ of the DFT in a moving window is a variable with the value of $m$. Using Euler's formula, (3-4) can be expressed in the form of a trigonometric function as follows:

$$V_k(m) = \sum_{n=m-N+1}^{m} v(n) \left[ \cos\left(2\pi k \frac{n-m-1}{N}\right) - i \cdot \sin\left(2\pi k \frac{n-m-1}{N}\right) \right] \quad (3\text{-}5)$$

Since $v_1(n)$ and $v_2(n)$ can be directly extracted by CH1 and CH2 of the SAC, respectively, $V_1$ and $V_2$ at time $t$ and specific frequency point $k$ can be obtained through (3-5). Thus, $Z_X$ at time $t$ and specific frequency point $k$ can be obtained by (3-3), and subsequently, $Z_{SUT}$ can be determined through embedding as explained.

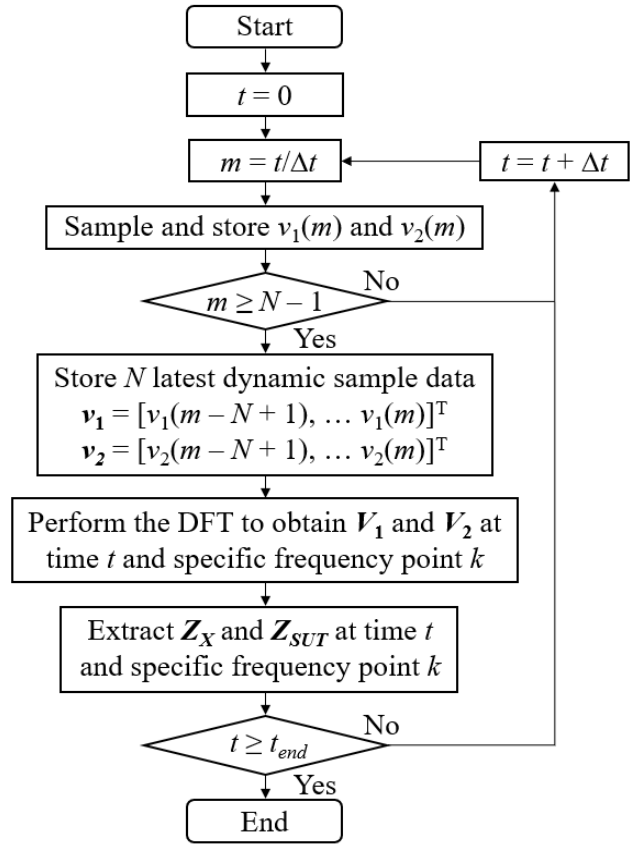

Fig. 3-3. Procedure of extracting $Z_{SUT}$ with the moving window DFT algorithm.

Fig. 3-3 describes the procedure of extracting $Z_{SUT}$ with the moving window DFT algorithm. The procedure starts with samples $v_1(m)$ and $v_2(m)$ when $t = 0$. $m$ corresponds to the $(m + 1)^{th}$ discrete sample during the entire measurement process, where $m = t/\Delta t$. When $m \geq N - 1$, the DFT is performed to obtain $V_1$ and $V_2$ at time $t$ and specific frequency point $k$. Obtaining of $V_1$ and $V_2$ at time $t$ and specific frequency point $k$ is based on the latest $N$ samples in the time domain, which is from $(m - N + 1)\Delta t$ to $m\Delta t$. Therefore, the $N$ samples are dynamically selected and are related to the sampling time point $t$. Thus, $Z_X$ at time $t$ and specific frequency point $k$ can be obtained by (3-3), and subsequently, $Z_{SUT}$ can be determined through deembedding as explained. After obtaining $Z_{SUT}$ at time $t$ and specific frequency point $k$, the next step is to judge whether the sampling time point $t$ has reached the last sampling time point ($t_{end}$). The procedure ends when $t \geq t_{end}$. Otherwise, resetting $t = t + \Delta t$, a new sampling cycle will commence and the same process is repeated until $t \geq t_{end}$.

For the moving window DFT algorithm, it is important to select the appropriate sampling rate $f_s$ (namely $1/\Delta t$) and the moving window size $N$ because they have a significant impact on the accuracy of the measurement results. Firstly, to ensure sufficient sampling in each cycle of the sinusoidal test signal, $f_s$ is suggested to set at ten times of $f_{sig}$. Therefore, $f_s = 10f_{sig}$, which provides ten sampling points in each cycle of the sinusoidal test signal. Besides, every DFT performed shall contain data of ten complete cycles of the sinusoidal test signal as per the rule of thumb to achieve good accuracy. Therefore, $N$ is set at 100. Moreover, for a time-variant SUT, the selection of $f_s$ depends on the frequency of impedance change ($f_{ch}$) of the time-variant SUT as well as the value of $N$. To achieve valid and accurate DFT results, the time interval of $N$ samples should be much smaller than $1/f_{ch}$, namely $N/f_s \ll 1/f_{ch}$. Thus, it is obtained that

$$f_s \gg N \cdot f_{\text{ch}} \tag{3-6}$$

Considering that the sampling rate ($f_s$) of a SAC in the market can be up to a few gigahertz [40], the proposed measurement setup has the ability to extract the time-variant online impedance of most electrical systems. In addition, since the window is shifted one sample at a time in the proposed moving window DFT algorithm, it shows a better ability to detect the dynamic changes of the time-variant online impedance of electrical systems as compared to the other DFT algorithms whose window is shifted multi-sample at a time.

## 3.2. Experimental Validation

For experimental validation, a National Instruments (NI) PXI platform is selected as the computer-controlled SGAS, as shown in Fig. 3-4. It consists of a signal generation card (PXI-5412), a two-channel signal acquisition card (PXI-5922), a computer interface card (PXI-8360), and a back panel (PXI-1031).

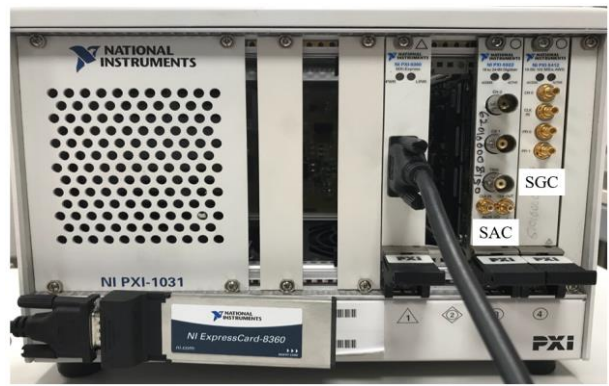

Fig. 3-4. A NI PXI platform chosen as the computer-controlled SGAS.

A computer with programmable software is interfaced with the PXI platform and performs the DFT algorithm as described in the flowchart given in Fig. 3-3. The PXI-5412 SGC generates a sinusoidal excitation signal with frequency of $f_{sig}$. The PXI-5922

SAC samples the excitation signal voltage at the input port of the IIP and the response signal voltage at the output port of the RIP, where $Z_{CH2} = 50\ \Omega\angle 0°$ [39]. The PXI-8360 card allows the user to control the PXI platform and Compact Peripheral Component Interconnect (PCI) systems from a computer using a fully transparent MXI-Express link. Since the maximum sampling rate of the PXI-5922 SAC is slightly over 5 MHz, $f_s$ is set at 5 MHz to evaluate the maximum $f_{ch}$, where the measurement setup can detect. Based on the aforementioned guideline, the signal frequency $f_{sig}$ is chosen as 500 kHz. Since this chapter focuses on the concept behind the proposed method, a SAC with a higher sampling rate should be selected for a time-variant SUT with a higher $f_{ch}$. Considering the size and electrical parameters of the SUTs in the experiment, the selected IIP and RIP are Tektronix CT1 current probes. $N_I$ and $N_R$ are precharacterized using a test jig in [30], and their respective *ABCD* parameters at 500 kHz are given as follows

$$N_I = \begin{bmatrix} 5.0025 + 0.0055i & 0.6073 + 0.8127i \\ 0.0981 - 0.0076i & 0.2129 + 0.0145i \end{bmatrix} \tag{3-7}$$

$$N_R = \begin{bmatrix} 0.2087 + 0.0081i & 0.5445 + 0.5481i \\ 0.0868 - 0.0079i & 5.0329 + 0.0082i \end{bmatrix} \tag{3-8}$$

### 3.2.1. Online Impedance Extraction of Time-Invariant Electrical System

This subsection aims to verify the ability of the proposed measurement setup to perform the online impedance extraction of a time-invariant electrical system. As shown in Fig. 3-5, a *RLC* circuit with a combination of passive components is constructed to emulate a time-invariant electrical system, where $R_1 = 4.7\ \Omega$, $R_2 = 51\ \Omega$, $L = 15\ \mu H$, $C_1 = 330$ pF, and $C_2 = 100$ nF. A 5-V DC power supply is used to supply power to the *RLC* circuit. Injecting a sinusoidal test signal with a frequency of 500 kHz into the *RLC* circuit, the online impedance of the *RLC* circuit was extracted using the proposed measurement setup. For verification, the online measurement results using the proposed measurement setup will be compared with the off-line measurement results of the same *RLC* circuit using a PM6306 RCL meter.

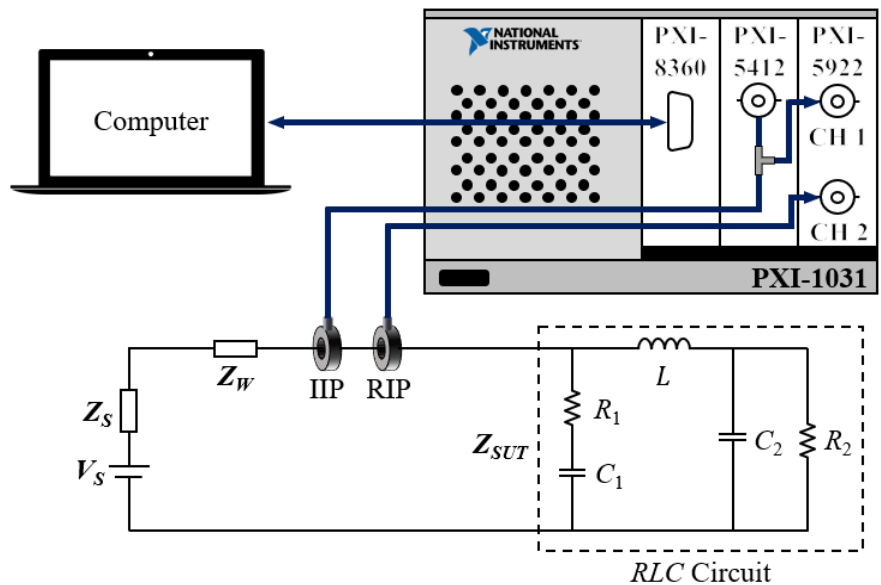

Fig. 3-5. Measurement setup for online impedance extraction of a *RLC* circuit.

Fig. 3-6 shows the magnitude and phase of the extracted impedance of the *RLC* circuit using the proposed measurement setup and the PM6306 RCL meter, where their means and standard deviations are presented in Table 3-1. Using the measurement results of the PM6306 RCL meter as the reference, the magnitude and phase measurement errors using the proposed measurement setup are calculated to be 0.2% and 1.1°, respectively. This shows that the measurement results using the proposed measurement setup are very consistent with those obtained from the PM6306 RCL meter. Therefore, it has demonstrated that the proposed measurement setup offers good measurement accuracy in the online impedance extraction of time-invariant electrical systems.

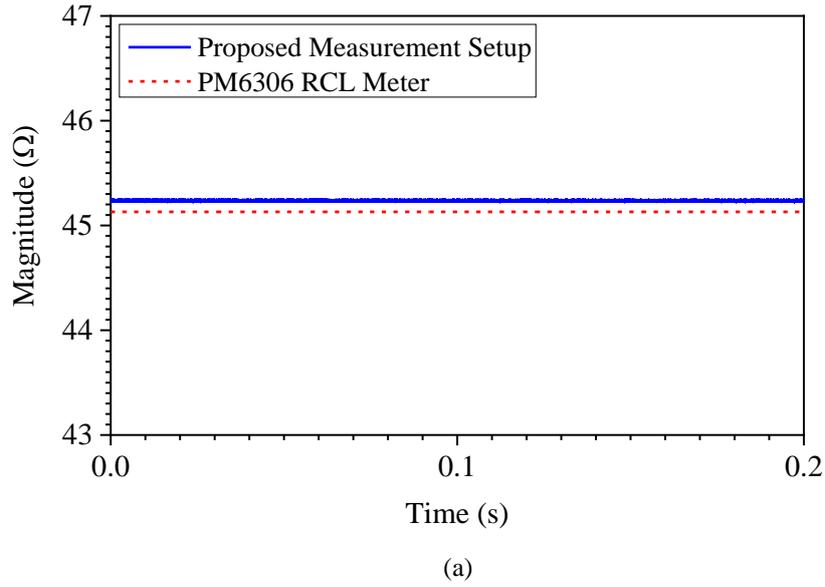

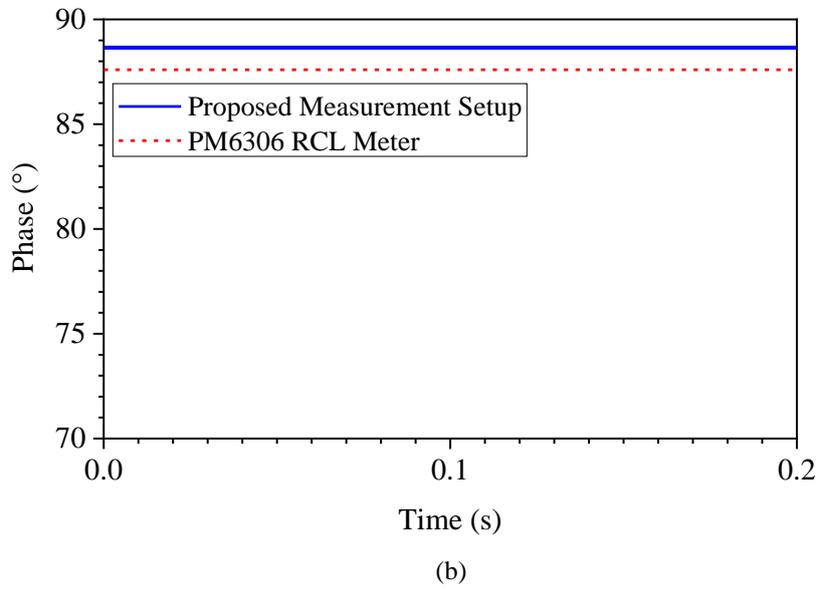

Fig. 3-6. (a) Magnitude and (b) phase of the extracted impedance of the *RLC* circuit using the proposed measurement setup and the PM6306 RCL meter.

Table 3-1. Comparison of magnitude and phase of the extracted impedance between the proposed measurement setup and the PM6306 RCL meter.

|  | Proposed Measurement Setup | | PM6306 RCL Meter | |
| --- | --- | --- | --- | --- |
|  | Mean | Standard Deviation | Mean | Standard Deviation |
| Magnitude ($\Omega$) | 45.24 | 0.0025 | 45.13 | 0.0040 |
| Phase (°) | 88.7 | 0.0032 | 87.6 | 0.0020 |

### 3.2.2. Online Impedance Extraction of Time-Variant Electrical System

This subsection is to demonstrate the ability of the proposed measurement setup for the online impedance extraction of a time-variant electrical system. To emulate a time-variant electrical system, a switching circuit with a power MOSFET shown in Fig. 3-7 is applied, where $R_1 = 1$ Ω, $R_2 = 51$ Ω, $L = 1$ mH, $C_1 = 4.7$ μF, and $C_2 = 47$ μF. The power MOSFET has a reverse diode and interelectrode capacitances. A gate drive signal with a period $T$ and a duty cycle $D = 0.5$ is connected to the gate of the power MOSFET to control its switching frequency, $f_{sw} = 1/T$. The equivalent impedance of the power MOSFET at the specified frequency is represented as $Z_{MOS}$ and its value depends on whether the power MOSFET is in the "on" or "off" state. $D_1$ is a diode with opposite switching states to the power MOSFET. A 1-V DC power supply is used to supply power to the switching circuit.

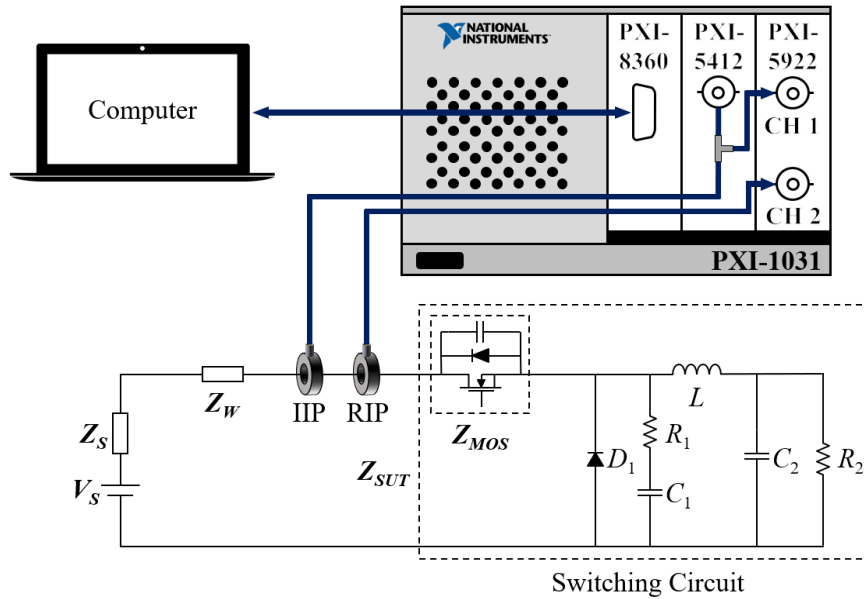

Fig. 3-7. Measurement setup for online impedance extraction of a switching circuit.

Thus, the switching circuit can be equated as two separate time-invariant circuits when the power MOSFET is "on" and "off", respectively. Considering that the impedance in steady-state of each switching cycle is the same with the impedance of the two separate time-invariant electrical circuits when the power MOSFET is always "on" or "off", the measurement result when the power MOSFET is always "on" or "off" can be taken as the reference impedance to verify the performance of the proposed measurement setup when the switching circuit is constantly switching "on" and "off". Since the proposed measurement setup has been verified to measure the online impedance of a time-invariant electrical system in Subsection 3.2.1, it is feasible to use the measured impedance when the power MOSFET is always "on" or "off" as the reference. In addition, for the switching circuit, $f_{ch}$ is the same as $f_{sw}$. As mentioned earlier, accurate impedance measurement of a time-variant electrical system needs to satisfy the requirement of (3-6). Considering that $N$ and $f_s$ have been selected as 100 and 5 MHz respectively, it is found that $f_{sw} \ll 50$ kHz. Therefore, experiments are conducted to verify it.

Fig. 3-8 to Fig. 3-10 show the extracted online impedances using the proposed measurement setup when $f_{sw}$ is 1 kHz, 10 kHz, and 50 kHz, respectively. The impedance when the power MOSFET is always "on" or "off" is indicated as the reference for comparison purposes. As explained, $f_{ch}$ should be much smaller than $f_s/N$ (namely 50 kHz) to achieve valid and accurate DFT results. Therefore, $f_{sw} = 1$ kHz yields a better result as compared to the case when $f_{sw} = 10$ kHz. By increasing $f_{sw}$ to 50 kHz, the impedance changes can no longer be captured because $f_{ch} = f_s/N$, causing the DFT algorithm unable to obtain the dynamic impedance changes of the switching circuit. Therefore, proper choices of $f_s$ and $N$ are critical to the accuracy of the measurement results. It should be explained that if the frequency of impedance change ($f_{ch}$) of a time-variant electrical system is larger than 50 kHz, a SAC with a higher sampling rate ($f_s$) can be used in this case.

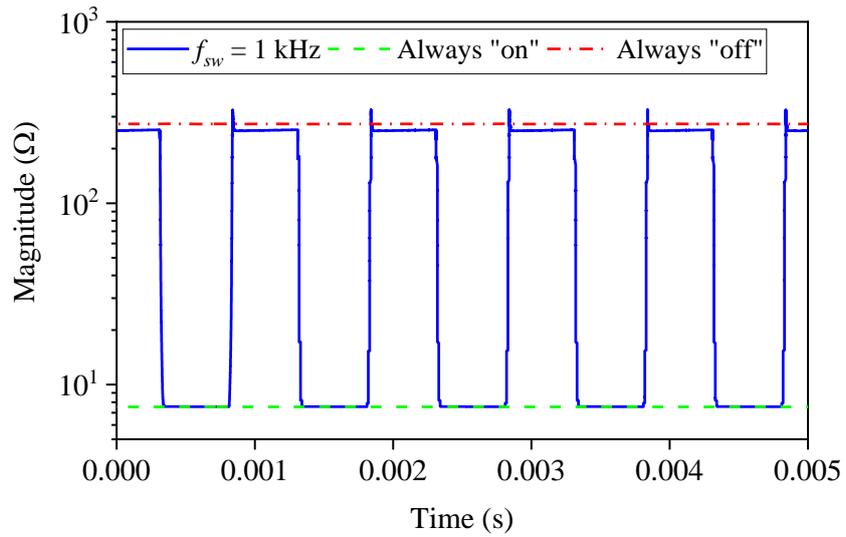

(a)

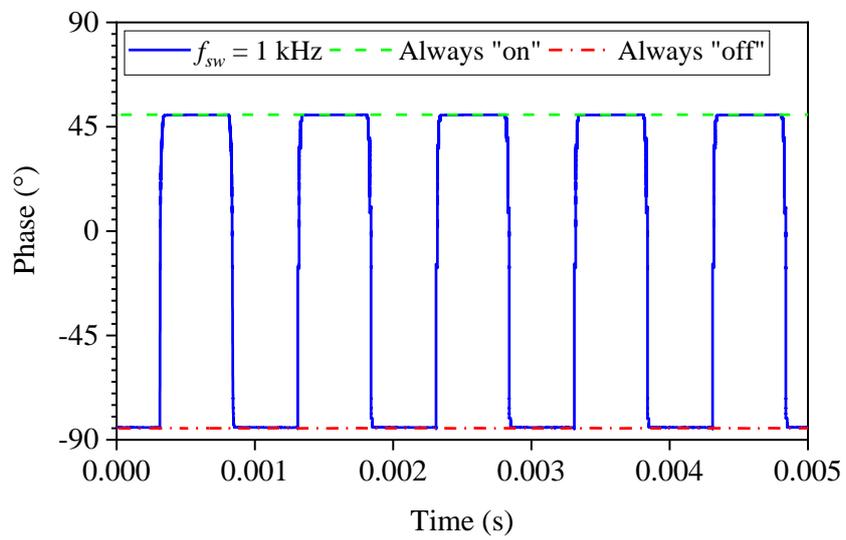

(b)

Fig. 3-8. (a) Magnitude and (b) phase of the extracted impedance of the switching circuit when $f_{sw}$ is 1 kHz.

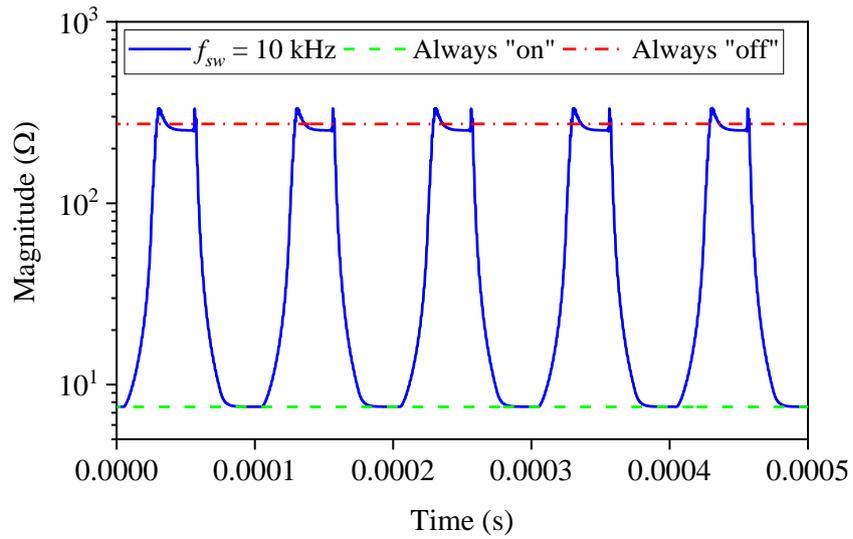

(a)

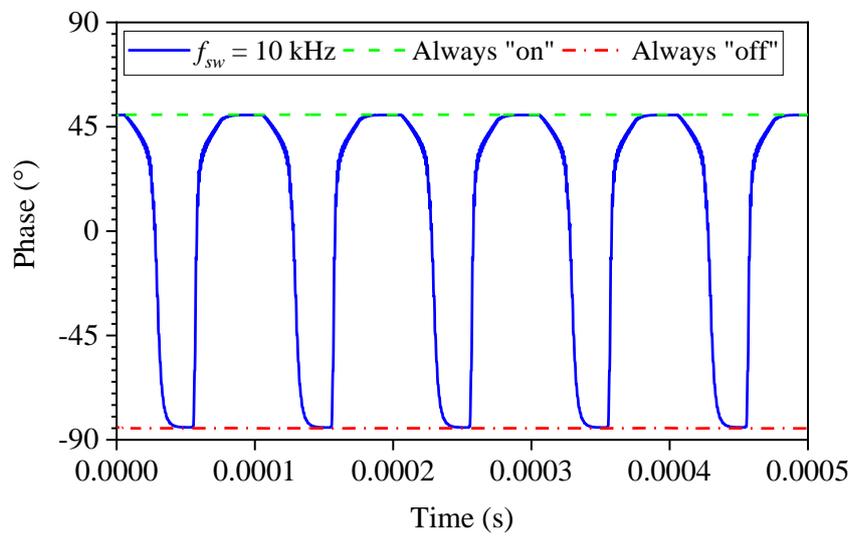

(b)

Fig. 3-9. (a) Magnitude and (b) phase of the extracted impedance of the switching circuit when $f_{sw}$ is 10 kHz.

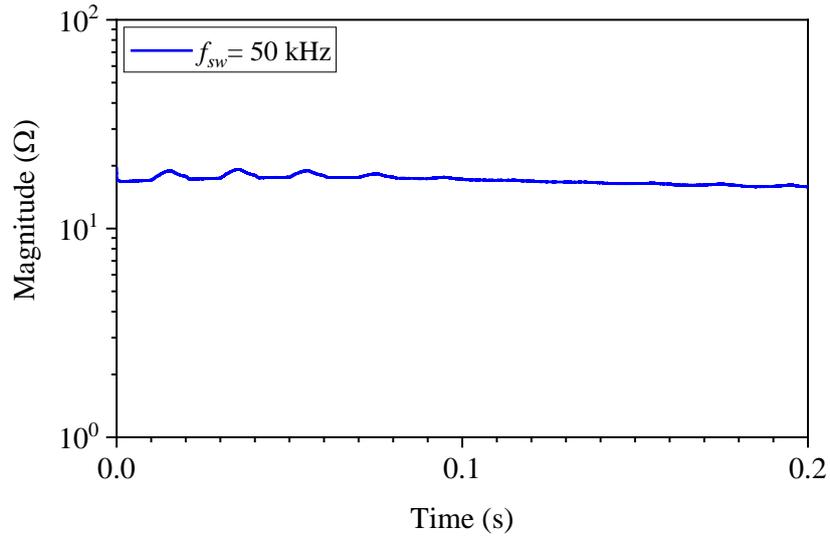

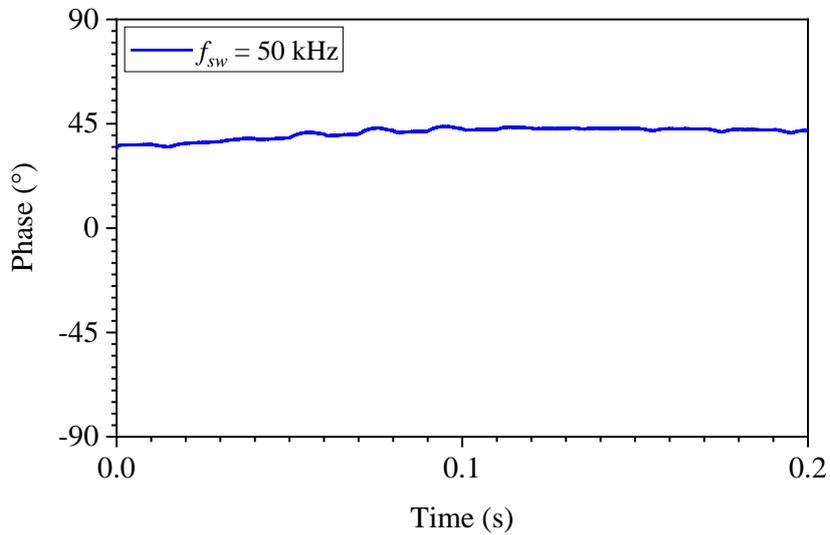

Fig. 3-10. (a) Magnitude and (b) phase of the extracted impedance of the switching circuit when $f_{sw}$ is 50 kHz.

Tables 3-2 and 3-3 show the steady-state magnitude and phase of the extracted impedance from Figs. 3-8 and 3-9, respectively. The magnitude and phase deviations are rather small when $f_{sw}$ is set at 1 kHz and 10 kHz. Using the measured results when the power MOSFET is always in the "on" or "off" state as the reference, the measurement errors of the magnitude and phase of the steady-state online impedance using the proposed measurement setup are 7.7% and 0.42°, respectively.

Table 3-2. The steady-state magnitude of the impedance from Figs. 3-8 and 3-9.

| (Ω) | $f_{sw}$ = 1 kHz, 10 kHz | | Reference Values | |
|---|---|---|---|---|
| | "on" | "off" | Always "on" | Always "off" |
| 1 kHz | 7.56 | 252.32 | 7.55 | 273.50 |
| 10 kHz | 7.58 | 265.95 | | |

Table 3-3. The steady-state phase of the impedance from Figs. 3-8 and 3-9.

| (°) | $f_{sw}$ = 1 kHz, 10 kHz | | Reference Values | |
|---|---|---|---|---|
| | "on" | "off" | Always "on" | Always "off" |
| 1 kHz | 50.12 | -84.64 | 50.22 | -85.00 |
| 10 kHz | 50.09 | -84.58 | | |

It should be noted that the extracted impedance at the switching edge has an oscillatory behaviour due to the interelectrode capacitances of the power MOSFET and the circuit loop inductance [42], [43]. From Figs. 3-8 and 3-9, the maximum transient impedances at the switching edge extracted using the proposed measurement setup when $f_{sw}$ = 1 kHz and 10 kHz are 326.81 Ω and 333.61 Ω, respectively. Using the measured time-domain peak transient voltage and current, the impedances computed are 357.16 Ω at $f_{sw}$ = 1 kHz and 378.13 Ω at $f_{sw}$ = 10 kHz. Taking the computed impedances as the reference, the deviations of the extracted maximum transient impedance when $f_{sw}$ = 1 kHz and 10 kHz are 8.5% and 11.8%, respectively.

## 3.3. Online Abnormality Detection of Time-Variant Electrical System

This section illustrates the ability of online impedance monitoring in the abnormality detection of a time-variant electrical system. The same switching circuit in Fig. 3-7 is used as a test case, where $f_s$ = 5 MHz, $f_{sig}$ = 500 kHz, $f_{sw}$ = 1 kHz, and $D$ = 0.5. First of all, the switching circuit starts operating at a normal temperature. Next, a heating element is applied on the power MOSFET to emulate its operation at a high temperature, and finally, the resulting online impedance of the switching circuit is re-measured.

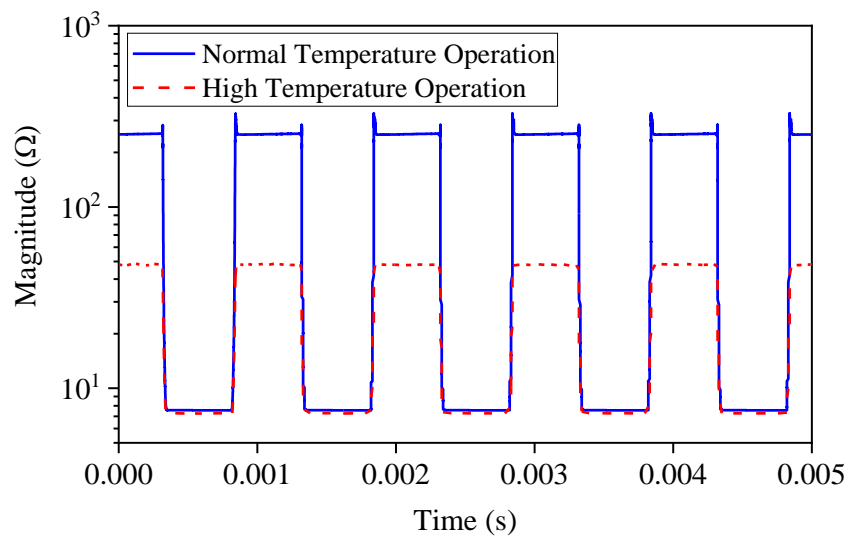

(a)

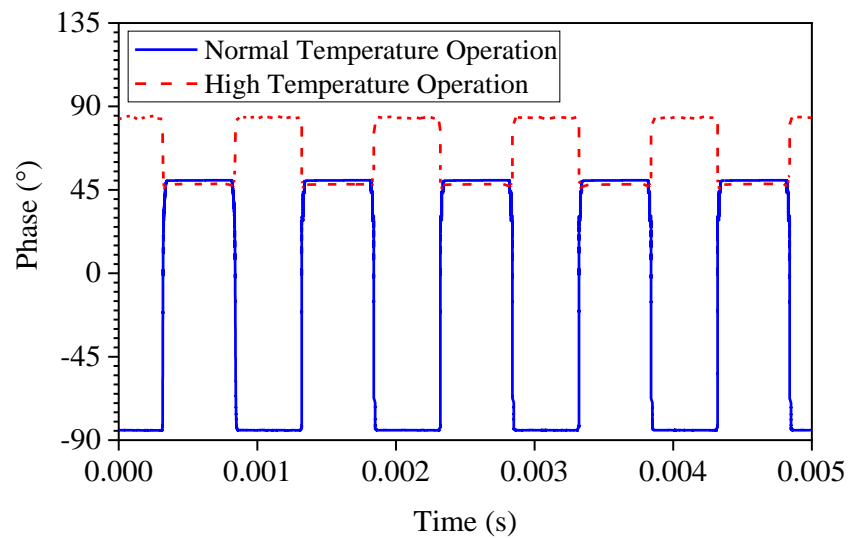

(b)

Fig. 3-11. (a) Magnitude and (b) phase of the extracted online impedance of the switching circuit when the power MOSFET operates at normal temperature and high temperature conditions.

The measurement results are shown in Fig. 3-11. By emulating the power MOSFET operated at the high temperature condition, the magnitude and phase values of the extracted online impedance when the power MOSFET is in the "off" state deviate from their original values operated at the normal temperature condition. This deviation has shown the ability of online impedance monitoring to detect the abnormal behavior of the power MOSFET due to overheating. The detection for the early abnormal behavior allows necessary remedial action to be taken before the electrical system deteriorates further to a catastrophic failure.

In this chapter, a novel measurement setup of the inductive coupling approach is proposed, which consists of a computer-controlled SGAS, an IIP, and a RIP. By using the cascaded two-port network concept and a moving window DFT algorithm, the proposed measurement setup can extract not only the time-invariant online impedance but also the time-variant online impedance of electrical systems. For verification, a switching circuit was chosen as a test case to validate the ability of this measurement setup to extract the online impedance of a time-variant electrical system. Such ability is useful to monitor the time-variant electrical system for abnormal behaviors that serve as the early signs of failures.

# Chapter 4 Eliminating the Effect of Probe-to-Probe Coupling on Measurement Accuracy

As mentioned in Chapter 2, the online impedance serves as a key parameter for assessing the operating status and health condition of many critical electrical systems. In Chapter 3, a novel measurement setup of the inductive coupling approach is proposed, which consists of a computer-controlled SGAS, an IIP, and a RIP. Combined with a moving window DFT algorithm, the proposed measurement setup can extract not only the time-invariant online impedance but also the time-variant online impedance of electrical systems. Before performing online impedance extraction of any electrical system, the IIP and RIP must be precharacterized using a VNA and a dedicated test jig, as shown in Fig. 2-9. Although the inductive coupling approach has been improved to handle time-variant online impedance extraction, the possible impact of the probe-to-probe coupling between the IIP and RIP on the accuracy of the extracted online impedance has not been investigated, especially when both probes are closely placed due to space constraints in some practical applications. Therefore, this chapter proposes a three-term calibration technique to eliminate potential errors caused by the probe-to-probe coupling in the extracted online impedance with the objective to achieve good measurement accuracy.

This chapter is organized as follows. Section 4.1 describes a comprehensive equivalent circuit model of the measurement setup based on a three-port network concept, in which the effect of the probe-to-probe coupling can be taken into account in the model. With the three-port network equivalent circuit model, Section 4.2 introduces the three-term calibration technique to eliminate the influence of the probe-to-probe coupling on the accuracy of the extracted online impedance. In Section 4.3, experiments to investigate the characteristics of the probe-to-probe coupling are carried out and the ability of the three-term calibration technique to eliminate the measurement error contributed by such coupling is demonstrated.

## 4.1. Three-Port Network Equivalent Circuit Model

In the previous inductively coupled online impedance extraction analysis, the probe-to-probe coupling between the IIP and RIP has not been considered. In some practical applications with space constraints, the IIP and RIP must be placed very close to each other and the potential measurement error caused by the probe-to-probe coupling could be significant and cannot be neglected. To account for the influence of the probe-to-probe coupling, based on the three-port network concept [32], a comprehensive equivalent circuit model of the proposed measurement setup in Fig. 3-1 is introduced, as shown in Fig. 4-1.

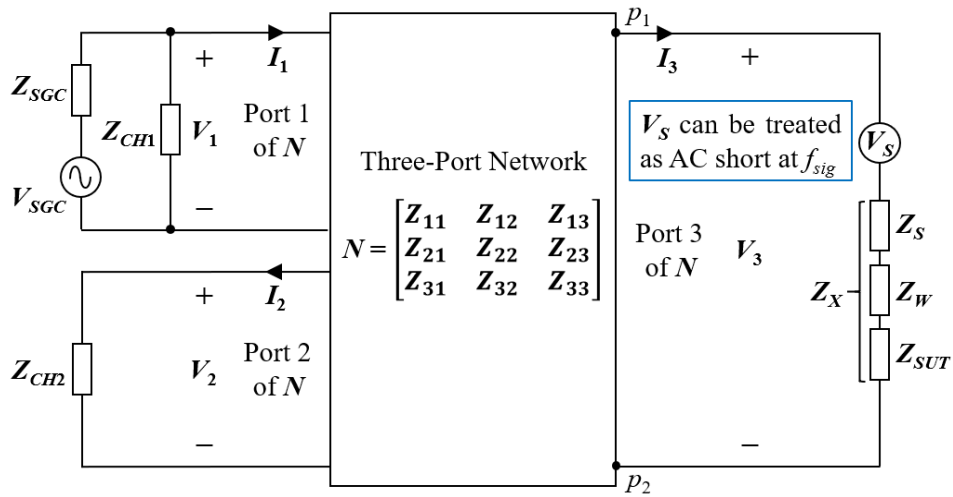

Fig. 4-1. Equivalent circuit model of Fig. 3-1 represented by a three-port network.

In Fig. 4-1, $V_{SGC}$ and $Z_{SGC}$ are the equivalent source voltage and internal impedance of the SGC, respectively. $Z_{CH1}$ and $Z_{CH2}$ are the input impedances of CH1 and CH2 of the SAC, respectively. $V_1$ and $I_1$ are the excitation signal voltage and current at the input port of the IIP, respectively; where $V_1$ can be directly extracted by CH1 of the SAC. $V_2$ and $I_2$ are the response signal voltage and current at the output port of the RIP, respectively; where $V_2$ can be directly extracted by CH2 of the SAC and $I_2 = V_2/Z_{CH2}$. $V_3$ is the induced signal voltage between $p_1$ and $p_2$, and $I_3$ is the induced signal current passing through $Z_X$ to be measured. As mentioned in Section 3.1, since the frequency ($f_{sig}$) of the sinusoidal test signal generated by the SGC is much higher than the frequency of the power source, $V_S$ can be treated as AC short at $f_{sig}$. Thus, $V_3$ and $I_3$ can be related by $I_3 = V_3/Z_X$. Therefore, the coupling between the inductive probe (IIP or RIP) and the respective wire being clamped, defined as the "probe-to-wire coupling", as well as the probe-to-probe coupling, can be fully described with a three-port network given by

$$N = \begin{bmatrix} Z_{11} & Z_{12} & Z_{13} \\ Z_{21} & Z_{22} & Z_{23} \\ Z_{31} & Z_{32} & Z_{33} \end{bmatrix} \quad (4\text{-}1)$$

where $Z_{ij}$ ($i$ and $j = 1, 2$ or $3$) is the impedance parameter of $N$. Port 1 of $N$ is the input port of the IIP; Port 2 of $N$ is the output port of the RIP; Port 3 of $N$ is the input port between $p_1$ and $p_2$. Therefore, the respective ports' voltages and currents are related by

$$\begin{bmatrix} V_1 \\ V_2 \\ V_3 \end{bmatrix} = \begin{bmatrix} Z_{11} & Z_{12} & Z_{13} \\ Z_{21} & Z_{22} & Z_{23} \\ Z_{31} & Z_{32} & Z_{33} \end{bmatrix} \begin{bmatrix} I_1 \\ I_2 \\ I_3 \end{bmatrix} \quad (4\text{-}2)$$

Since $I_2 = V_2/Z_{CH2}$ and $I_3 = V_3/Z_X$, (4-2) can be re-written as

$$\begin{bmatrix} V_1 \\ V_2 \\ V_3 \end{bmatrix} = \begin{bmatrix} Z_{11} & Z_{12} & Z_{13} \\ Z_{21} & Z_{22} & Z_{23} \\ Z_{31} & Z_{32} & Z_{33} \end{bmatrix} \begin{bmatrix} I_1 \\ V_2/Z_{CH2} \\ V_3/Z_X \end{bmatrix} \quad (4\text{-}3)$$

Dividing $V_2$ at both sides of (4-3), it becomes

$$\begin{bmatrix} V_1/V_2 \\ 1 \\ V_3/V_2 \end{bmatrix} = \begin{bmatrix} Z_{11} & Z_{12} & Z_{13} \\ Z_{21} & Z_{22} & Z_{23} \\ Z_{31} & Z_{32} & Z_{33} \end{bmatrix} \begin{bmatrix} I_1/V_2 \\ 1/Z_{CH2} \\ (V_3/V_2)/Z_X \end{bmatrix} \quad (4\text{-}4)$$

By solving (4-4), $V_1/V_2$, $V_3/V_2$, and $I_1/V_2$ can be represented in terms of $Z_{ij}$, $Z_{CH2}$, and $Z_X$; where $V_1/V_2$ can be expressed as

$$\frac{V_1}{V_2} = \frac{a_1 \cdot Z_X + a_2}{Z_X + a_3} \quad (4\text{-}5)$$

where

$$a_1 = \frac{Z_{11}}{Z_{21}} \cdot \left(1 - \frac{Z_{22}}{Z_{CH2}}\right) + \frac{Z_{12}}{Z_{CH2}} \quad (4\text{-}6)$$

$$a_2 = \left(Z_{13} - \frac{Z_{11}Z_{23}}{Z_{21}}\right) \cdot \left[\frac{Z_{31}}{Z_{21}} \cdot \left(1 - \frac{Z_{22}}{Z_{CH2}}\right) + \frac{Z_{32}}{Z_{CH2}}\right] \\ - \left(Z_{33} - \frac{Z_{31}Z_{23}}{Z_{21}}\right) \cdot \left[\frac{Z_{11}}{Z_{21}} \cdot \left(1 - \frac{Z_{22}}{Z_{CH2}}\right) + \frac{Z_{12}}{Z_{CH2}}\right] \quad (4\text{-}7)$$

$$a_3 = \frac{Z_{31}Z_{23}}{Z_{21}} - Z_{33} \quad (4\text{-}8)$$

The detailed derivation of (4-5) is given as follows. From (4-4), we obtain

$$\frac{V_1}{V_2} = Z_{11} \cdot \frac{I_1}{V_2} + \frac{Z_{12}}{Z_{CH2}} + \frac{Z_{13}}{Z_X} \cdot \frac{V_3}{V_2} \quad (4\text{-}9)$$

$$1 = Z_{21} \cdot \frac{I_1}{V_2} + \frac{Z_{22}}{Z_{CH2}} + \frac{Z_{23}}{Z_X} \cdot \frac{V_3}{V_2} \quad (4\text{-}10)$$

$$\frac{V_3}{V_2} = Z_{31} \cdot \frac{I_1}{V_2} + \frac{Z_{32}}{Z_{CH2}} + \frac{Z_{33}}{Z_X} \cdot \frac{V_3}{V_2} \quad (4\text{-}11)$$

From (4-10), $I_1/V_2$ can be expressed by

$$\frac{I_1}{V_2} = \frac{1}{Z_{21}}\left(1 - \frac{Z_{22}}{Z_{CH2}} - \frac{Z_{23}}{Z_X} \cdot \frac{V_3}{V_2}\right) \tag{4-12}$$

Substituting (4-12) into (4-11), $V_3/V_2$ has the expression given by

$$\frac{V_3}{V_2} = \frac{\frac{Z_{31}}{Z_{21}}\left(1 - \frac{Z_{22}}{Z_{CH2}}\right) + \frac{Z_{32}}{Z_{CH2}}}{1 + \frac{Z_{31}Z_{23}}{Z_{21}Z_X} - \frac{Z_{33}}{Z_X}} \tag{4-13}$$

Substituting (4-12) and (4-13) into (4-9), the final expression of $V_1/V_2$ is obtained and given by (4-5). Finally, $Z_X$ can be denoted as a function of $V_1/V_2$ as follows

$$Z_X = \frac{a_3 \cdot \frac{V_1}{V_2} - a_2}{-\frac{V_1}{V_2} + a_1} \tag{4-14}$$

## 4.2. Three-Term Calibration for Error Elimination

In this section, a three-term calibration technique is introduced and elaborated. By performing this calibration, the probe-to-probe coupling between the IIP and RIP can be evaluated and its effect can be eliminated. From (4-14), $Z_X$ can be determined by $V_1/V_2$ once $a_1$, $a_2$, and $a_3$ are known, where $V_1$ and $V_2$ can be directly extracted through the measurement setup in Fig. 3-1. By observing (4-6), (4-7), and (4-8), for a given SAC with a specific $Z_{CH2}$, the values of $a_1$, $a_2$, and $a_3$ depend solely on the impedance parameters of $N$ (i.e. $Z_{ij}$), which are determined by the physical properties of the inductive probes and the test signal frequency $f_{sig}$. Besides, the spacing between the inductive probes, denoted as $d$, will also affect $Z_{ij}$ because the probe-to-probe coupling varies with $d$. To determine the values of $a_1$, $a_2$, and $a_3$, $Z_X$ can be replaced with three known but different conditions before the actual online measurement. Since the usual industrial calibration standards adopt open, short, and 50-Ω resistive load for the three conditions, we adopt the same calibration standards. Based on (4-14), when Port 3 of $N$ in Fig. 4-1 is left opened ($Z_X = \infty$ and this condition is represented as $R_\infty$), then

$$a_1 = \left.\frac{V_1}{V_2}\right|_{R_\infty} \tag{4-15}$$

When Port 3 of $N$ in Fig. 4-1 is shorted ($Z_X = 0\ \Omega$ and this condition is represented as $R_0$), it leads to

$$a_2 = a_3 \cdot \left.\frac{V_1}{V_2}\right|_{R_0} \tag{4-16}$$

When Port 3 of $N$ in Fig. 4-1 is terminated with a 50-$\Omega$ resistive load ($Z_X = 50\ \Omega$ and this condition is represented as $R_{50}$), one can obtain

$$R_{50} = \frac{a_3 \cdot \left.\frac{V_1}{V_2}\right|_{R_{50}} - a_2}{-\left.\frac{V_1}{V_2}\right|_{R_{50}} + a_1} \tag{4-17}$$

In (4-15), (4-16), and (4-17), $V_1/V_2|_t$ is the ratio of the voltages measured at the input port of the IIP and the output port of the RIP, where "$t$" represents the termination, either $R_\infty$, $R_0$, or $R_{50}$. Since $V_1/V_2|_t$ can be directly extracted through the measurement setup, from (4-15), (4-16), and (4-17), $a_2$ and $a_3$ can be determined by (4-18) and (4-19), respectively.

$$a_2 = R_{50} \cdot \left.\frac{V_1}{V_2}\right|_{R_0} \cdot \frac{\left(\left.\frac{V_1}{V_2}\right|_{R_\infty} - \left.\frac{V_1}{V_2}\right|_{R_{50}}\right)}{\left(\left.\frac{V_1}{V_2}\right|_{R_{50}} - \left.\frac{V_1}{V_2}\right|_{R_0}\right)} \tag{4-18}$$

$$a_3 = R_{50} \cdot \frac{\left(\frac{V_1}{V_2}\Big|_{R_\infty} - \frac{V_1}{V_2}\Big|_{R_{50}}\right)}{\left(\frac{V_1}{V_2}\Big|_{R_{50}} - \frac{V_1}{V_2}\Big|_{R_0}\right)} \tag{4-19}$$

Substituting (4-15), (4-18), and (4-19) into (4-14), $Z_X$ can be determined as follows

$$Z_X = R_{50} \cdot \frac{\left(\frac{V_1}{V_2}\Big|_{R_\infty} - \frac{V_1}{V_2}\Big|_{R_{50}}\right)\left(\frac{V_1}{V_2} - \frac{V_1}{V_2}\Big|_{R_0}\right)}{\left(\frac{V_1}{V_2}\Big|_{R_{50}} - \frac{V_1}{V_2}\Big|_{R_0}\right)\left(\frac{V_1}{V_2}\Big|_{R_\infty} - \frac{V_1}{V_2}\right)} \tag{4-20}$$

It should be noted that the prerequisite of obtaining $Z_X$ from (4-20) is that the signals $V_1|_t$ and $V_2|_t$ are measurable. Since $V_1|_t$ is the signal of the injecting side, it is well above the measurement system's noise floor with good signal-to-noise ratio (SNR). Under the condition that $Z_X = 0$ or $R_{50}$, $V_2|_{R_0}$ and $V_2|_{R_{50}}$ are usually well above the measurement system's noise floor with good SNR. Under the condition that $Z_X = \infty$, if the probe-to-probe coupling is strong, $V_2|_{R_\infty}$ will still be measurable. Thus, $Z_X$ can be extracted from (4-20). If the probe-to-probe coupling is weak, $V_2|_{R_\infty}$ will be lower than the measurement system's noise floor and therefore is very weak to affect the measurement. Thus, the effect of the probe-to-probe coupling on the accuracy of the extracted online impedance is negligible, and (3-3) will be accurate to extract $Z_X$. To simplify the expression of (3-3), it is rewritten as

$$Z_X = k \cdot \frac{V_1}{V_2} + b \tag{4-21}$$

where

$$k = \frac{1}{A_I(C_R + D_R/Z_{CH2})} \tag{4-22}$$

$$b = -\frac{A_R + B_R/Z_{CH2}}{C_R + D_R/Z_{CH2}} - \frac{B_I}{A_I} \qquad (4\text{-}23)$$

In addition to the characterization technique shown in Fig. 2-9, the proposed calibration technique can also be used to determine the values of $k$ and $b$. When $Z_X$ is replaced with a short ($Z_X = 0\ \Omega$), then (4-21) becomes

$$0 = k \cdot \left.\frac{V_1}{V_2}\right|_{R_0} + b \qquad (4\text{-}24)$$

Similarly, when $Z_X$ is replaced with a 50-$\Omega$ resistive load ($Z_X = R_{50}$), then (4-21) will be

$$R_{50} = k \cdot \left.\frac{V_1}{V_2}\right|_{R_{50}} + b \qquad (4\text{-}25)$$

By solving (4-24) and (4-25), $k$ and $b$ can be obtained by

$$k = \frac{R_{50}}{\left(\left.\frac{V_1}{V_2}\right|_{R_{50}} - \left.\frac{V_1}{V_2}\right|_{R_0}\right)} \qquad (4\text{-}26)$$

$$b = \frac{-R_{50} \cdot \left.\frac{V_1}{V_2}\right|_{R_0}}{\left(\left.\frac{V_1}{V_2}\right|_{R_{50}} - \left.\frac{V_1}{V_2}\right|_{R_0}\right)} \qquad (4\text{-}27)$$

Substituting (4-26) and (4-27) into (4-21), $Z_X$ can finally be determined by

$$Z_X = \frac{R_{50} \cdot \left(\frac{V_1}{V_2} - \left.\frac{V_1}{V_2}\right|_{R_0}\right)}{\left(\left.\frac{V_1}{V_2}\right|_{R_{50}} - \left.\frac{V_1}{V_2}\right|_{R_0}\right)} \qquad (4\text{-}28)$$

Fig. 4-2 summarizes the measurement process according to the proposed calibration

technique and is briefly described as follows:

- If $V_2|_{R_\infty}$ > noise floor of the measurement system, it means that $V_2|_{R_\infty}$ is measurable. Therefore, (4-20) should be used to determine $Z_X$.

- If $V_2|_{R_\infty}$ < noise floor of the measurement system, it means that the probe-to-probe coupling is weak and can be neglected. Therefore, (4-28) can be applied to determine $Z_X$.

- After extracting $Z_X$, $Z_{SUT}$ can be determined through deembedding $(Z_S + Z_W)$ from $Z_X$.

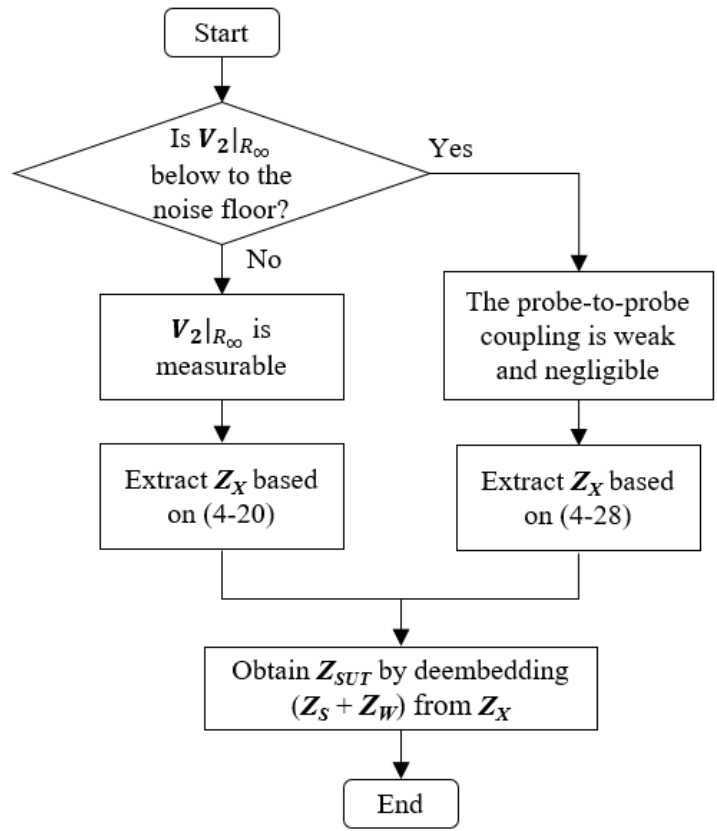

Fig. 4-2. Online impedance measurement process based on the three-term calibration technique.

## 4.3. Experimental Validation

Fig. 4.3 shows an actual measurement setup used for experimental validation. The computer-controlled SGAS is a National Instruments (NI) PXI platform, which consists of a signal generation card (PXI-5412), a two-channel signal acquisition card (PXI-5922), and an embedded controller (PXIe-8840). A Solar 9144-1N inductive probe is selected as the IIP and a Solar 9134-1 inductive probe is selected as the RIP. It should be noted that the selection of the IIP and RIP must consider the maximum current passing through the SUT to avoid the core saturation of the inductive probes. Since the accuracy of the measurement depends largely on the performance of the calibration, a calibration fixture with N-type short and N-type 50-Ω terminators are selected for the premeasurement calibration, as shown in Fig. 4-3.

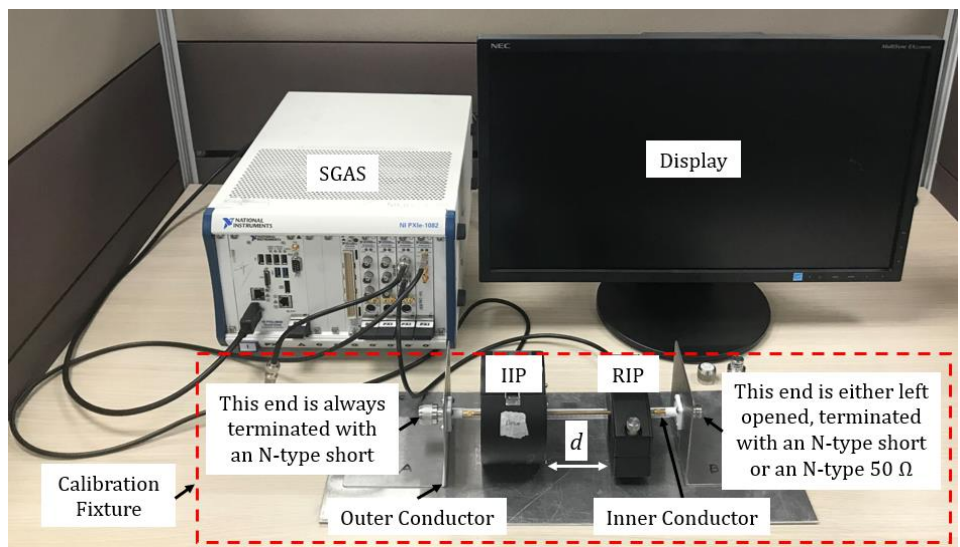

Fig. 4-3. Measurement setup and calibration fixture for validation.

### 4.3.1. Calibration for Measurement Setup

When using the fixture to calibrate the measurement setup, the IIP and RIP are clamped onto the inner conductor of the fixture. One end of the fixture is connected with an N-type short termination (HP 11512A) so that the inner conductor can be shorted with the outer conductor of the fixture at one end. The other end of the fixture is either left opened, terminated with an N-type short (HP 11512A), or terminated with an N-type 50 Ω (KARN-50-18+) to establish one of the three conditions. Thus, the inner conductor, outer conductor, and N-type terminators at both ends of the fixture form a circuit loop. In reality, it is impossible to have a perfect short and 50-Ω resistive load. Therefore, the short and 50-Ω resistive load must be as close as possible to their respective ideal values in the frequency range of interest. Fig. 4-4 shows the measured impedances of the $R_0$ and $R_{50}$ calibration conditions using an impedance analyzer (Agilent 4294A) in the frequency range of interest from 20 kHz to 1 MHz.

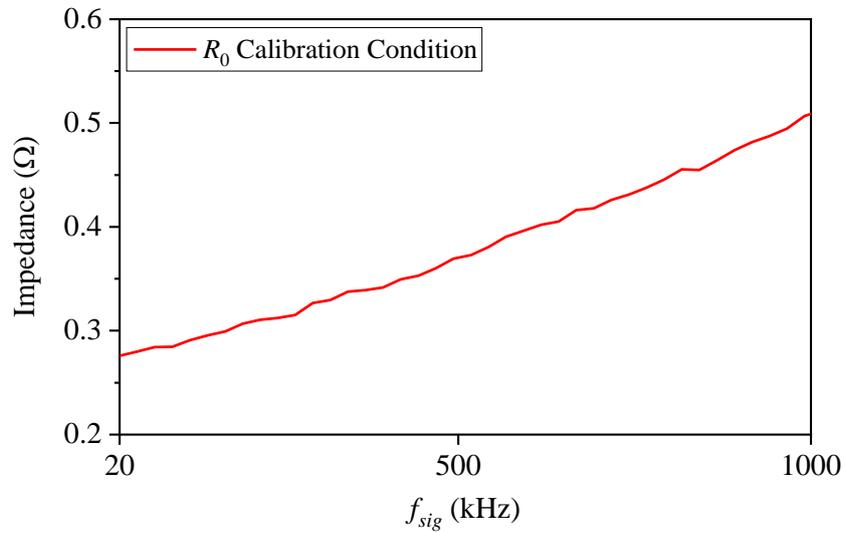

(a)

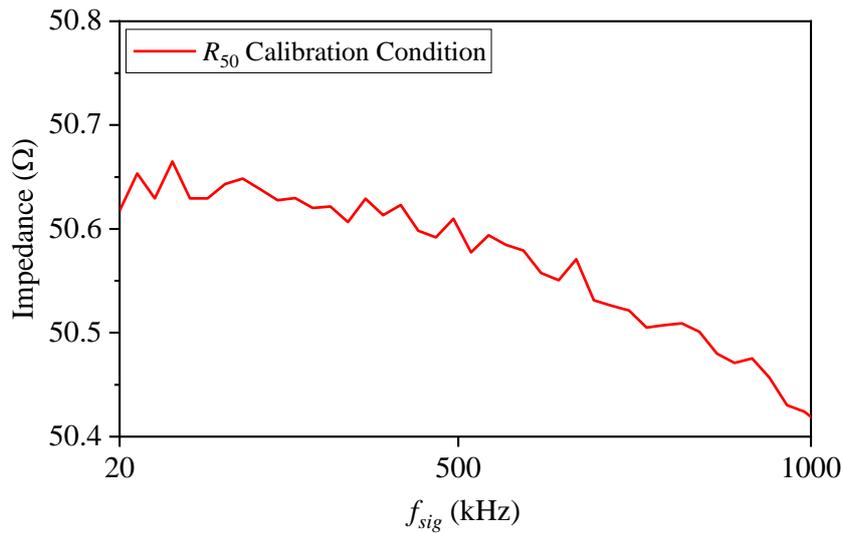

(b)

Fig. 4-4. Measured impedances of (a) the $R_0$ calibration condition and (b) the $R_{50}$ calibration condition.

As can be seen in Fig. 4-4, the measured impedance of the $R_0$ calibration condition varies between 0.2 Ω and 0.5 Ω from 20 kHz to 1 MHz, which can be regarded as a short. Also, the measured impedance of the $R_{50}$ calibration condition varies between 50.4 Ω and 50.7 Ω from 20 kHz to 1 MHz, which is very close to 50 Ω.

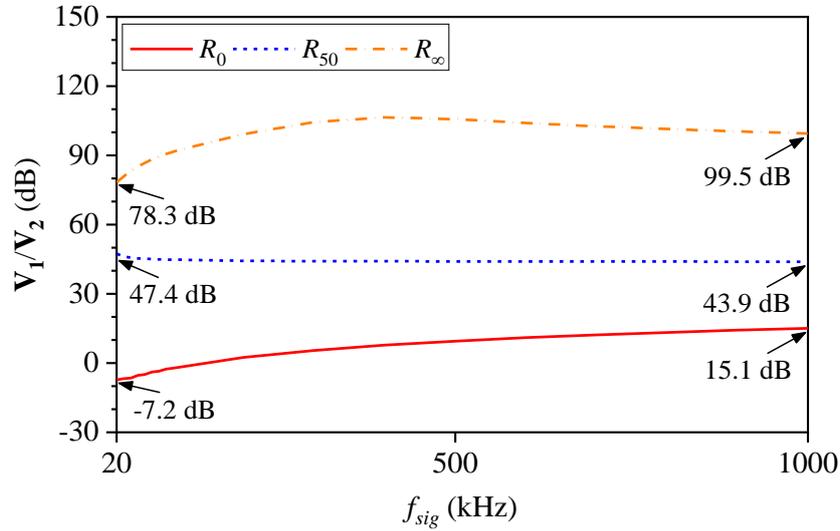

(a)

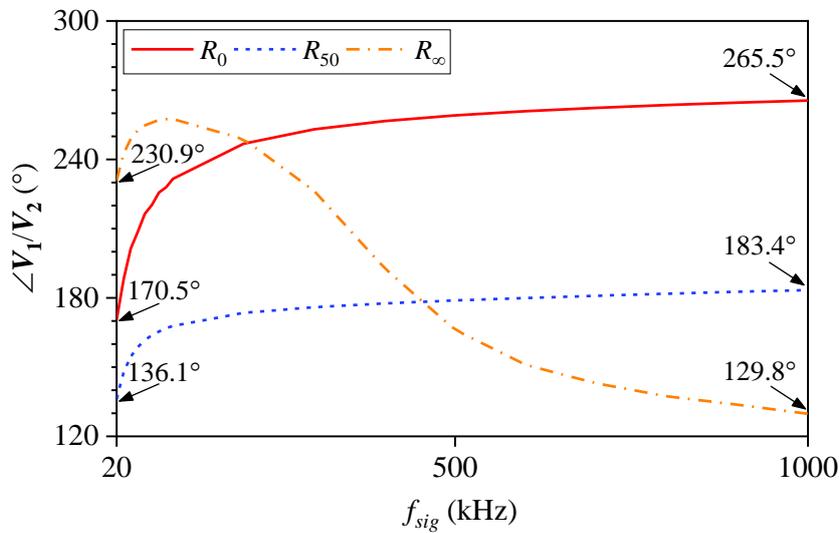

(b)

Fig. 4-5. (a) Magnitude and (b) phase of $V_1/V_2|_t$ when $d = 0$ mm and $f_{sig}$ varies from 20 kHz to 1 MHz.

As mentioned in Section 4.2, once the inductive probes are selected, the values of $a_1$, $a_2$, and $a_3$ mainly depend on the spacing ($d$) between the inductive probes and the sinusoidal test signal frequency ($f_{sig}$). Therefore, it is obtained from (4-5) that the values of $V_1/V_2|_t$ are also influenced by $d$ and $f_{sig}$, where "$t$" represents the termination, either $R_\infty$, $R_0$, or $R_{50}$. Fig. 4-5 shows the measured $V_1/V_2|_t$ when $d = 0$ mm (strongest coupling) and $f_{sig}$ varies from 20 kHz to 1 MHz. In addition, Fig. 4-6 shows the measured $V_1/V_2|_t$ when $f_{sig}$ is fixed at 100 kHz and $d$ varies from 0 to 50 mm.

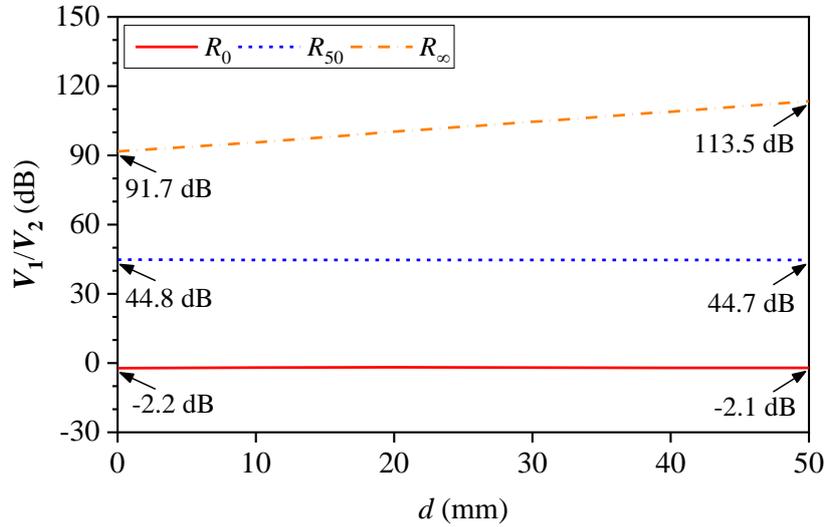

(a)

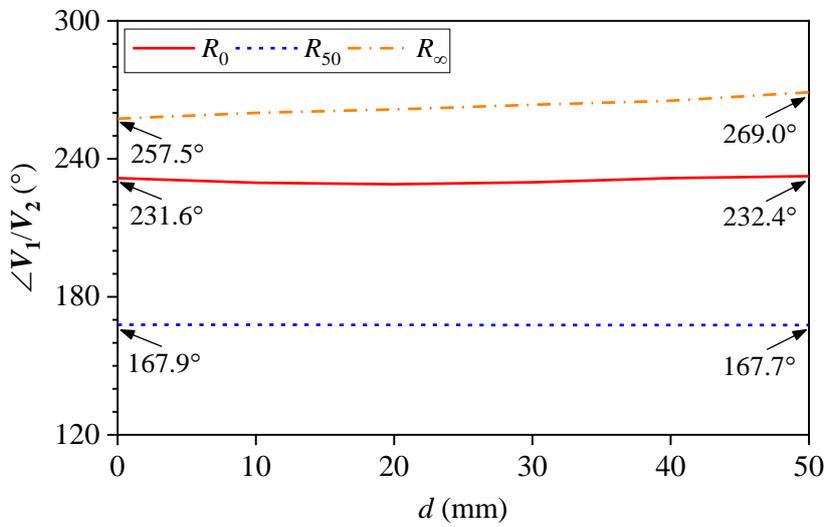

(b)

Fig. 4-6. (a) Magnitude and (b) phase of $V_1/V_2|_t$ when $f_{sig} = 100$ kHz and $d$ varies from 0 to 50 mm.

As can be seen in Figs. 4-5 and 4-6, $V_1/V_2$ varies with the change of $Z_X$. Fig. 4-5 shows that when $Z_X$ varies from short ($R_0$) to open ($R_\infty$), there is a significant difference between $V_1/V_2|_{R_\infty}$ and $V_1/V_2|_{R_0}$ in the given frequency range (85.5 dB at $f_{sig} = 20$ kHz and 84.4 dB at $f_{sig} = 1$ MHz). Fig. 4-6 also shows that for a given measurement setup and a specific $f_{sig}$ = 100 kHz, $V_1/V_2|_{R_\infty}$ increases with increasing $d$ (91.7 dB when $d$ = 0 mm to 113.5 dB when $d$ = 50 mm), which is expected since the probe-to-probe coupling weakens with increasing $d$ and $V_1/V_2|_{R_\infty}$ increases correspondently. However, it is also observed that $V_1/V_2|_{R_0}$ and $V_1/V_2|_{R_{50}}$ remain rather constant with increasing $d$. This is because when $Z_X$ is short ($R_0$) or 50 Ω ($R_{50}$), the probe-to-wire coupling, which is the intended (or desired) coupling, is much stronger than the probe-to-probe coupling, which is the unintended (or undesired) coupling.

### 4.3.2. Online Impedance Extraction of *RLC* Circuits

To verify the online measurement, a set of *RLC* circuits with circuit configuration shown in Fig. 4-7 are selected to emulate different SUT. The *RLC* circuits are powered by a 20-V DC power supply via the wiring connection, as shown in Fig. 4-8. The total length of the wiring connection is less than 1 meter, which is much shorter than the wavelength of the sinusoidal test signal when $f_{sig}$ ≤ 1 MHz. Four *RLC* circuits with different *R*, *L*, and *C* components are selected and named as SUT1, SUT2, SUT3, and SUT4.

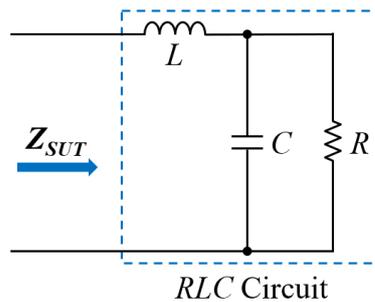

Fig. 4-7. Circuit configuration of the *RLC* circuits.

The values and tolerances of these components are shown in Table 4-1. The rationale for the selected components' values is to provide a sufficiently wide impedance range of $Z_{SUT}$. Fig. 4-8 shows the measurement setup and one of the SUTs. Firstly, the impedance of the SUT is measured directly using an impedance analyzer (Agilent 4294A) offline as a reference for comparison. Subsequently, the online measured impedances with and without the proposed three-term calibration technique are compared with that measured offline.

Table 4-1. Values and tolerances of the selected components for experiments in Section 4.3.

| SUT | R | L | C |
|---|---|---|---|
| 1 | 0.2 kΩ, ±1% | 0.82 µH, ±10% | 8200 pF, ±5% |
| 2 | 2.2 kΩ, ±1% | 8.2 µH, ±10% | 820 pF, ±20% |
| 3 | 8.2 kΩ, ±1% | 82 µH, ±5% | 150 pF, ±20% |
| 4 | 20 kΩ, ±1% | 820 µH, ±5% | 82 pF, ±5% |

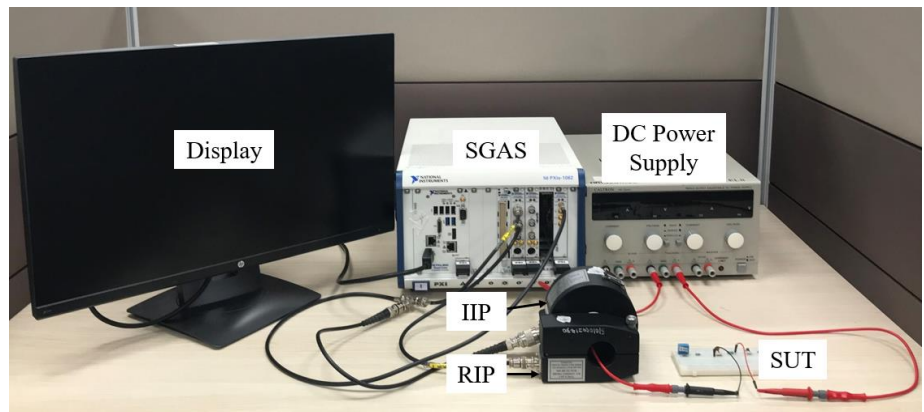

Fig. 4-8. Measurement setup and one of the SUTs.

Table 4-2 shows the comparison of the measured impedances when $f_{sig}$ = 100 kHz and $d$ = 0 mm. Using the offline measured impedances as a reference, Fig. 4-9 shows the deviation of the online measured impedances based on the inductive coupling approach with and without (w/o) the proposed calibration (Cal.) technique for varying online impedance values. From Table 4-2 and Fig. 4-9, the deviation with and without the proposed calibration technique is negligible when $Z_{SUT}$ is 139 Ω. As mentioned earlier, this is because the probe-to-wire coupling is much stronger than the probe-to-probe coupling for the given case. However, when $Z_{SUT}$ is relatively large (6.22 kΩ and 13.3 kΩ), the magnitude error ($\Delta|Z_{SUT}|$) and the phase error ($\Delta\angle Z_{SUT}$) without the proposed

calibration technique are more than 10% and 30°, respectively. This is because the probe-to-wire coupling weakens with increasing $Z_{SUT}$. When $Z_{SUT}$ is relatively large, the probe-to-probe coupling is comparable to the probe-to-wire coupling so that the probe-to-probe coupling causes a significant error in the measured online impedance. However, if the probe-to-probe coupling is taking into consideration via the proposed calibration technique, the measurement accuracy improves significantly with $\Delta|Z_{SUT}|$ and $\Delta\angle Z_{SUT}$ kept within 3.0% and 1.6°, respectively, for all four emulated SUTs with the impedance range from a few hundreds of ohms to a few tens of kiloohms.

Table 4-2. Comparison of the measured impedances of the SUTs in Table 4-1 when $f_{sig}$ = 100 kHz and $d$ = 0 mm.

| SUT | Offline Measurement (Reference) | | Online Measurement w/o Proposed Cal. Technique | | Online Measurement with Proposed Cal. Technique | |
|---|---|---|---|---|---|---|
| | $|Z|$(kΩ) | $\angle Z$(°) | $|Z|$(kΩ) | $\angle Z$(°) | $|Z|$(kΩ) | $\angle Z$(°) |
| 1 | 0.139 | -45.5 | 0.138 | -44.8 | 0.137 | -45.2 |
| 2 | 1.53 | -45.7 | 1.65 | -40.6 | 1.51 | -45.5 |
| 3 | 6.22 | -38.6 | 7.12 | -7.16 | 6.27 | -37.7 |
| 4 | 13.3 | -45.2 | 15.8 | 26.3 | 13.7 | -43.6 |

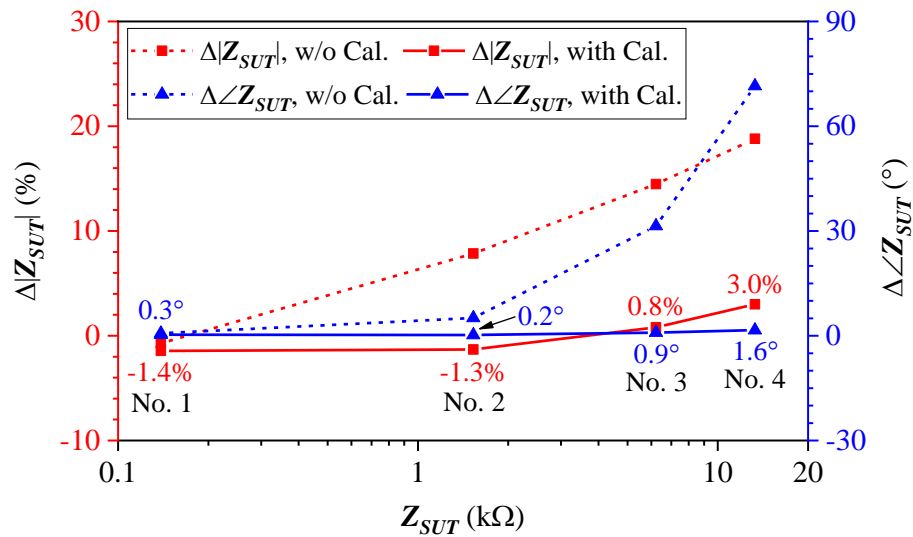

Fig. 4-9. Calculated measurement errors from Table 4-2 when $f_{sig}$ = 100 kHz and $d$ = 0 mm.

Table 4-3 compares the measured impedances of the four SUTs when $f_{sig}$ = 100 kHz and $d$ = 80 mm. Based on the measurement process shown in Fig. 4-2, $V_2|_{R_\infty}$ is found to be lower than the noise floor of the SGAS and therefore the probe-to-probe coupling between the IIP and RIP is weak and negligible. Similarly, using the offline measured impedance as a reference, Fig. 4-10 shows the measurement error of the inductive coupling approach with and without the proposed calibration technique.

Table 4-3. Comparison of the measured impedances of the SUTs in Table 4-1 when $f_{sig}$ = 100 kHz and $d$ = 80 mm.

| SUT | Offline Measurement (Reference) | | Online Measurement w/o Proposed Cal. Technique | | Online Measurement with Proposed Cal. Technique | |
|---|---|---|---|---|---|---|
| | $|Z|$(kΩ) | ∠$Z$(°) | $|Z|$(kΩ) | ∠$Z$(°) | $|Z|$(kΩ) | ∠$Z$(°) |
| 1 | 0.139 | -45.5 | 0.137 | -45.3 | 0.138 | -45.2 |
| 2 | 1.53 | -45.7 | 1.49 | -46.0 | 1.50 | -45.9 |
| 3 | 6.22 | -38.6 | 6.10 | -39.3 | 6.05 | -38.4 |
| 4 | 13.3 | -45.2 | 13.0 | -46.3 | 13.1 | -46.3 |

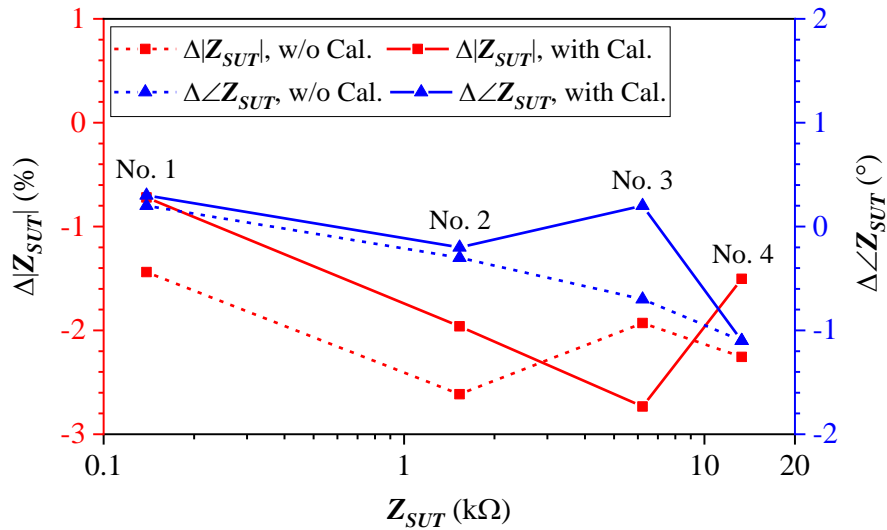

Fig. 4-10. Calculated measurement errors from Table 4-3 when $f_{sig}$ = 100 kHz and $d$ = 80 mm.

According to Table 4-3 and Fig. 4-10, the measured online impedances with and without the proposed calibration technique show rather good accuracy (Δ$|Z_{SUT}|$ kept within 3% and Δ∠$Z_{SUT}$ kept within 1.1°) for the given SUT impedance range. It confirms that when $V_2|_{R_\infty}$ is lower than the SGAS's noise floor, the probe-to-probe coupling between the IIP and RIP is weak and has negligible impact on the accuracy of the online impedance measurement.

In this chapter, a three-term calibration technique of the inductive coupling approach is proposed to deembed the online impedance measurement error caused by the probe-to-probe coupling between the IIP and RIP. Compared with the conventional equivalent circuit model based on the cascaded two-port network concept, a comprehensive equivalent circuit model based on the three-port network concept is introduced, in which the effect of the probe-to-probe coupling can be taken into account in the model. With the three-port network equivalent circuit model, in-depth analysis based on the three-term calibration technique to deembed the effect of the probe-to-probe coupling becomes possible. In this way, the accuracy of the online impedance measurement can be preserved even the inductive probes are placed very close together.

# Chapter 5 Measurement Setup Consideration Under Significant Electrical Noise and Power Surges

In Chapter 3, a computer-controlled SGAS-based inductively coupled measurement setup has been reported, which can extract not only the time-invariant but also the time-variant online impedance of electrical systems. In Chapter 4, a three-term calibration technique has been proposed to eliminate the effect of the probe-to-probe coupling on the accuracy of the extracted online impedance. However, due to some practical issues, the measurement setup in Fig. 3-1 cannot be directly applied to an electrical system with significant electrical noise and power surges. For example, the significant electrical noise in the electrical system can considerably mask the injecting test signal and the power surges can cause expected damage to the measurement instruments. To overcome these practical implementation issues, the measurement setup in Fig. 3-1 has to improve its SNR by adding a power amplifier and to enhance its ruggedness by incorporating surge protection devices. Therefore, an improved measurement setup incorporating signal amplification and surge protection modules is introduced in this chapter.

This chapter is organized as follows. Section 5.1 introduces the improved measurement setup that incorporates signal amplification and surge protection modules. In Section 5.2, a comprehensive equivalent circuit model of this measurement setup is introduced. In Section 5.3, experiments are conducted for verification and evaluation.

## 5.1. Measurement Setup Incorporating Signal Amplification and Surge Protection Modules

Fig. 5-1 shows the measurement setup incorporating signal amplification and surge

protection modules for online impedance extraction of a SUT, where the SUT is powered by either a DC or a low-frequency (e.g. 50/60 Hz) AC power source through the wiring connection. The measurement setup consists of a computer-controlled SGAS, a power amplifier (PA), two attenuators (AT1 and AT2), two surge protectors (SP1 and SP2), an IIP, and a RIP.

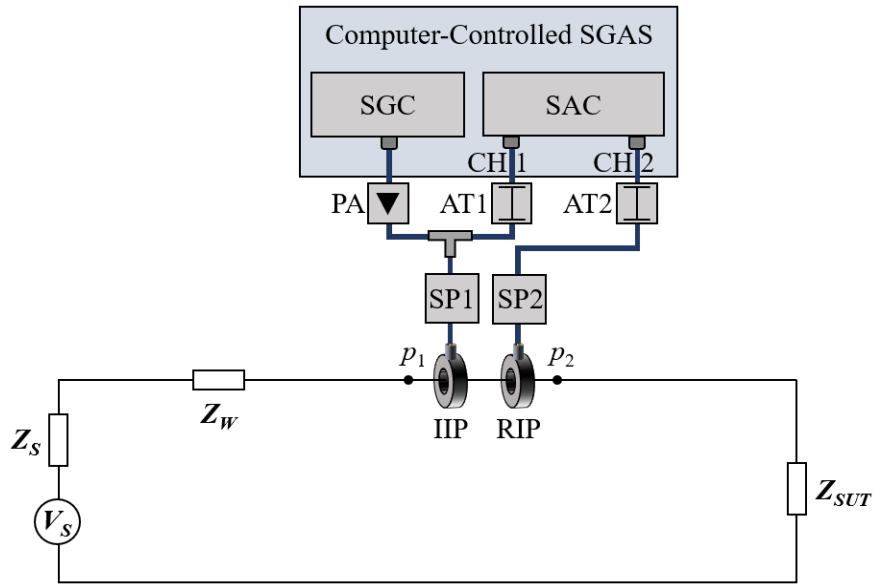

Fig. 5-1. Measurement Setup incorporating signal amplification and surge protection modules.

To extract the online impedance of the SUT ($Z_{SUT}$), a SGC in the SGAS generates a sinusoidal excitation (test) signal of known frequency ($f_{sig}$). The excitation signal is amplified by the PA and then is injected into the connecting wire through the IIP, and the RIP monitors the response signal. The IIP and RIP are clamped onto the connecting wire between $p_1$ and $p_2$ with a spacing $d$. Channel 1 (CH1) of the SAC in the SGAS measures the excitation signal voltage at the output port of the AT1, and channel 2 (CH2) of the SAC measures the response signal voltage at the output port of the AT2. AT1 and AT2 are required to ensure the measured signal voltage levels are within the permissible input range of the SAC. SP1 and SP2 serve to protect the measurement instruments from power surges in the electrical system. By doing so, the measurement setup can be applied to an electrical system with significant electrical noise and power surges and therefore the scope of application of the inductive coupling approach has been expanded. In Fig. 5-1, $V_S$ and $Z_S$ are the equivalent source voltage and impedance of the power source, respectively. $Z_W$ is the equivalent impedance of the wiring connection except for the part being clamped by the IIP and RIP. $Z_X$ is the resultant impedance seen at $p_1$-$p_2$, namely $Z_X = Z_S + Z_W + Z_{SUT}$.

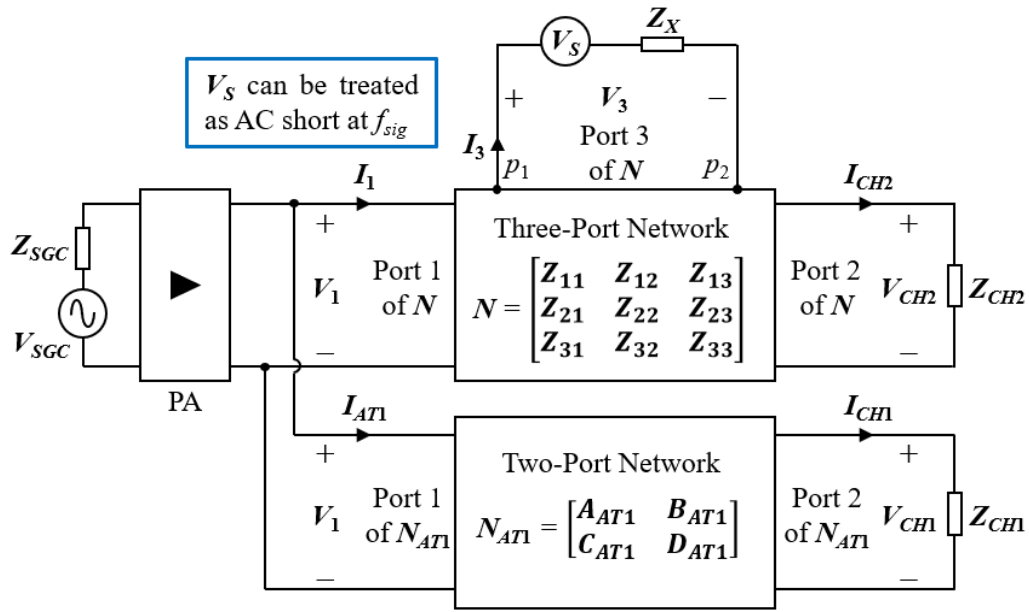

Fig. 5-2. Equivalent circuit model of Fig. 5-1 represented by two-port and three-port networks.

## 5.2. Equivalent Circuit Model based on Two-Port and Three-Port Network Concepts

Based on two-port and three-port network concepts [32], Fig. 5-2 shows an equivalent circuit model of Fig. 5-1, where $V_{SGC}$ and $Z_{SGC}$ are the equivalent source voltage and internal impedance of the SGC, respectively; $Z_{CH1}$ and $Z_{CH2}$ are the input impedances of CH1 and CH2 of the SAC, respectively; $V_1$ and $I_1$ are the excitation signal voltage and current at the input port of the SP1, respectively; $V_{CH2}$ and $I_{CH2}$ are the response signal voltage and current at the output port of the AT2, respectively. $V_{CH2}$ can be directly measured by CH2 of the SAC and $I_{CH2} = V_{CH2}/Z_{CH2}$. $V_3$ is the induced signal voltage between $p_1$ and $p_2$, and $I_3$ is the induced signal current passing through $Z_X$ to be extracted. With the PA, the test signal can be amplified so that the induced signal level is much higher than the noise level in the electrical system even if the noise level is significant. Since the frequency ($f_{sig}$) of the sinusoidal test signal is much higher than the frequency of the power source, $V_S$ can be treated as AC short at $f_{sig}$. Thus, $V_3$ and $I_3$ can be related by $I_3 = V_3/Z_X$. $N$ is a three-port network, in which Port 1 is the input port of the SP1; Port 2 is the output port of the AT2; Port 3 is the input port between $p_1$ and $p_2$. Thus, the respective ports' voltages and currents of $N$ are related by

$$\begin{bmatrix} V_1 \\ V_{CH2} \\ V_3 \end{bmatrix} = \begin{bmatrix} Z_{11} & Z_{12} & Z_{13} \\ Z_{21} & Z_{22} & Z_{23} \\ Z_{31} & Z_{32} & Z_{33} \end{bmatrix} \begin{bmatrix} I_1 \\ I_{CH2} \\ I_3 \end{bmatrix} \quad (5\text{-}1)$$

where $Z_{ij}$ ($i$ and $j$ = 1, 2 or 3) is the impedance parameter of $N$. Since $I_{CH2} = V_{CH2}/Z_{CH2}$ and $I_3 = V_3/Z_X$, dividing $V_{CH2}$ at both sides of (5-1), it becomes

$$\begin{bmatrix} V_1/V_{CH2} \\ 1 \\ V_3/V_{CH2} \end{bmatrix} = \begin{bmatrix} Z_{11} & Z_{12} & Z_{13} \\ Z_{21} & Z_{22} & Z_{23} \\ Z_{31} & Z_{32} & Z_{33} \end{bmatrix} \begin{bmatrix} I_1/V_{CH2} \\ 1/Z_{CH2} \\ (V_3/V_{CH2})/Z_X \end{bmatrix} \quad (5\text{-}2)$$

By solving (5-2), $V_1/V_{CH2}$, $V_3/V_{CH2}$, and $I_1/V_{CH2}$ can be represented in terms of $Z_{ij}$, $Z_{CH2}$, and $Z_X$; and $V_1/V_{CH2}$ can be expressed as

$$\frac{V_1}{V_{CH2}} = \frac{a_1 \cdot Z_X + a_2}{Z_X + a_3} \quad (5\text{-}3)$$

where

$$a_1 = \frac{Z_{11}}{Z_{21}} \cdot \left(1 - \frac{Z_{22}}{Z_{CH2}}\right) + \frac{Z_{12}}{Z_{CH2}} \quad (5\text{-}4)$$

$$a_2 = \left(Z_{13} - \frac{Z_{11}Z_{23}}{Z_{21}}\right) \cdot \left[\frac{Z_{31}}{Z_{21}} \cdot \left(1 - \frac{Z_{22}}{Z_{CH2}}\right) + \frac{Z_{32}}{Z_{CH2}}\right]$$
$$- \left(Z_{33} - \frac{Z_{31}Z_{23}}{Z_{21}}\right) \cdot \left[\frac{Z_{11}}{Z_{21}} \cdot \left(1 - \frac{Z_{22}}{Z_{CH2}}\right) + \frac{Z_{12}}{Z_{CH2}}\right] \quad (5\text{-}5)$$

$$a_3 = \frac{Z_{31}Z_{23}}{Z_{21}} - Z_{33} \quad (5\text{-}6)$$

In addition, AT1 can be equated to a two-port network $N_{AT1}$, as shown in Fig. 5-2. Thus, the respective ports' voltages and currents of $N_{AT1}$ are related by

$$\begin{bmatrix} V_1 \\ I_{AT1} \end{bmatrix} = \begin{bmatrix} A_{AT1} & B_{AT1} \\ C_{AT1} & D_{AT1} \end{bmatrix} \begin{bmatrix} V_{CH1} \\ I_{CH1} \end{bmatrix} \quad (5\text{-}7)$$

where $A_{AT1}$, $B_{AT1}$, $C_{AT1}$, and $D_{AT1}$ are the transmission parameters of $N_{AT1}$; $I_{AT1}$ is the test signal current at the input port of the AT1; $V_{CH1}$ and $I_{CH1}$ are the excitation signal voltage and current at the output port of the AT1, respectively. $V_{CH1}$ can be directly measured by CH1 of the SAC and $I_{CH1} = V_{CH1}/Z_{CH1}$. By solving (5-7), $V_1$ and $V_{CH1}$ are related by

$$V_1 = (A_{AT1} + B_{AT1}/Z_{CH1}) \cdot V_{CH1} \quad (5\text{-}8)$$

Substituting (5-8) into (5-3), the former takes the form

$$\frac{V_{CH1}}{V_{CH2}} = \frac{a_4 \cdot Z_X + a_5}{Z_X + a_3} \quad (5\text{-}9)$$

where

$$a_4 = \frac{a_1}{A_{AT1} + B_{AT1}/Z_{CH1}} \quad (5\text{-}10)$$

$$a_5 = -\frac{a_2}{A_{AT1} + B_{AT1}/Z_{CH1}} \tag{5-11}$$

Finally, $Z_X$ can be represented as a function of $V_{CH1}/V_{CH2}$ as follows

$$Z_X = \frac{a_3 \cdot \dfrac{V_{CH1}}{V_{CH2}} - a_5}{-\dfrac{V_{CH1}}{V_{CH2}} + a_4} \tag{5-12}$$

Once $a_3$, $a_4$, and $a_5$ are known, $Z_X$ can be determined from (5-12) where $V_{CH1}$ and $V_{CH2}$ are directly measured by CH1 and CH2 of the SAC of the measurement setup, respectively. By observing (5-4), (5-5), (5-6), (5-10), and (5-11), $a_3$, $a_4$, and $a_5$ depend only on the impedance parameters of $N$, the transmission parameters of $N_{AT1}$, and the input impedances of CH1 and CH2 of the SAC (i.e. $Z_{CH1}$ and $Z_{CH2}$). Since $N$, $N_{AT1}$, $Z_{CH1}$, and $Z_{CH2}$ at a given frequency are only determined by the inherent characteristics of instruments of the measurement setup, $a_3$, $a_4$, and $a_5$ at a given frequency remain unchanged for a specific measurement setup. To determine $a_3$, $a_4$, and $a_5$ of a given measurement setup at $f_{sig}$, the three-term calibration technique by sequentially replacing $Z_X$ with three known but different conditions can be conducted before the actual online measurement. Since the three-term calibration technique has been well described in Section 4.2, it will not be repeated here. After determining $a_3$, $a_4$, and $a_5$ according to the three-term calibration technique, $Z_X$ can be obtained from (5-12) through the measurement of $V_{CH1}$ and $V_{CH2}$ using the measurement setup. Thus, $Z_{SUT}$ can be extracted through deembedding ($Z_S + Z_W$) from $Z_X$.

## 5.3. Experimental Validation

Table 5-1 shows the details of the instruments of an actual measurement setup for experimental validation. To verify the online measurement ability of this measurement

setup, a set of *RLC* circuits with the same circuit configuration shown in Fig. 4-7 are selected to emulate different SUT. The *RLC* circuits are powered by a 20-V DC power supply through the wiring connection. The total length of the wiring connection is less than 1 meter. Four *RLC* circuits with different values of *R*, *L*, and *C* are selected and named as SUT1, SUT2, SUT3, and SUT4. The values and tolerances of these components are shown in Table 5-2.

Table 5-1. Details of the instruments (inst.) of the actual measurement setup incorporating signal amplification and surge protection modules.

| Inst. | Descriptions | Inst. | Descriptions |
|---|---|---|---|
| SGAS | NI PXI Platform | PA | Mini-Circuits LZY-22+ |
| SGC | NI PXI-5412 | AT1 | Pasternack PE7025-20 |
| SAC | NI PXI-5922 | AT2 | AIM-Cambridge 27-9300-20 |
| IIP | Solar 9144-1N | SP1 | Phoenix Contact C-UFB |
| RIP | Solar 9134-1 | SP2 | Phoenix Contact C-UFB |

Table 5-2. Values and tolerances of the selected components for experiments in Section 5-3.

| SUT | *R* | *L* | *C* |
|---|---|---|---|
| 1 | 0.2 kΩ, ±1% | 2.2 µH, ±5% | 100 nF, ±10% |
| 2 | 0.5 kΩ, ±1% | 6.8 µH, ±5% | 15 nF, ±10% |
| 3 | 2 kΩ, ±1% | 15 µH, ±5% | 1 nF, ±10% |
| 4 | 4.7 kΩ, ±1% | 33 µH, ±5% | 0.33 nF, ±10% |

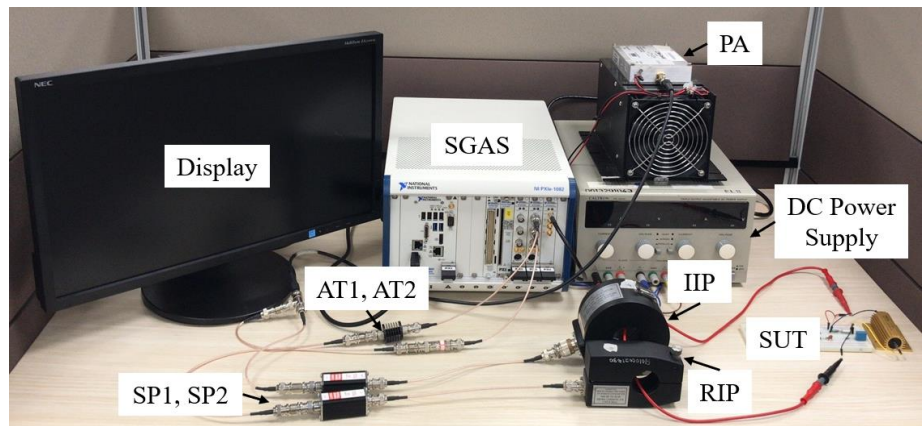

Fig. 5-3. Actual measurement setup incorporating signal amplification and surge protector modules.

Fig. 5-3 shows the actual measurement setup and one of the SUTs. Firstly, the impedances of the SUTs are directly measured using an impedance analyzer (Agilent 4294A) offline as a reference for comparison. Subsequently, the online measured impedances using this measurement setup are compared with that measured offline. Table 5-3 shows the comparison of the measured impedances when $f_{sig}$ = 100 kHz and $d$ = 0 mm. Using the offline measured impedances as a reference, Fig. 5-4 shows the deviation of the online measured impedances using this measurement setup. Similarly, Table 5-4 shows the comparison of the measured impedances when $f_{sig}$ = 50 kHz and $d$ = 20 mm, and Fig. 5-5 shows the deviation of the online measured impedances using this measurement setup.

Table 5-3. Comparison of the measured impedances of the SUTs in Table 5-2 when $f_{sig}$ = 100 kHz and $d$ = 0 mm.

| SUT | Offline Measurement (Reference) | | Online Measurement using the measurement setup in Fig. 5-3 | |
|---|---|---|---|---|
| | $|Z|(\Omega)$ | $\angle Z(°)$ | $|Z|(\Omega)$ | $\angle Z(°)$ |
| 1 | 15.48 | -84.7 | 15.32 | -84.9 |
| 2 | 96.59 | -77.9 | 95.51 | -77.6 |
| 3 | 1272 | -50.1 | 1284 | -49.7 |
| 4 | 3332 | -44.3 | 3376 | -43.4 |

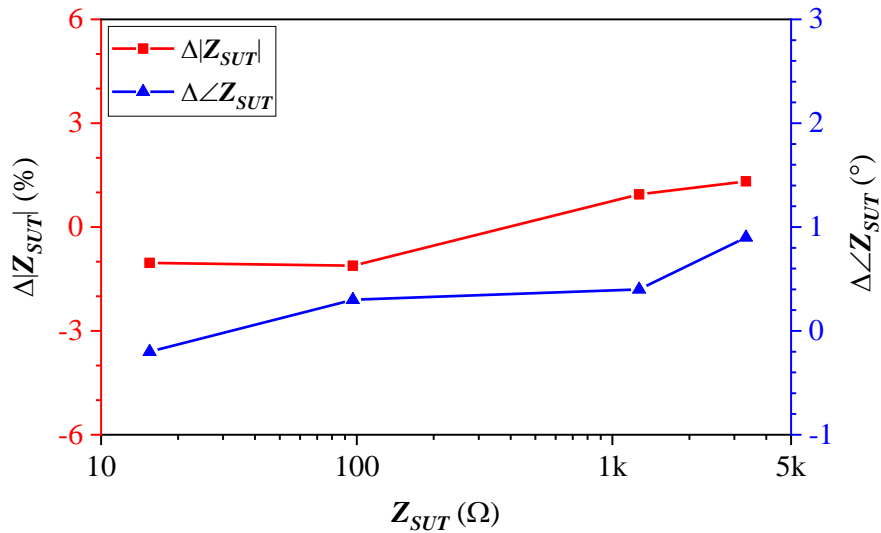

Fig. 5-4. Calculated measurement errors from Table 5-3 when $f_{sig}$ =100 kHz and $d$ = 0 mm.

Table 5-4. Comparison of the measured impedances of the SUTs in Table 5-2 when $f_{sig}$ = 50 kHz and $d$ = 20 mm.

| SUT | Offline Measurement (Reference) | | Online Measurement using the measurement setup in Fig. 5-3 | |
|---|---|---|---|---|
| | $|Z|(\Omega)$ | $\angle Z(°)$ | $|Z|(\Omega)$ | $\angle Z(°)$ |
| 1 | 32.72 | -80.1 | 32.45 | -80.2 |
| 2 | 188.2 | -67.4 | 186.5 | -67.2 |
| 3 | 1715 | -31.0 | 1732 | -30.6 |
| 4 | 4208 | -26.1 | 4313 | -25.2 |

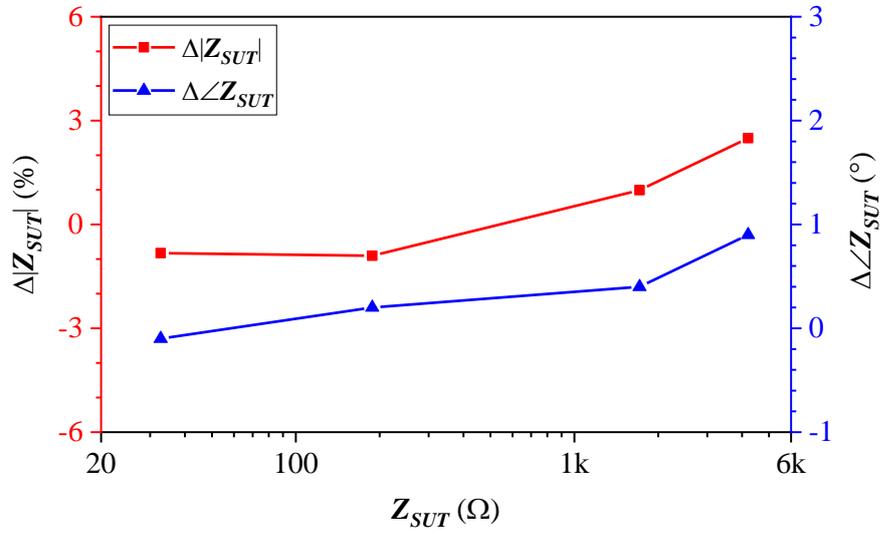

Fig. 5-5. Calculated measurement errors from Table 5-4 when $f_{sig}$ = 50 kHz and $d$ = 20 mm.

From Figs. 5-4 and 5-5, the measured online impedances show rather good accuracy ($\Delta|Z_{SUT}|$ kept within 3% and $\Delta\angle Z_{SUT}$ kept within 1°) for all four SUTs. Therefore, the accuracy of the measurement setup for online impedance extraction has been validated.

In this chapter, an inductively coupled measurement setup incorporating signal amplification and surge protection modules is introduced. Based on two-port and three-port network concepts, the equivalent circuit model of this measurement setup has been described and its ability to extract the online impedance of an electrical system has been verified by experiments. By doing so, this measurement setup can be used to extract the online impedance of an electrical system with significant electrical noise and power surges. Thus, the scope of application of the inductive coupling approach has been expanded. In the next chapter, a practical application case to extract the online impedance of an electrical system with significant electrical noise and power surges using this measurement setup will be elaborated.

# Chapter 6 Application in Stator Fault Detection of Inverter-Fed Induction Motor

With the advancement of both the fast-switching power electronics and the pulse width modulation (PWM) control techniques, the inverters have become highly popular for variable torque and speed controls of induction motors (IMs) because of their excellent conversion efficiency and control efficiency [44]. Just like any other electrical assets, the IM is also subject to faults due to thermal, electrical, ambient, and mechanical stresses. Among the common faults of the IM, bearing, stator and rotor faults are three major ones that account for 41%, 36% and 9% of the total faults of the IM, respectively [45]. The stator faults usually result from the stator winding insulation breakdown. Such a breakdown can lead to an inter-turn short-circuit (ITSC) fault, which in turn will result in a greater current flowing through the shorted turns and generate excessive heat in adjacent turns. Thus, an ITSC fault can quickly develop into a more severe fault, such as a coil-to-coil, phase-to-phase, or phase-to-ground short-circuit fault, and eventually results in a complete failure of IM [46]. The failure is financially deleterious because of the asset replacement and the loss of revenue associated with unscheduled IM downtime. Therefore, effective detection of stator ITSC faults is necessary to avoid catastrophic failures.

A variety of ITSC fault detection methods have been reported and in general, they can be classified into two categories: the model-based methods [47]-[49] and the signal-based methods [50]-[57]. The model-based methods require a rather accurate numerical model of the faulty IM to identify the ITSC faults, which can be computationally intensive and might not always be feasible. The signal-based methods monitor one or more signals and detect the ITSC faults by analyzing these monitored signals. These signals can be the motor current/voltage [50], [51], the stray flux [52], the air-gap torque [53], the vibration [54], the temperature [55], the instantaneous power [56], and the sequence component impedance matrix [57].

Although the above-mentioned signal-based methods have achieved effective detection of ITSC faults under specific operation conditions, there are still a few aspects not considered. For examples, the motor current, the stray flux, the air-gap torque, the vibration and the temperature monitoring can be affected by the faults other than stator ITSC faults, such as bearing or rotor faults [53], [58]-[61]; thus, the reliability of the ITSC fault detection can be an issue. In addition, the voltage sensors used in the measurement setups to monitor the motor voltage [51], the instantaneous power [56], and the sequence component impedance matrix [57], require physical electrical connection to the energized IM for online ITSC fault detection. This physical electrical connection can pose electrical safety hazards to the service personnel who maintains the instruments on-site. Also, the voltage sensors used in the measurement setups are subject to dielectric and thermal stresses, which require regular maintenance and replacement; and result in unnecessary downtime of the IM. Recently, several artificial intelligence (AI) techniques [62]-[66] have been combined with the aforementioned signal-based methods to improve the ITSC fault detection but these require a large set of training data obtained from different operating states and conditions to ensure good detection reliability and robustness.

In this chapter, a novel signal-based method for the detection of the stator ITSC faults in the inverter-fed IM is proposed. The proposed method detects the ITSC faults through online monitoring of the common-mode (CM) impedance of the IM ($Z_{CM,IM}$) at a specific frequency of interest (FOI). The inductively coupled measurement setup shown in Fig. 5-1 is adopted to monitor $Z_{CM,IM}$ by tracking a known externally injected radio-frequency (RF) sinusoidal signal in the IM. Thus, the monitored $Z_{CM,IM}$ will not be affected by the supply voltage quality (e.g. unbalanced supply voltage). Also, the measurement setup requires no physical electrical connection to the IM, and the clamp-on inductive probes used in the measurement setup can be easily mounted on or removed from the power cable connected between the IM and inverter. Therefore, it simplifies the on-site implementation without shutting down the IM. In addition, the measurement setup has incorporated signal amplification and surge protection modules to enhance its ruggedness in the harsh industrial environment where significant electrical noise and power surges are present. Furthermore, by carefully selecting the FOI to monitor $Z_{CM,IM}$, it eliminates the need of monitoring $Z_{CM,IM}$ for a wide frequency range, thereby shortening the response time for ITSC fault detection. Fast response time is crucial for quick decisions to be made before an ITSC fault develops into a catastrophic failure. Through online monitoring of $Z_{CM,IM}$ at the FOI, the proposed method can detect stator ITSC faults with good confidence. Also, this method is reliable and robust for stator ITSC fault detection because it is invariant to motor load and speed variation, as well as bearing and rotor faults.

The organization of this chapter is as follows. Section 6.1 looks into the influence of ITSC faults on $Z_{CM,IM}$ and elaborates the process of selecting an appropriate FOI that is sensitive to the variation of $Z_{CM,IM}$ due to the ITSC faults. Section 6.2 describes the measurement setup for online monitoring of $Z_{CM,IM}$ at the FOI. Using an IM as a test case, Section 6.3 illustrates the reliability and robustness of the proposed method for ITSC fault detection under various conditions.

## 6.1. ITSC Fault Analysis and FOI Selection

### 6.1.1. ITSC Fault Influence on CM Impedance of IM

Before analyzing the influence of stator ITSC faults on $Z_{CM,IM}$, the definition of $Z_{CM,IM}$ is briefly introduced. Fig. 6-1 shows a comprehensive CM equivalent circuit model of a typical 3-phase IM with all the relevant electrical elements [67].

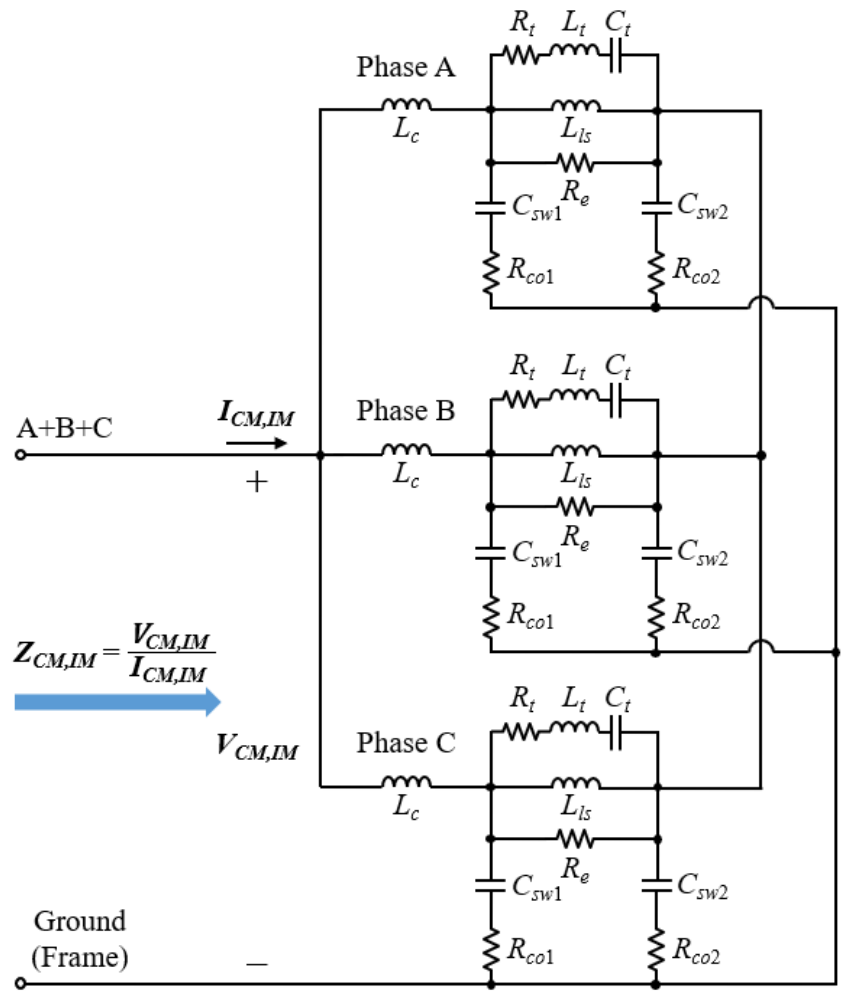

Fig. 6-1. A comprehensive CM equivalent circuit model of a typical 3-phase IM.

Regardless of the actual stator winding connection, the equivalent model can take the form of a star-connected configuration with the various elements calculated based on available measurements. Therefore, the model is universal and remains effective for any stator winding configuration (e.g. star, delta, series, or parallel) [67]. As shown in Fig.

6-1, $C_{sw1}$ and $C_{sw2}$ are the equivalent winding-to-frame parasitic capacitances. $R_{co1}$ and $R_{co2}$ are the equivalent copper resistances including the skin- and proximity-effects. $L_{ls}$ is the stator winding leakage inductance. $R_e$ is the eddy-current losses of the stator core. $R_t$, $L_t$, and $C_t$ represent the interturn parasitic effects of the stator windings. $L_c$ is the combined inductance of the IM's internal feed conductors to the stator windings and the stator winding leakage inductance of the first few turns. $Z_{CM,IM}$ is defined as $V_{CM,IM}/I_{CM,IM}$, where $V_{CM,IM}$ and $I_{CM,IM}$ are the CM voltage and current present at the IM, respectively.

To analyze the influence of ITSC faults on $Z_{CM,IM}$, Fig. 6-2 shows an ITSC fault that occurs in one of the phases of an IM. Such a fault results in a very low contact impedance between the shorted turns ($Z_{ITSC}$) and it has an influence on $Z_{CM,IM}$. It should be noted that besides the ITSC faults, other stator faults like coil-to-coil, phase-to-phase, and phase-to-ground short-circuit faults also have an effect on $Z_{CM,IM}$. Since these more serious stator faults arise from ITSC faults [46], this chapter focuses on the detection of stator ITSC faults. Herein, the analysis looks into the relative change in $Z_{CM,IM}$ due to the ITSC faults. To avoid the long measurement time to monitor $Z_{CM,IM}$ across a wide frequency band, monitoring $Z_{CM,IM}$ at a single frequency is preferred for faster response should an ITSC fault occurs. To ensure reliable detection of the ITSC faults, the selection process of an appropriate FOI to monitor $Z_{CM,IM}$ will be explained in the following section.

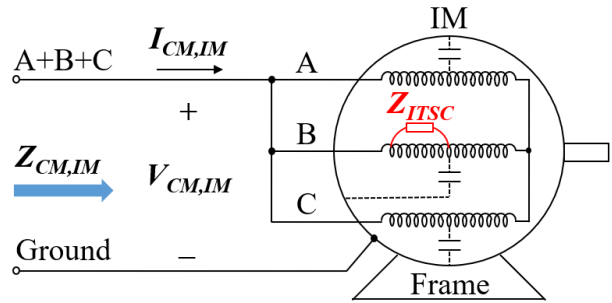

Fig. 6-2. An ITSC fault that occurs in one of the phases of an IM.

## 6.1.2. FOI Selection

It is reported that the IM's load and speed have negligible effects on $Z_{CM,IM}$ in the mid-to-high frequency range [67]. Once the FOI is selected within this frequency range, the ITSC faults can be detected via monitoring $Z_{CM,IM}$ with good confidence under the varying load and speed conditions. The following process is adopted to determine this frequency range.

i. Measure the differential-mode (DM) impedance of the IM ($Z_{DM,IM}$) offline using an impedance analyzer shown in Fig. 6-3 over a wide frequency range, say, up to a few megahertz.

ii. Determine the frequency where $Z_{DM,IM}$ exhibits a self-resonance. The self-resonant frequency is the start frequency of the mid-to-high frequency range of the IM and from this frequency onwards, the IM's load and speed have negligible effects on its impedance characteristics [67].

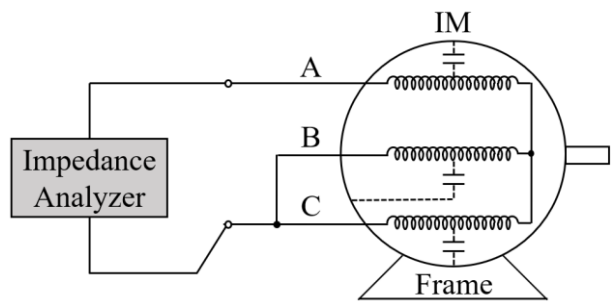

Fig. 6-3. Offline measurement of $Z_{DM,IM}$ using an impedance analyzer.

Using a 1/2-hp IM (TECO AEAV1902R500FUS) as an illustration, Fig. 6-4 shows the frequency response of $Z_{DM,IM}$ of the healthy IM measured offline from 40 Hz to 5 MHz with an impedance analyzer (Agilent 4294A). As observed in Fig. 6-4, $Z_{DM,IM}$ exhibits a self-resonant frequency at 42 kHz. Therefore, the mid-to-high frequency range of the IM starts from 42 kHz.

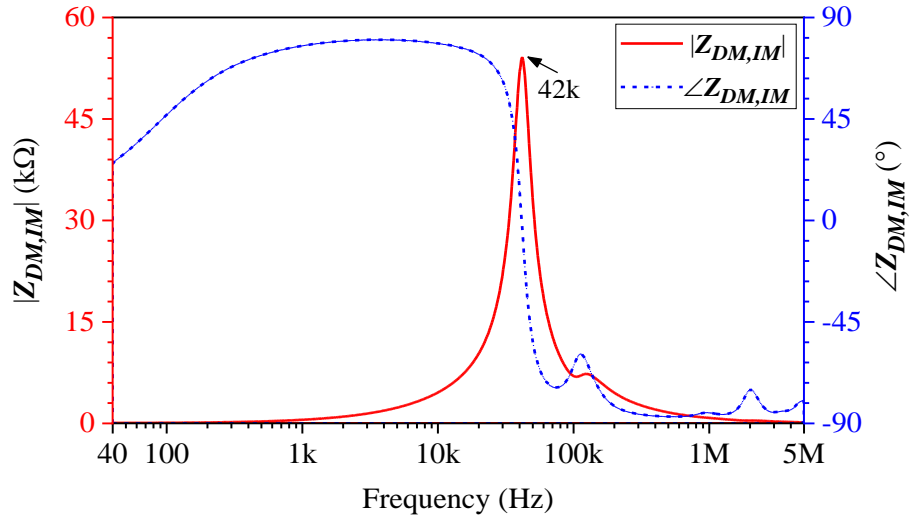

Fig. 6-4. Frequency response of $Z_{DM,IM}$ of the healthy TECO IM measured offline from 40 Hz - 5 MHz.

Once the mid-to-high frequency range of the IM is determined, the next step is to select the FOI within this frequency range as follows:

i. Measure $Z_{CM,IM}$ of the healthy IM offline in the mid-to-high frequency range using an impedance analyzer, as shown in Fig. 6-5.

ii. Measure $Z_{CM,IM}$ of the IM of the same model with ITSC faults of different severity levels offline in the same frequency range as (i) using an impedance analyzer.

iii. Compute the relative change in $Z_{CM,IM}$ caused by the ITSC faults and select a frequency where the relative change is significant even for low severity ITSC faults. The selected frequency is defined as the FOI.

The relative change in $Z_{CM,IM}$ at each frequency is defined and computed as follows:

$$\text{Relative Change} = \frac{|Z_{CM,IM}(\text{faulty}) - Z_{CM,IM}(\text{healthy})|}{|Z_{CM,IM}(\text{healthy})|} \times 100\% \qquad (6\text{-}1)$$

The criterion for judging whether the relative change is significant or not is 5%, which will be explained in Section 6.3. To emulate the ITSC faults with different severity levels, the connections of the stator windings of the IM are modified with various tapping points, as shown in Fig. 6-6. These tapping points are provided on the adjacent stator windings in each phase of the IM.

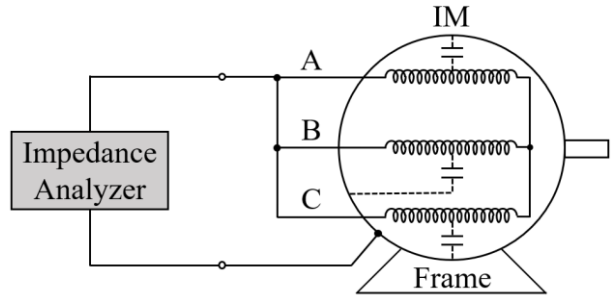

Fig. 6-5. Offline measurement of $Z_{CM,IM}$ using an impedance analyzer.

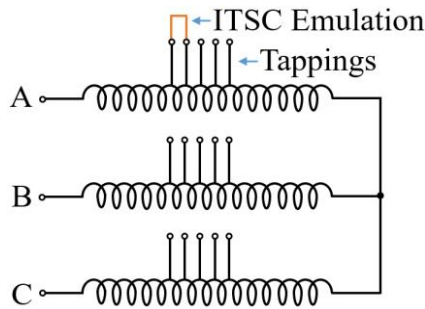

Fig. 6-6. Schematic of the emulated ITSC faults.

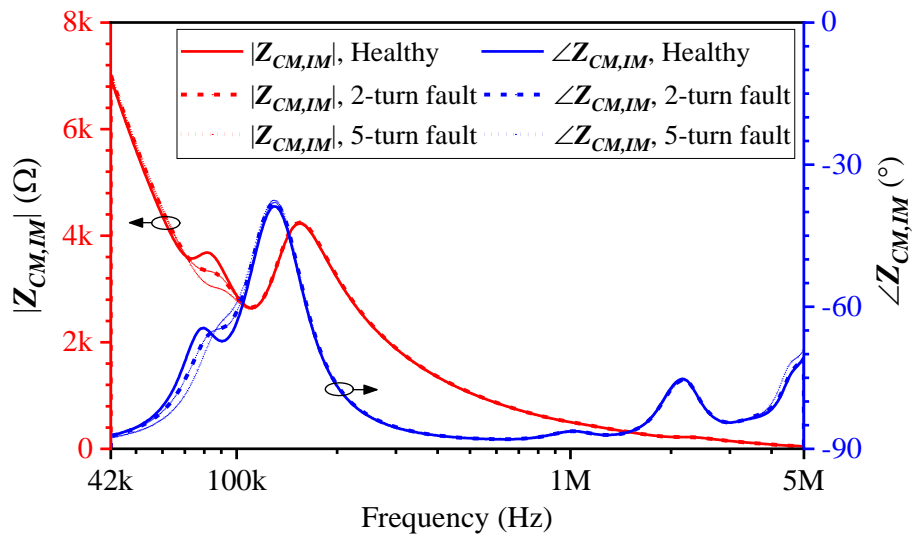

Fig. 6-7. Frequency responses of $Z_{CM,IM}$ of the healthy and faulty TECO IM measured offline from 42 kHz to 5 MHz.

Fig. 6-7 shows the measured offline frequency responses of $Z_{CM,IM}$ of the TECO IM

under the healthy condition and with emulated ITSC faults from 42 kHz to 5 MHz. Taking the frequency response of $Z_{CM,IM}$ of the healthy IM as a reference, Fig. 6-8 shows the relative change in $Z_{CM,IM}$ caused by two emulated ITSC faults of different severity levels, the 2-turn and 5-turn ITSC faults. As observed in Fig. 6-8, the relative change caused by the 2-turn fault is already noticeable (> 5%) between 66 kHz and 96 kHz and it is as high as 9.6% at 80 kHz. For the 5-turn fault, an increase in the relative change across almost the entire frequency range is clearly observed, especially between 66 kHz and 96 kHz, where the highest relative change increases to 17.2%. The observation in results has shown that a more severe ITSC fault causes a higher relative change. Therefore, for the given IM, a frequency between 66 kHz and 96 kHz can be selected as the FOI to monitor $Z_{CM,IM}$ for the detection of ITSC faults even the faults are at low severity levels (e.g. 2-turn fault).

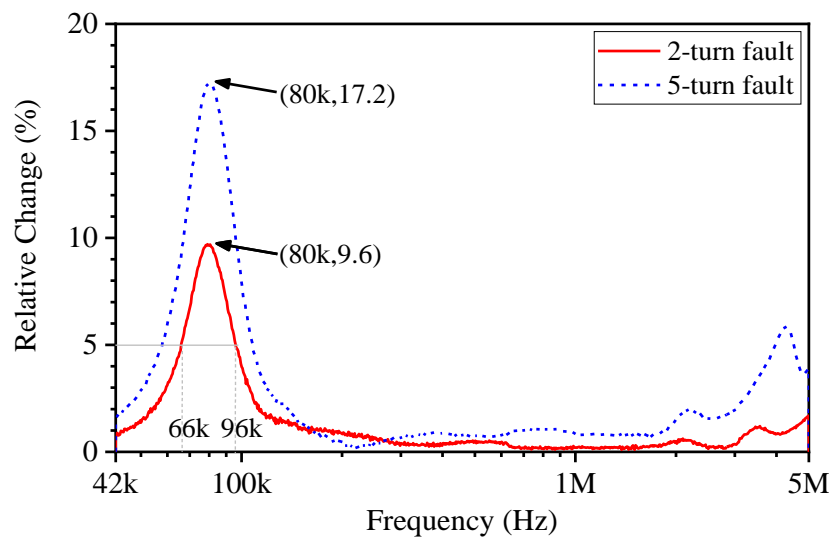

Fig. 6-8. Relative change in $Z_{CM,IM}$ of the TECO IM caused by ITSC faults with two fault severity levels (i.e. 2-turn fault and 5-turn fault).

## 6.2. Online Monitoring of IM's CM Impedance

The earlier section has shown that the ITSC faults can be detected by monitoring the relative change of $Z_{CM,IM}$. Also, the procedure to select an appropriate FOI for an IM to monitor its $Z_{CM,IM}$ has been described. This section elaborates on the measurement setup for online monitoring of $Z_{CM,IM}$ at the FOI. As mentioned in Chapter 2, there are mainly three online impedance extraction approaches; namely the capacitive coupling approach, the voltage-current measurement approach, and the inductive coupling approach. Among them, the measurement setup of the inductive coupling approach does not have any physical electrical connection to the energized system under test and therefore can be implemented with ease. Since the inverter used in the motor drive system (MDS) generates significant EMI noise and the MDS often experiences power surges [46], [68], the inductively coupled measurement setup incorporating signal amplification and surge protection modules is adopted for online monitoring of $Z_{CM,IM}$.

Fig. 6-9 shows the construction of a typical inverter-fed MDS and the CM current path in the MDS. A typical inverter-fed MDS consists of a variable-frequency drive (VFD) and an IM with 3-phase power cable in between. Usually, the AC input of the VFD is equipped with a built-in EMI filter to meet the EMI regulatory requirement [69]. The switching of power semiconductor devices in the VFD is the source of the CM noise at the AC output, which results in a CM current path formed by the 3-phase power cable buddle, the parasitic capacitance between the IM stator windings and its frame, and the ground return path.

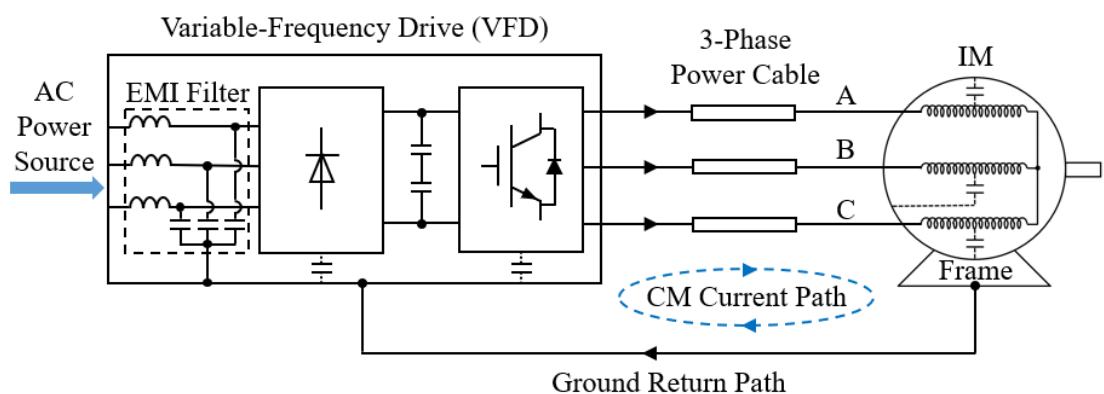

Fig. 6-9. A typical inverter-fed MDS with the CM current path.

Fig. 6-10(a) shows the measurement setup for online monitoring of $Z_{CM,IM}$. As mentioned in Section 5.1, it consists of a computer-controlled signal generation and acquisition system (SGAS), a power amplifier (PA), two attenuators (AT1 and AT2), two surge protectors (SP1 and SP2), a clamp-on injecting inductive probe (IIP), and a clamp-on receiving inductive probe (RIP). To monitor $Z_{CM,IM}$ online, the IIP and RIP are clamped onto the 3-phase power cable with the clamping position denoted as $p_1$-$p_2$.

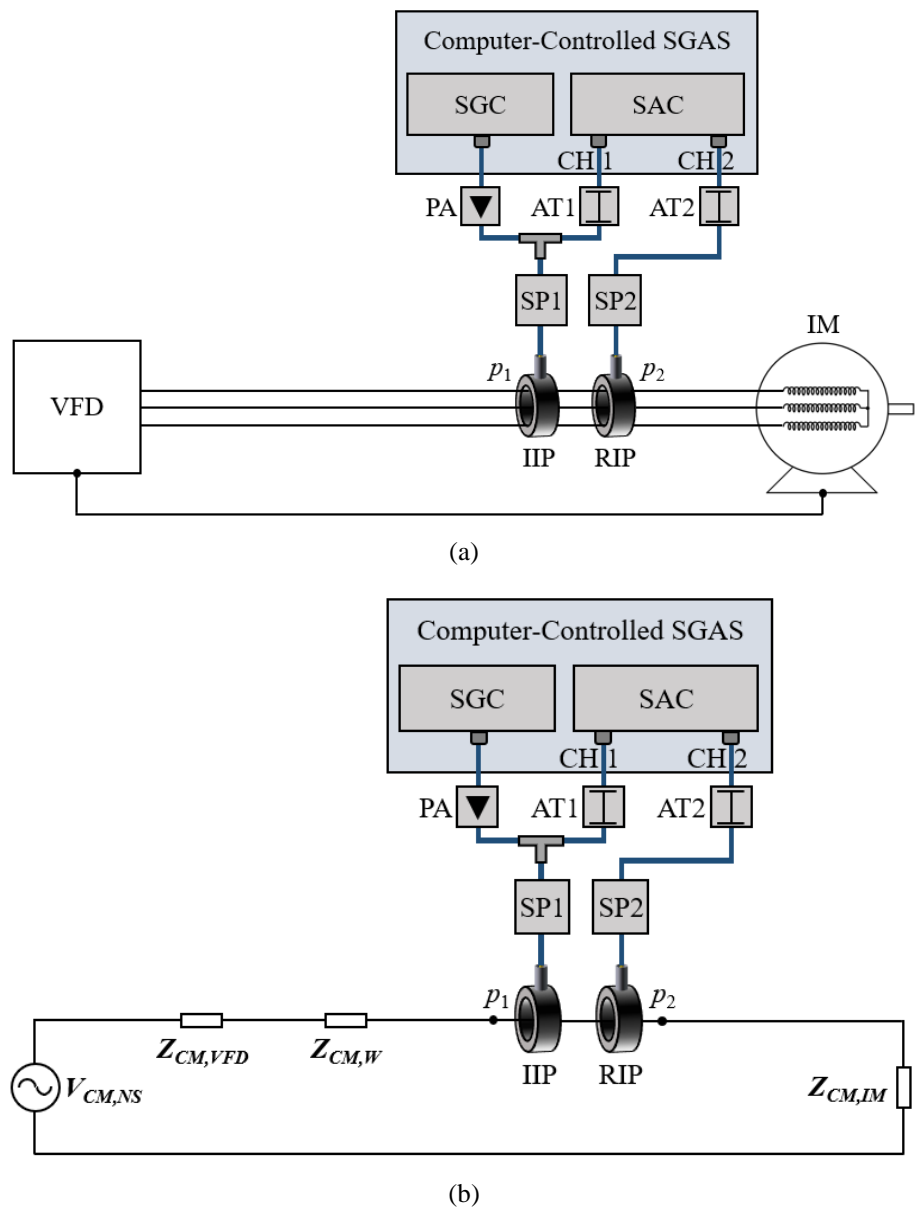

Fig. 6-10. (a) Measurement setup for online monitoring of $Z_{CM,IM}$; (b) CM equivalent circuit of the inverter-fed MDS with the measurement setup.

Fig. 6-10(b) shows the CM equivalent circuit of the MDS, in which $V_{CM,NS}$ is the equivalent CM noise source due to the switching of the VFD; $Z_{CM,VFD}$ is the equivalent CM impedance of the VFD; $Z_{CM,W}$ is the equivalent CM loop impedance of the 3-phase power cable and the ground return path. Therefore, the resultant impedance $Z_X$ seen at $p_1$-$p_2$ is given by:

$$Z_X = Z_{CM,VFD,CM} + Z_{CM,W} + Z_{CM,IM} \qquad (6\text{-}2)$$

Since the physical length of the power cable is much shorter than the wavelength of the FOI, the equivalent circuit concerned can be treated as a lumped circuit model [18]. As described in Section 5.2, for this measurement setup, once the three-term calibration technique is performed before online measurement, $Z_X$ can be directly extracted through this measurement setup. It should be explained that with the PA, the test signal can be amplified so that the level of the induced signal in the CM path of the MDS can be much higher than the level of the noise at the FOI to achieve adequate SNR (> 40 dB). Since the three-term calibration technique and the principle of extracting $Z_X$ have been well described in Section 4.2 and Section 5.2, respectively, they will not be repeated here. After obtaining $Z_X$, $Z_{CM,IM}$ can be extracted online once the online ($Z_{CM,VFD} + Z_{CM,W}$) is determined as follows:

i. Measure $Z_X$ of the healthy MDS online using the measurement setup. It should be noted that $Z_X$ at the FOI is unaffected by varying load and speed conditions, which will be verified in Section 6.3.

ii. Measure $Z_{CM,IM}$ of the healthy IM offline using an impedance analyzer, as shown in Fig. 6-5. Since the offline measured $Z_{CM,IM}$ can be treated as the online $Z_{CM,IM}$ at the FOI [67], [70], the online ($Z_{CM,VFD} + Z_{CM,W}$) can be obtained by subtracting the offline measured $Z_{CM,IM}$ from the online measured $Z_X$.

Noted that the online $Z_{CM,VFD}$ is mainly determined by the parasitic capacitances between the switching devices and the heat sink in the VFD, as well as by the capacitors (linked between each phase and the ground) in the built-in EMI filter [71]. Therefore, the online $Z_{CM,VFD}$ at a given frequency remains almost unchanged for a specific VFD under different operation modes [71], [72]. Besides, the online $Z_{CM,W}$ at a given frequency also remains almost unchanged for a specific 3-phase power cable and ground return path. Thus, once the online ($Z_{CM,VFD} + Z_{CM,W}$) at the FOI is determined and using the $Z_{CM,IM}$ at the FOI of the healthy IM as a reference, any changes of $Z_{CM,IM}$ due to the ITSC faults can be detected through online monitoring.

## 6.3. Experimental Validation

The same IM (TECO AEAV1902R500FUS) is used as the motor under test for experimental validation. The ratings of the IM are shown in Table 6-1. A VFD (TECO L510s) is used to control the IM and the length of the 3-phase power cable is 1 meter. The details of instruments of the actual measurement setup are given in Table 6-2.

Table 6-1. Ratings of the 3-phase IM under test.

| Rating | Value | Rating | Value |
|---|---|---|---|
| Power | 1/2-hp | Frequency | 50 Hz |
| Voltage | 220-240 V (Δ) | Pole | 2 |
| Current | 1.85-1.70 A (Δ) | Speed | 2765 RPM |

Table 6-2. Details of instruments (inst.) of the actual measurement setup.

| Inst. | Descriptions | Inst. | Descriptions |
|---|---|---|---|
| SGAS | NI PXI Platform | PA | Mini-Circuits ZHL-32A-S+ |
| SGC | NI PXI-5412 | AT1 | Pasternack PE7025-20 |
| SAC | NI PXI-5922 | AT2 | AIM-Cambridge 27-9300-6 |
| IIP | Solar 9144-1N | SP1 | Bussmann BSPD5BNCDI |
| RIP | Solar 9134-1 | SP2 | Bussmann BSPD5BNCDI |

Fig. 6-11 shows the actual measurement setup with the MDS under test. An adjustable friction load is driven by the IM with a belt. The ITSC faults are emulated by connecting additional low-impedance contacts (< 1 Ω) between the selected tappings as shown in Fig. 6-6. The response time with the measurement setup and the LabVIEW program to monitor and detect the ITSC faults is found to be within 1 s.

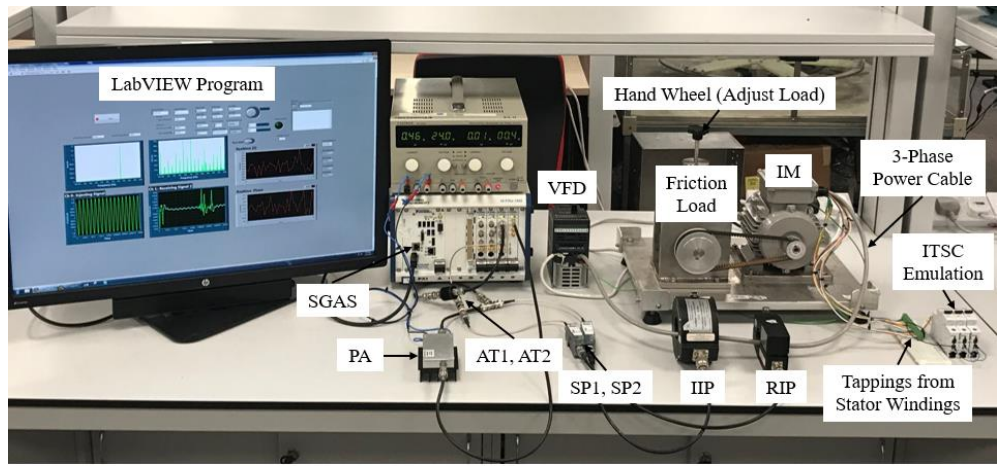

Fig. 6-11. Actual measurement setup with the MDS under test.

As mentioned in Subsection 6.1.2, to ensure reliable detection of ITSC faults, the FOI to monitor $Z_{CM,IM}$ of the IM is between 66 kHz and 96 kHz. For practical considerations, the FOI has to also be selected with care so that the background noise in the MDS at the FOI is relatively low to achieve the best SNR. Fig. 6-12 shows the noise current spectrum from 66 kHz to 96 kHz in the MDS under test. It shows that the frequencies with relatively low noise can be identified. It is observed from the enlarged image of Fig. 6-12 that the noise level at 82 kHz is rather low at 2.4 µA. Therefore, the FOI is selected to be 82 kHz to monitor $Z_{CM,IM}$ for online ITSC fault detection of the IM. As mentioned in Section 6.2, the SNR in the MDS is required to be larger than 40 dB for accurate extraction of $Z_{CM,IM}$. Fig. 6-13 shows the current spectrum around 82 kHz in the MDS under test with the present of the injected RF sinusoidal excitation signal at 82 kHz, which is clearly visible at 796 µA with an excitation signal source at 600 mV. With the background noise of 2.4 µA at 82 kHz, the induced signal achieves 50 dB SNR for online monitoring of $Z_{CM,IM}$. The induced 796 µA is too small to affect the normal operation of the MDS.

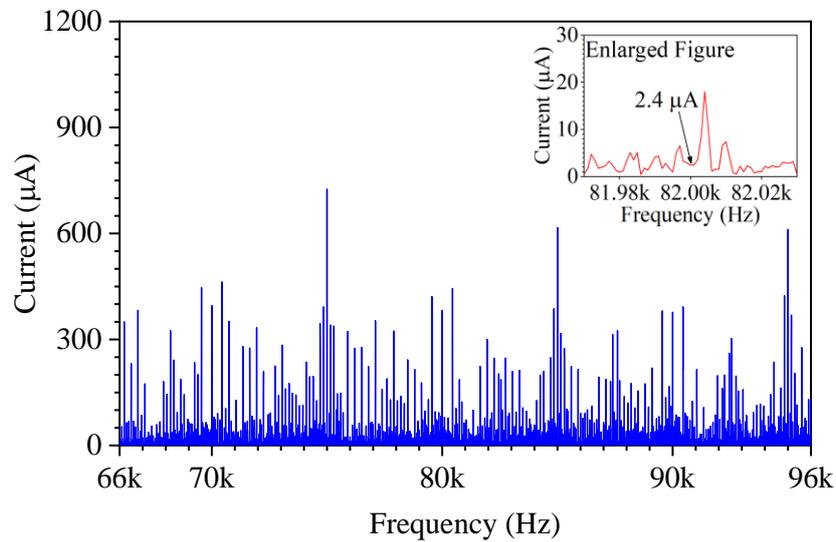

Fig. 6-12. Noise current spectrum from 66 kHz to 96 kHz in the MDS under test.

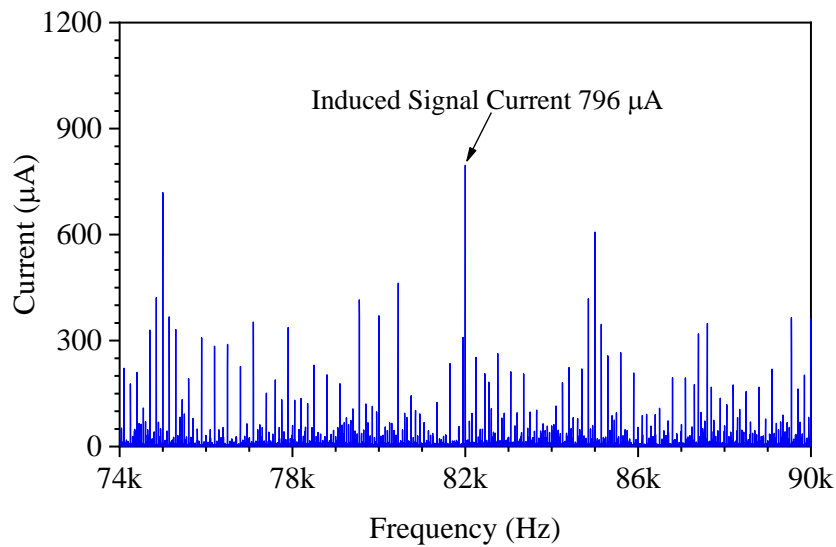

Fig. 6-13. Current spectrum around 82 kHz in the MDS under test when injecting an RF sinusoidal excitation signal (600 mV at 82 kHz) into the MDS by using the measurement setup.

Injecting the RF sinusoidal excitation signal (600 mV at 82 kHz) into the MDS under test with the measurement setup, Figs. 6-14 and 6-15 show the measured $Z_X$ of the healthy MDS at varying load (nominal speed) and varying speed conditions, respectively. It should be noted that all $Z_X$, $Z_{CM,IM}$, and $(Z_{CM,VFD} + Z_{CM,W})$ in this section refer to the impedance at the FOI (i.e. 82 kHz). As can be seen in Figs. 6-14 and 6-15, $Z_X$ remains unchanged to load and speed variations, even during the speed transient (ST).

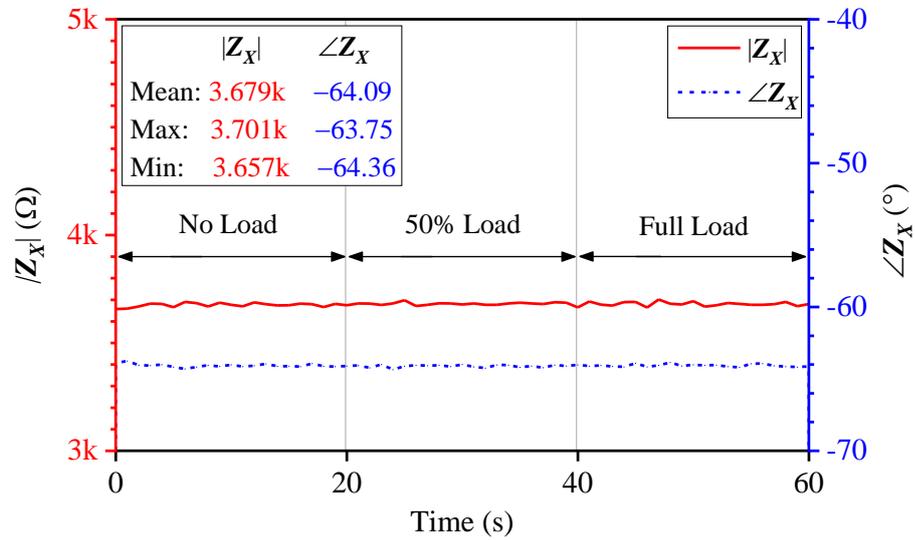

Fig. 6-14. $Z_X$ of the healthy MDS under test measured online with the measurement setup at varying load conditions (nominal speed).

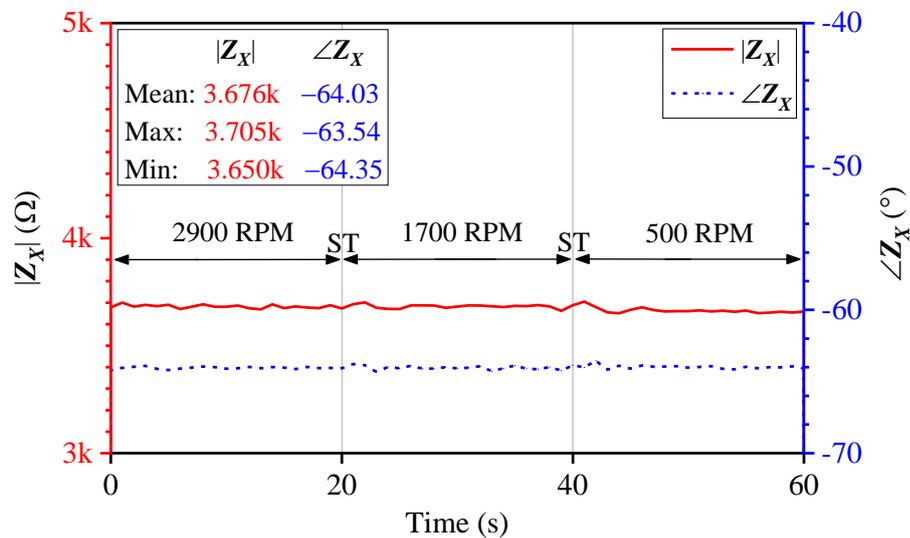

Fig. 6-15. $Z_X$ of the healthy MDS under test measured online with the measurement setup at varying speed conditions.

Using the measurement configuration shown in Fig. 6-5, the measured $Z_{CM,IM}$ of the healthy IM is 3.682 k$\Omega\angle$-64.98°. It should be noted that $(Z_{CM,VFD} + Z_{CM,W})$ is very small at 82 kHz due to the short length of the 3-phase power cable (i.e. 1 m) and the built-in EMI filter of the VFD [69]. Since $(Z_{CM,VFD} + Z_{CM,W}) \ll Z_{CM,IM}$, $Z_X$ can be approximated as $Z_{CM,IM}$ for the MDS under test. Therefore, the relative change in $Z_X$ is very much the same as the relative change in $Z_{CM,IM}$.

Fig. 6-16 shows the relative change in $Z_{CM,IM}$ of the IM at varying load conditions (nominal speed) when the IM is in the healthy condition and in the 2-turn ITSC fault condition, respectively. It is observed that the relative change is unaffected by the load variations. Also, when the IM is in the 2-turn ITSC fault condition, the relative change remains almost unchanged at about 9.4%. However, there is a very small relative change ($< 1\%$) when the IM is in the healthy condition. From Fig. 6-14, such a minor relative change is caused by the slight normal fluctuation in the monitored impedance. Fig. 6-17 shows the relative change in $Z_{CM,IM}$ of the IM at varying speed conditions when the IM is in the healthy condition and in the 2-turn ITSC fault condition, respectively. Again, the relative change is unaffected by the speed variations. Also, under the 2-turn ITSC fault condition, the relative change still remains almost unchanged at 9.4%. Besides, the minor relative change caused by the slight normal fluctuation in the monitored impedance is still kept within 1%. From Figs. 6-16 and 6-17, it is found that the relative change in $Z_{CM,IM}$ at the FOI is unaffected by the load and speed variations. In addition, since the minor relative change due to the slight normal fluctuation in the monitored impedance is kept within 1%, the threshold can be determined to be 1% for the ITSC fault detection. Based on the 1% threshold, the relative change can be treated as significant when it is larger than 5%. Furthermore, even the IM with a 2-turn ITSC fault has shown a rather noticeable relative change of 9.4%, which indicates that the proposed method has good sensitivity to detect the ITSC faults that are less severe. Table 6-3 shows the relative change in $Z_{CM,IM}$ of the IM for the ITSC faults with different severity levels, which increases with higher ITSC fault severity.

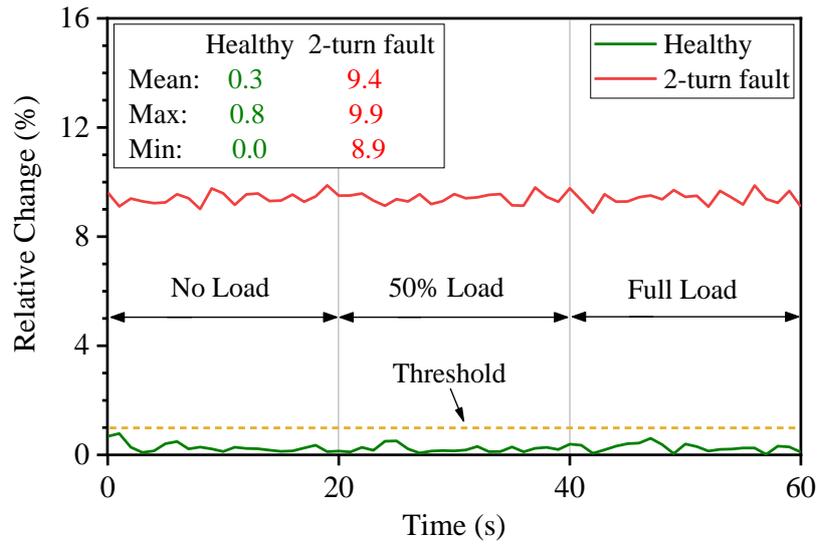

Fig. 6-16. Relative change in $Z_{CM,IM}$ of the IM (healthy and 2-turn fault) monitored online with the measurement setup at varying load conditions (nominal speed).

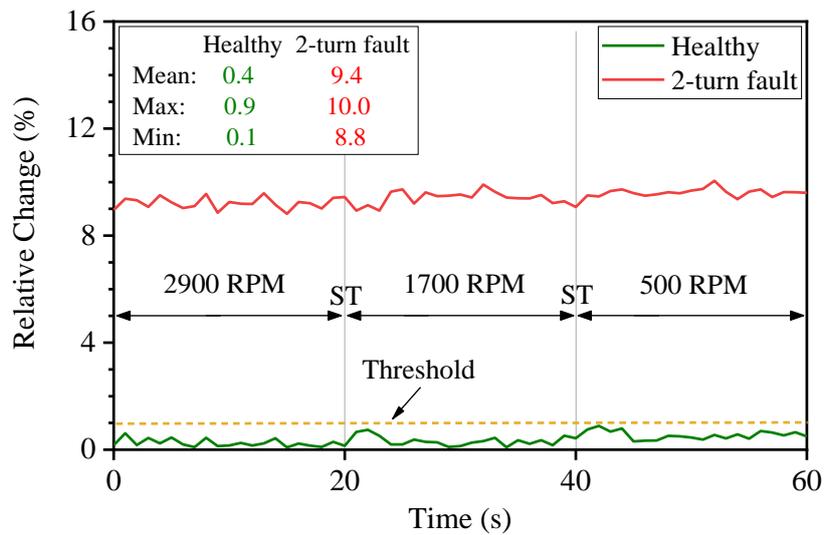

Fig. 6-17. Relative change in $Z_{CM,IM}$ of the IM (healthy and 2-turn fault) monitored online with the measurement setup at varying speed conditions.

Table 6-3. Relative change in $Z_{CM,IM}$ of the IM for different ITSC fault severities

| Fault Severity Level | Relative Change in $Z_{CM,IM}$ | | |
|---|---|---|---|
| | Mean Value | Maximum Value | Minimum Value |
| 2-turn fault | 9.4% | 10.0% | 8.8% |
| 3-turn fault | 10.5% | 11.2% | 10.1% |
| 4-turn fault | 12.1% | 12.9% | 11.5% |
| 5-turn fault | 17.4% | 18.1% | 16.6% |

As mentioned earlier, besides the stator faults, the rotor and bearing faults are another two of the most prevalent types of the IM faults. It is generally believed that the bearing faults are developed from the general roughness [46], and the rotor cage fault is a typical rotor fault mode [45], [73]. As a proof of concept, the abnormal friction of bearing and the rotor cage fault of the IM are emulated (see Fig. 6-18) to check their possible influence on the detection of the ITSC faults. Fig. 6-19 shows the relative change in $Z_{CM,IM}$ of the IM for the abnormal friction of bearing and the rotor cage fault. It is found that the relative changes in $Z_{CM,IM}$ for both types of faults are well within 1%, which is negligible to have any impact on the ITSC fault detection.

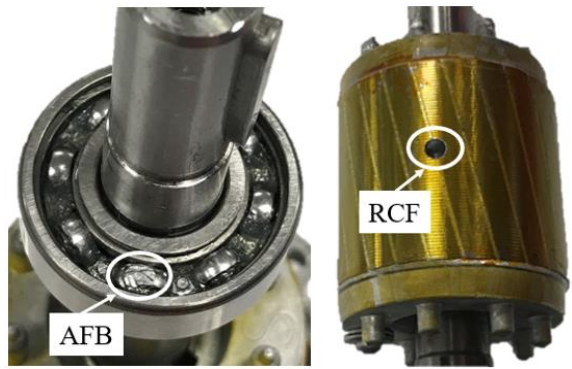

Fig. 6-18. Emulated abnormal friction of bearing (AFB) and emulated rotor cage fault (RCF) of the IM.

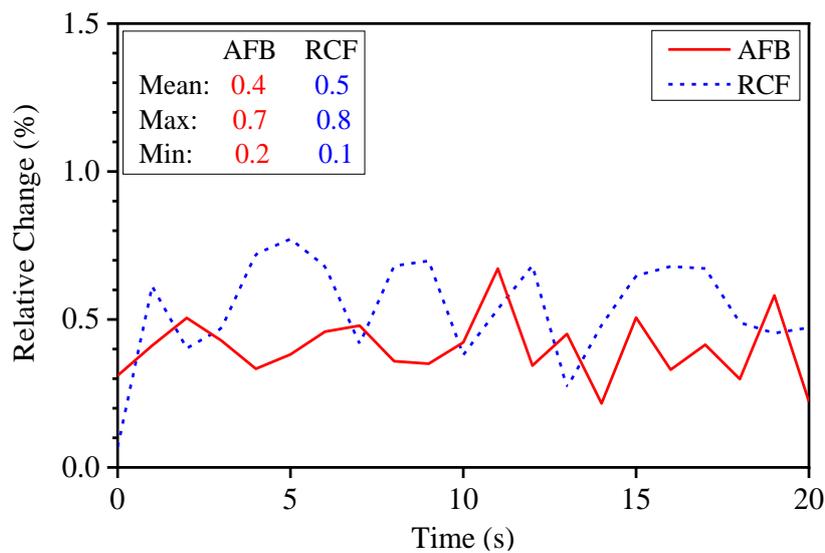

Fig. 6-19. Relative change in $Z_{CM,IM}$ of the IM at the conditions of the AFB and the RCF.

Based on these findings, the proposed method has shown its reliability and robustness for the stator ITSC fault detection under varying load and speed conditions, as well as under rotor fault and bearing fault conditions.

In this chapter, a novel signal-based method for the detection of stator ITSC faults in the inverter-fed IM has been described and developed. The measurement setup of the proposed method does not have any physical electrical connection to the energized IM under test and therefore eliminates electrical safety hazards. In addition, the clamp-on IIP and RIP used in the measurement setup can be easily mounted on or removed from the power cable connected to the IM and therefore facilitates ease of implementation. By monitoring the CM impedance at a carefully selected FOI, it is efficient to detect the ITSC faults to enable fast remedial response. This method has been experimentally verified to be reliable and robust for the ITSC fault detection, as it is invariant to motor load and speed variations, as well as bearing and rotor faults.

# Chapter 7 Application in Voltage-Dependent Capacitance Extraction of SiC Power MOSFET

Compared to the conventional power semiconductor devices, the wide-bandgap (WBG) power semiconductor devices such as the silicon carbide (SiC) power MOSFETs have lower on-resistance, higher switching frequency, and higher rated junction temperature, and therefore are gaining more popularity [74]-[77]. However, the higher switching frequency has resulted in undesirable oscillations at the rising and falling edges due to the resonance of the circuit loop inductances and the interelectrode capacitances of the SiC power MOSFET [78], [79]. Such oscillations can reduce the reliability of the system due to their associated EMI [80], [81]. To evaluate and mitigate the negative effects in the early design phase, the values of the interelectrode capacitances of the SiC power MOSFET must be obtained [82]-[84]. Since the interelectrode capacitances are voltage-dependent [85], knowing their values in the entire voltage operating range is required.

There are several methods reported in the literature to extract the voltage-dependent capacitances. One of them is using 3-D full-wave simulation tools, such as the finite element method, to compute the capacitances [86]. This method is possible only if the full details of the geometries and material properties of the SiC power MOSFET are known, which may not always be readily available since intellectual property issues. Another method is using the impedance analyzer [20], in which many additional accessories, such as DC blocking capacitors, AC blocking inductors, various resistors, and protection diodes are necessary for safe and efficient online measurements. A time-domain reflectometry method has also been reported [85], but proper isolation is required to ensure that the measurement setup is disconnected from the ground. Besides, this method is only limited to extract the output capacitance of the power MOSFET. A three-probe measurement method has also been reported [87], but the parasitic components of the probes and the probe-to-probe coupling between them are neglected during the measurement process, which may affect the accuracy of the measured capacitances. In addition to the aforementioned methods, a power device analyzer (Keysight B1505A) is also available on the market. It requires to connect a specific bias-tee to extract the voltage-dependent capacitance under high voltage biasing conditions. Since the bias-tee has a physical electrical connection to the high voltage DC power supply, additional protection devices are required for safety hazard compliance. To extract the voltage-dependent capacitances of the SiC power MOSFET in a simple and accurate manner, this chapter proposes an alternative approach based on the inductive coupling approach. The measurement setup of this approach does not have any physical electrical connection to the SiC power MOSFET that is biased by high DC voltage, thereby eliminating potential safety hazards without introducing extra protection devices. Besides, the measurement setup is simple for on-site implementation.

This chapter is organized as follows. Section 7.1 introduces an equivalent circuit model of the SiC power MOSFET. Section 7.2 describes the principle and process of extracting the voltage-dependent capacitances of the SiC power MOSFET. Using a 1.2 kV SiC power MOSFET as a test case, Section 7.3 verifies the ability of the proposed method to extract the values of the capacitances over a wide range of voltage biasing.

## 7.1. Equivalent Circuit Model of SiC Power MOSFET

Fig. 7-1 shows a small-signal equivalent circuit model of the SiC power MOSFET in the off-state, where $d$, $g$, and $s$ are the drain, gate, and source terminals, respectively. $C_{gd}$, $C_{gs}$, and $C_{ds}$ are the interelectrode capacitances; $L_d$, $L_g$, and $L_s$ are the terminal

inductances; $R_d$, $R_g$, and $R_s$ are the terminal resistances.

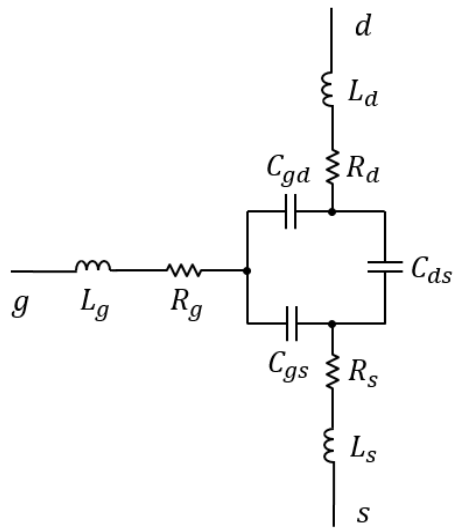

Fig. 7-1. Small-signal equivalent circuit model of the SiC power MOSFET in the off-state.

The terminal inductances are typically a few nanohenry [88]-[90] and the interelectrode capacitances are typically smaller than a few nanofarad [91]-[94]. Therefore, in the frequency range of $f_{sig} \leq 1$ MHz, the impedance of the SiC power MOSFET in the off-state is dominated by its interelectrode capacitances, in which the terminal inductances are negligible during the capacitances extraction process [20], [87]. Thus, the equivalent circuit model in Fig. 7-1 can be simplified to that shown in Fig. 7-2.

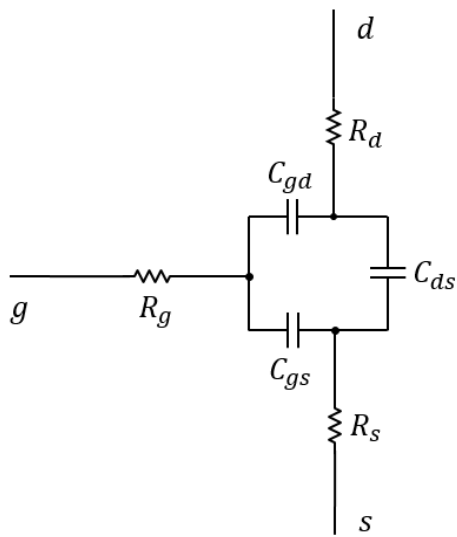

Fig. 7-2. Simplified equivalent circuit model of Fig. 7-1 when $f_{sig} \leq 1$ MHz.

Fig. 7-3 shows the star-connected equivalent circuit model of Fig. 7-2. By using delta-to-star circuit transformation, the star-connected capacitances $C_{dn}$, $C_{gn}$, and $C_{sn}$ are determined by:

$$C_{dn} = C_{gd} + C_{ds} + C_{gd}C_{ds}/C_{gs} \tag{7-1}$$

$$C_{gn} = C_{gd} + C_{gs} + C_{gd}C_{gs}/C_{ds} \tag{7-2}$$

$$C_{sn} = C_{gs} + C_{ds} + C_{gs}C_{ds}/C_{gd} \tag{7-3}$$

Based on the star-connected capacitances, the equivalent star-connected impedances $\boldsymbol{Z_{dn}}$, $\boldsymbol{Z_{gn}}$, and $\boldsymbol{Z_{sn}}$ are obtained by:

$$\boldsymbol{Z_{dn}} = R_d + 1/(j\omega C_{dn}) \tag{7-4}$$

$$\boldsymbol{Z_{gn}} = R_g + 1/(j\omega C_{gn}) \tag{7-5}$$

$$\boldsymbol{Z_{sn}} = R_s + 1/(j\omega C_{sn}) \tag{7-6}$$

where $\omega = 2\pi f_{sig}$.

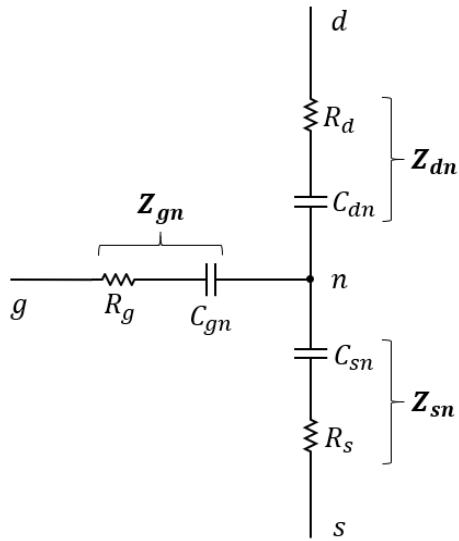

Fig. 7-3. Star-connected equivalent circuit model of Fig. 7-2.

## 7.2. Principle of Voltage-Dependent Capacitance Extraction

From (7-4) to (7-6), once $Z_{dn}$, $Z_{gn}$, and $Z_{sn}$ are obtained, $C_{dn}$, $C_{gn}$, and $C_{sn}$ can be determined at a static functionality. Furthermore, $C_{gd}$, $C_{gs}$, and $C_{ds}$ can be obtained based on the determined $C_{dn}$, $C_{gn}$, and $C_{sn}$ by (7-7) to (7-9).

$$C_{gd} = C_{gn}C_{dn}/(C_{dn} + C_{gn} + C_{sn}) \tag{7-7}$$

$$C_{gs} = C_{gn}C_{sn}/(C_{dn} + C_{gn} + C_{sn}) \tag{7-8}$$

$$C_{ds} = C_{dn}C_{sn}/(C_{dn} + C_{gn} + C_{sn}) \tag{7-9}$$

Fig. 7-4 shows the test circuit to extract the equivalent terminal impedances, in which a capacitor ($C_1$) is placed in parallel with the DC power supply to provide an AC short circuit to the test signal ($f_{sig}$). $Z_{ext1}$, $Z_{ext2}$, and $Z_{ext3}$ are the equivalent impedances of the external wires, which can be measured offline with an impedance analyzer. The selected measurement setup is shown in Fig. 3-1, in which the signal amplification and surge protection modules are not incorporated because there is no significant electrical noise and power surges in the test circuit.

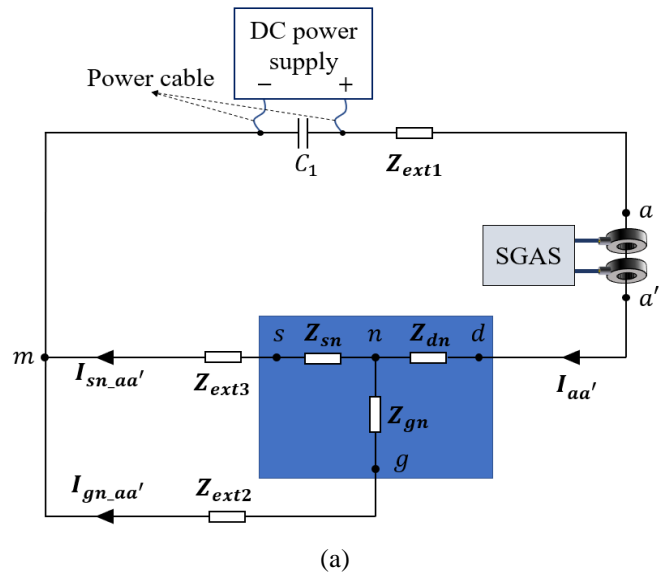

(a)

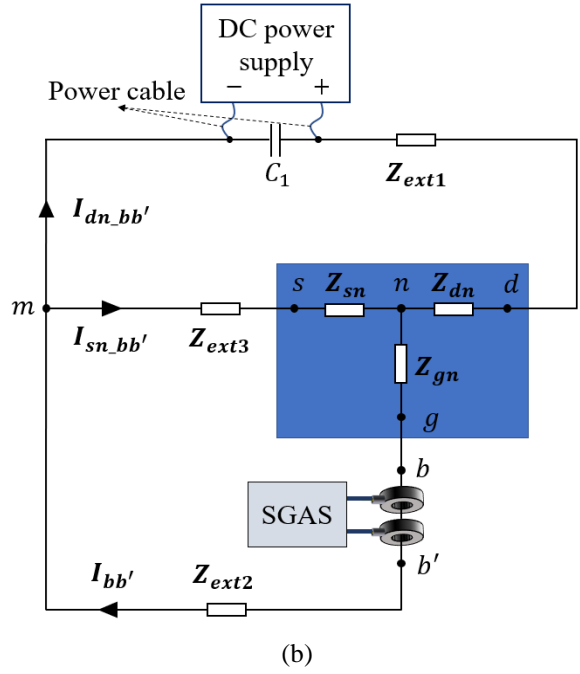

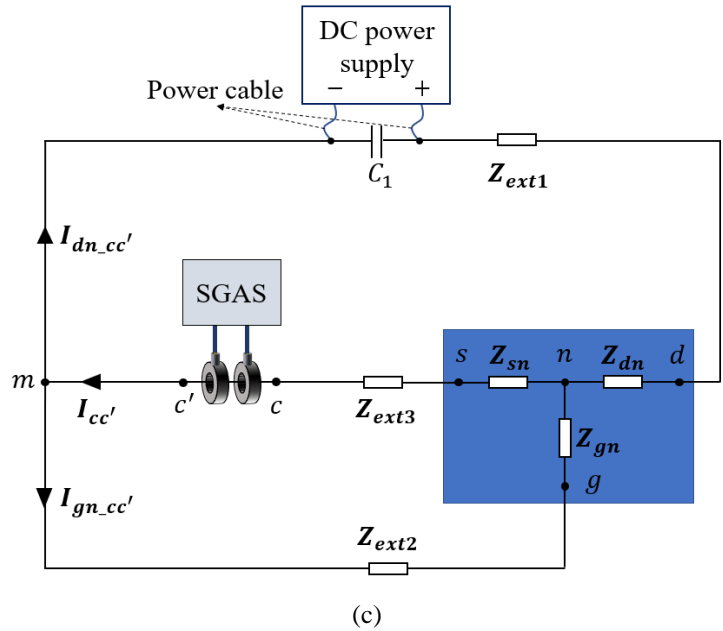

Fig. 7-4. Test circuit to extract $Z_{dn}$, $Z_{gn}$, and $Z_{sn}$ with the inductive probes mounted between (a) position $a$-$a'$, (b) position $b$-$b'$, and (c) position $c$-$c'$.

In Fig. 7-4(a), the IIP and RIP are mounted between position $a$-$a'$. A test signal of frequency $f_{sig}$ and current $I_{aa'}$ is injected into the test circuit from the SGC through the IIP. When the test signal current passes through point $n$, it is divided into two paths, one through terminal $s$ ($I_{sn\_aa'}$) and another via terminal $g$ ($I_{gn\_aa'}$). Based on Kirchhoff's current law (KCL), $I_{aa'}$ is given by

$$I_{aa'} = I_{gn\_aa'} + I_{sn\_aa'} \tag{7-10}$$

Thus, the resultant impedance measured by the measurement setup when the IIP and RIP are mounted between position $a$-$a'$ is given by

$$Z_{aa'} = V_{aa'}/I_{aa'} \tag{7-11}$$

where $V_{aa'} = V_{aa'1} - V_{aa'2}$; $V_{aa'1}$ and $V_{aa'2}$ represent the test signal voltages at the output port of the IIP and the input port of the RIP, respectively. Furthermore, from Fig. 7-4(a)

$$I_{gn\_aa'} = V_{nm\_aa'}/(Z_{ext2} + Z_{gn}) \tag{7-12}$$

$$I_{sn\_aa'} = V_{nm\_aa'}/(Z_{ext3} + Z_{sn}) \tag{7-13}$$

$$V_{aa'} = I_{aa'}(Z_{ext1} + Z_{dn}) + V_{nm\_aa'} \tag{7-14}$$

where $V_{nm\_aa'}$ is the test signal voltage between point *n* and *m*. Substituting (7-14) into (7-11), $Z_{aa'}$ can be re-written as

$$Z_{aa'} = Z_{ext1} + Z_{dn} + V_{nm\_aa'}/I_{aa'} \tag{7-15}$$

Substituting (7-12) and (7-13) into (7-10), $I_{aa'}$ can be re-written as

$$I_{aa'} = V_{nm\_aa'}/(Z_{ext2} + Z_{gn}) + V_{nm\_aa'}/(Z_{ext3} + Z_{sn}) \tag{7-16}$$

Substituting (7-16) into (7-15), $Z_{aa'}$ can be finally extracted as

$$Z_{aa'} = Z_{ext1} + Z_{dn} + \frac{(Z_{ext2} + Z_{gn})(Z_{ext3} + Z_{sn})}{(Z_{ext2} + Z_{gn}) + (Z_{ext3} + Z_{sn})} \tag{7-17}$$

Similarly, Fig. 7-4(b) and Fig. 7-4(c) show the current paths when the IIP and RIP are mounted between position $b$-$b'$ and between position $c$-$c'$, respectively. Based on KCL, (7-18) and (7-19) are obtained

$$I_{bb'} = I_{dn\_bb'} + I_{sn\_bb'} \tag{7-18}$$

$$I_{cc'} = I_{dn\_cc'} + I_{gn\_cc'} \tag{7-19}$$

Therefore, the resultant impedances measured by the measurement setup when the IIP and RIP are mounted between position $b$-$b'$ ($Z_{bb'}$) and between position $c$-$c'$ ($Z_{cc'}$) are extracted as follows

$$Z_{bb'} = Z_{ext2} + Z_{gn} + \frac{(Z_{ext1} + Z_{dn})(Z_{ext3} + Z_{sn})}{(Z_{ext1} + Z_{dn}) + (Z_{ext3} + Z_{sn})} \tag{7-20}$$

$$Z_{cc'} = Z_{ext3} + Z_{sn} + \frac{(Z_{ext1} + Z_{dn})(Z_{ext2} + Z_{gn})}{(Z_{ext1} + Z_{dn}) + (Z_{ext2} + Z_{gn})} \tag{7-21}$$

From (7-17), (7-20), and (7-21), $Z_{dn}$, $Z_{gn}$, and $Z_{sn}$ can be obtained via (7-24) to (7-26):

$$\Delta = Z_{aa'}^2 Z_{bb'}^2 - 2Z_{aa'}^2 Z_{bb'} Z_{cc'} + Z_{aa'}^2 Z_{cc'}^2 - 2Z_{aa'} Z_{bb'}^2 Z_{cc'} \\ -2Z_{aa'} Z_{bb'} Z_{cc'}^2 + Z_{bb'}^2 Z_{cc'}^2 \tag{7-22}$$

$$\Lambda = -2Z_{aa'} Z_{bb'} Z_{cc'} \tag{7-23}$$

$$Z_{dn} = \frac{\Lambda(Z_{aa'} Z_{bb'} + Z_{aa'} Z_{cc'} - Z_{bb'} Z_{cc'})}{\Delta} - Z_{ext1} \tag{7-24}$$

$$Z_{gn} = \frac{\Lambda(Z_{aa'} Z_{bb'} - Z_{aa'} Z_{cc'} + Z_{bb'} Z_{cc'})}{\Delta} - Z_{ext2} \tag{7-25}$$

$$Z_{sn} = \frac{\Lambda(-Z_{aa'} Z_{bb'} + Z_{aa'} Z_{cc'} + Z_{bb'} Z_{cc'})}{\Delta} - Z_{ext3} \tag{7-26}$$

As mentioned earlier, $Z_{aa'}$, $Z_{bb'}$, and $Z_{cc'}$ can be directly extracted online using the measurement setup. $Z_{ext1}$, $Z_{ext2}$, and $Z_{ext3}$ can be measured offline using an impedance analyzer. After $Z_{dn}$, $Z_{gn}$, and $Z_{sn}$ are obtained, $C_{dn}$, $C_{gn}$, and $C_{sn}$ can be determined by (7-4) to (7-6). Furthermore, $C_{gd}$, $C_{gs}$, and $C_{ds}$ can be derived based on the extracted $C_{dn}$, $C_{gn}$, and $C_{sn}$ through (7-7) to (7-9). Based on the obtained $C_{dn}$, $C_{gn}$, and $C_{sn}$, the input capacitance ($C_{iss}$), output capacitance ($C_{oss}$), and reverse transfer capacitance ($C_{rss}$) of the SiC power MOSFET can also be extracted via (7-27) to (7-29).

$$C_{iss} = C_{gs} + C_{gd} \qquad (7\text{-}27)$$

$$C_{oss} = C_{ds} + C_{gd} \qquad (7\text{-}28)$$

$$C_{rss} = C_{gd} \qquad (7\text{-}29)$$

## 7.3. Experimental Validation

Fig. 7-5 shows an actual measurement setup for experimental validation. The computer-controlled SGAS is a National Instruments (NI) PXI platform, which consists of a signal generation card (PXI-5412), a two-channel signal acquisition card (PXI-5122), a computer interface card (PXI-8360), and a back panel (PXI-1031). A laptop with programmable software was connected to the SGAS for instrument control. Two Tektronix CT1 current probes were selected as the IIP and RIP. A Cree SiC power MOSFET (C2M0080120D, 1.2 kV) was selected as the test sample for extracting its voltage-dependent capacitances at varying drain-source voltage ($V_{ds}$), where it was controlled by a Xantrex DC power supply (XDC 600-10, 0 to 600 V). Besides, an impedance analyzer (Agilent 4294A) was used to extract the values of $Z_{ext1}$, $Z_{ext2}$, and $Z_{ext3}$ offline. To verify the proposed method, the capacitances specified in the manufacturer's datasheet are used as a reference for comparison. Since the datasheet

only gives the values of $C_{iss}$, $C_{oss}$, and $C_{rss}$ at the frequency of 1 MHz, $f_{sig}$ is set to 1 MHz for ease of comparison. Besides, $C_1$ was selected as 1 μF (TDK B32656S7105K561, 1.25 kV). $\boldsymbol{Z_{ext1}}$, $\boldsymbol{Z_{ext2}}$, and $\boldsymbol{Z_{ext3}}$ at 1 MHz were measured by the Agilent impedance analyzer, and their respective impedance values are 0.4422 Ω∠77.4°, 0.2053 Ω∠79.1°, and 0.0493 Ω∠54.8°.

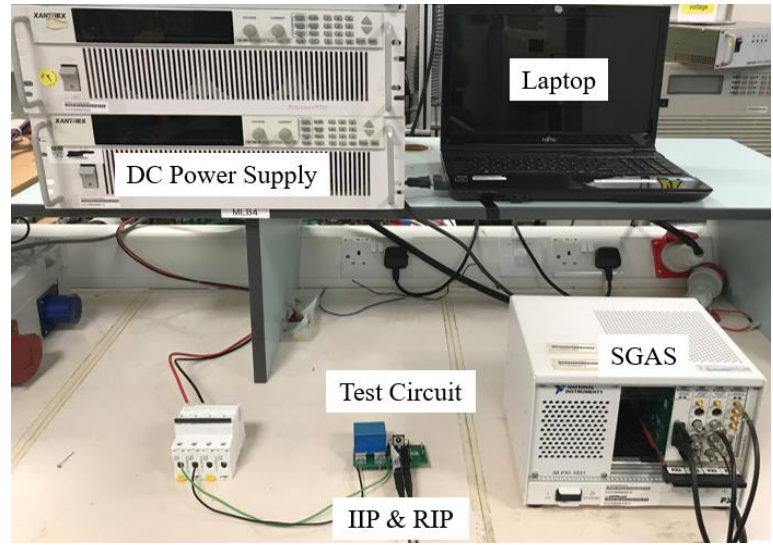

Fig. 7-5. Measurement setup to extract the voltage-dependent capacitances of the Cree SiC power MOSFET.

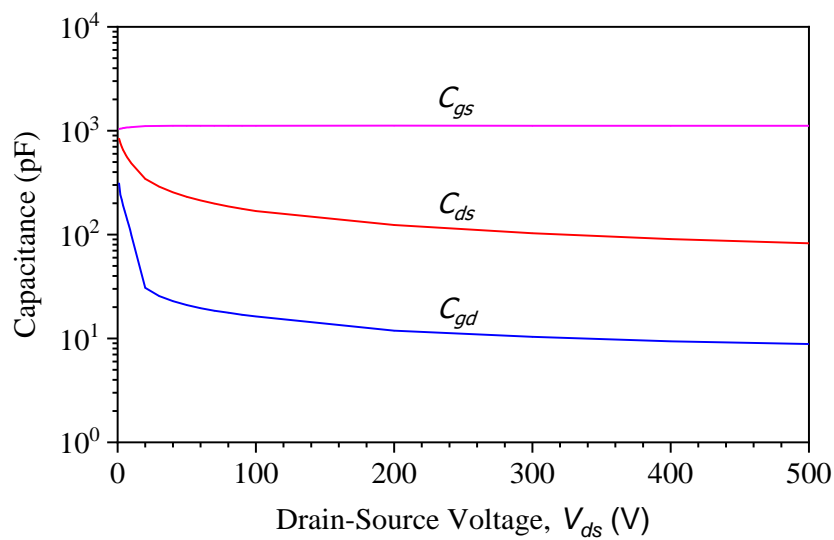

Fig. 7-6. Extracted interelectrode capacitances of the Cree SiC power MOSFET through the proposed method at the frequency of 1 MHz.

Fig. 7-6 shows the extracted interelectrode capacitances of the Cree SiC power

MOSFET with $V_{ds}$ varying from 0 to 500 V. Since the datasheet only gives the values of $C_{iss}$, $C_{oss}$, and $C_{rss}$, the corresponding capacitances are derived from the extracted interelectrode capacitances for ease of comparison. Fig. 7-7 shows the values of $C_{iss}$, $C_{oss}$, and $C_{rss}$ extracted via the proposed method and those provided by the datasheet. Based on the 23 measurement points at different $V_{ds}$, Table 7-1 lists the measurement errors. As observed in Fig. 7-7 and Table 7-1, the capacitances extracted by the proposed method are very consistent with those provided by the datasheet, which is a good indication of the validity of this method to extract the voltage-dependent capacitances of the SiC power MOSFET with a reasonable accuracy for a wide range of voltage biasing.

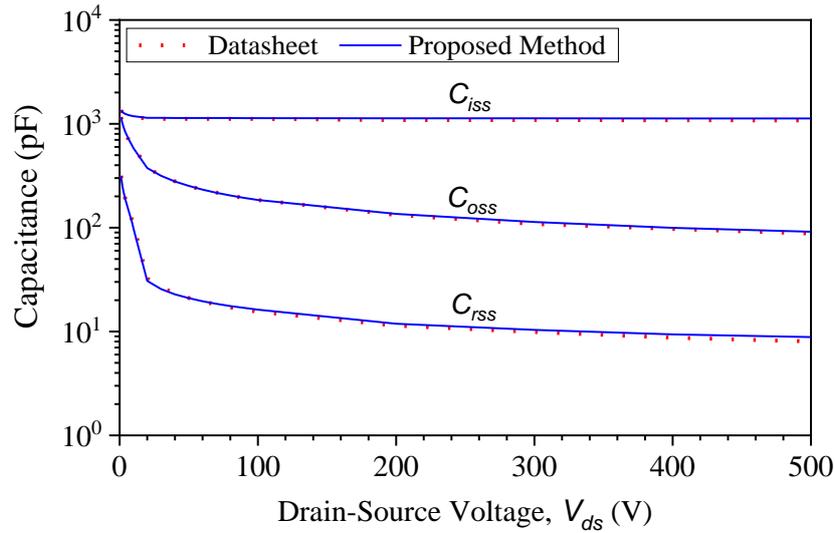

Fig. 7-7. Comparison of the extracted $C_{iss}$, $C_{oss}$, and $C_{rss}$ via the proposed method and those provided by the datasheet.

Table 7-1. Errors of the extracted capacitances for 23 measurement points shown in Fig. 7-7.

| Capacitance | $C_{iss}$ | $C_{oss}$ | $C_{rss}$ |
| --- | --- | --- | --- |
| Mean absolute error | 1.9% | 1.5% | 3.7% |
| Standard deviation of absolute error | 1.2% | 1.4% | 2.7% |
| Maximum absolute error | 4.3% | 4.4% | 9.8% |

Fig. 7-8 shows the values of $C_{iss}$, $C_{oss}$, and $C_{rss}$ obtained by the proposed method when $f_{sig}$ is 100 kHz, 1MHz, and 10 MHz. The values of $C_{iss}$, $C_{oss}$, and $C_{rss}$ at 100 kHz are slightly different from those extracted at 1 MHz. However, the values of $C_{iss}$, $C_{oss}$, and $C_{rss}$ at 10 MHz have a larger deviation from those extracted at 100 kHz or 1 MHz. The larger deviation of these extracted capacitances at 10 MHz is expected because of the effect of the terminal inductances ($L_d$, $L_g$, and $L_s$). When $f_{sig} \leq 1$ MHz, the impedances of the terminal inductances are much smaller than those of the interelectrode capacitances. However, when $f_{sig}$ increases to 10 MHz, the impedances of the terminal inductances become significant and cannot be ignored, which explains the larger deviation of the obtained capacitances from those extracted at 100 kHz and 1 MHz. Therefore, when the proposed method is used to extract the voltage-dependent capacitances of the SiC power MOSFET, $f_{sig}$ should not exceed 1 MHz to ensure measurement accuracy.

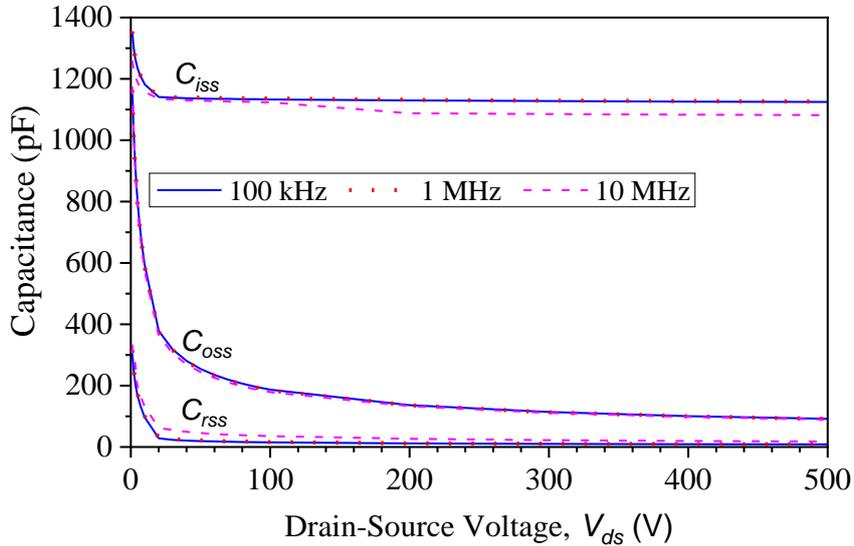

Fig. 7-8. Comparison of the values of $C_{iss}$, $C_{oss}$, and $C_{rss}$ obtained by the proposed method at the frequency of 100 kHz, 1 MHz, and 10 MHz.

In this chapter, a novel method to extract the voltage-dependent capacitances of the SiC power MOSFET based on the inductive coupling approach has been described and verified. The proposed measurement setup can extract the voltage-dependent capacitances of the SiC power MOSFET under its actual high voltage biasing condition without any physical electrical connection to the energized power MOSFET and therefore it is safe and relatively simple to set up. A 1.2 kV SiC power MOSFET was selected as the test case, and the accuracy of the extracted capacitances has been verified with the datasheet provided by the manufacturers. With the voltage-dependent capacitances of the SiC power MOSFET, its equivalent circuit model can be represented to evaluate and mitigate potential EMI issues in the early design phase of the power converter.

# Chapter 8 Conclusion and Future Work

## 8.1. Conclusion

Given the significance and importance of the online impedance of many critical electrical systems, a comprehensive literature review on the existing online impedance extraction approaches has been covered. In view of the limitations of the existing approaches, this thesis has proposed several further improvements on the measurement setup of the inductive coupling approach.

Firstly, by using a computer-controlled signal generation and acquisition system (SGAS), a clamp-on injecting inductive probe (IIP), and a clamp-on receiving inductive probe (RIP) as the basic measurement setup, combined with a moving window discrete Fourier transform (DFT) algorithm, the proposed measurement setup has the ability to extract not only the time-invariant online impedance but also the time-variant online impedance of electrical systems.

Following that, based on the three-port network concept, a comprehensive equivalent circuit model of the proposed measurement setup is introduced, in which the effect of the probe-to-probe coupling between the IIP and RIP can be taken into account in the model. With the three-port network equivalent circuit model, a three-term calibration technique is proposed so that the influence of the probe-to-probe coupling can be evaluated quantitatively and compensated, which helps enhance the accuracy of online impedance measurement. By doing so, the accuracy of the extracted online impedance can be preserved even the IIP and RIP are placed very close together.

Then, by incorporating signal amplification and surge protection modules into the measurement setup, the signal-to-noise ratio (SNR) can be improved and the damage caused by the power surges can be avoided. Thus, it expands the scope of application of this approach into the electrical systems with significant electrical noise and power surges.

Finally, with the improved measurement setup and the associated theories, the application aspects of this approach are demonstrated. The first application is the detection of the stator inter-turn short-circuit (ITSC) faults in the inverter-fed induction motor (IM) through IM's online common-mode (CM) impedance monitoring. By selecting a 1/2-hp IM as a motor under test, it has been verified by experiments that the proposed method can achieve excellent sensitivity to detect ITSC faults even the faults are at low severity levels. Also, this method is reliable and robust to detect ITSC faults because it is not affected by motor load and speed variations, as well as rotor and bearing faults. The second application is to measure the voltage-dependent capacitances of the silicon carbide (SiC) power metal-oxide-semiconductor field-effect transistor (MOSFET) in a non-contact and simple manner. By using a 1.2 kV SiC power MOSFET as the test sample, the accuracy of the extracted voltage-dependent capacitances has been verified with the datasheet provided by manufacturers. With the extracted voltage-dependent capacitances of a SiC power MOSFET, its equivalent circuit model can be represented for the evaluation and mitigation of the potential electromagnetic interference (EMI) issues in the early design phase of a power converter.

## 8.2. Future Work

The proposed measurement setup and the related theories behind have shown its advantages and potential in the online impedance extraction of electrical systems. The future works that are worth exploring are summarized as follows:

- Multitone signal injection and processing techniques can be explored so that the online impedance of an electrical system at multiple frequency points can be simultaneously measured so that the operation status and health condition of the electrical system can be better evaluated.

- 3-D full-wave electromagnetic model of the injecting and receiving inductive probes should be developed so that these probes can be evaluated in terms of their performance, thereby selecting or customizing the most appropriate probes in a given application.

- Based on the extracted online impedance and other useful information of an electrical system, combined with the artificial intelligence (AI) algorithms to classify various types of faults of the electrical system for preventive maintenance purposes.

# Author's Publications

## Journal Publications

1. Z. Zhao, K. Y. See, E. K. Chua, A. S. Narayanan, W. Chen, and A. Weerasinghe, "Time-variant in-circuit impedance monitoring based on the inductive coupling method," *IEEE Trans. Instrum. Meas.,* vol. 68, no. 1, pp. 169-176, Jan. 2019.

2. Z. Zhao, K. Y. See, W. Wang, E. K. Chua, A. Weerasinghe, Z. Yang, and W. Chen, "Voltage-dependent capacitance extraction of SiC power MOSFETs using inductively coupled in-circuit impedance measurement technique," *IEEE Trans. Electromagn. Compat.*, vol. 61, no. 4, pp. 1322-1328, Aug. 2019.

3. Z. Zhao, A. Weerasinghe, W. Wang, E. K. Chua, and K. Y. See, "Eliminating the effect of probe-to-probe coupling in inductive coupling method for in-circuit impedance measurement," *IEEE Trans. Instrum. Meas.*, vol. 70, 2021, Art no. 1000908.

4. Z. Zhao, F. Fan, W. Wang, Y. Liu, and K. Y. See, "Detection of stator inter-turn short-circuit faults in inverter-fed induction motors by online common-mode impedance monitoring," *IEEE Trans. Instrum. Meas.*, accepted.

**Conference Publications**

1. Z. Zhao, K. Y. See, E. K. Chua, A. S. Narayanan, A. Weerasinghe, and W. Chen, "Extraction of voltage-dependent capacitances of SiC device through inductive coupling method," in *Proc. IEEE Int. Symp. Electromagn. Compat. IEEE Asia-Pacific Symp. Electromagn. Compat. (EMC&APEMC)*, Singapore, 2018, pp. 1301–1304. (Best Student Paper Award)

2. Z. Zhao, Y. Liu, K. Y. See, W. Wang, E. K. Chua, A. S. Narayanan, Arjuna Weerasinghe, Ivan Christian, "Extraction of loop inductances of SiC half-bridge power module using an improved two-port network method," in *Proc. Annu. Conf. IEEE Ind. Electron. Soc. (IECON)*, Washington, DC, USA, 2018, pp. 1204–1208.

3. Z. Zhao, K. Y. See, E. K. Chua, A. S. Narayanan, A. Weerasinghe, Z. Yang, K. Tan, "Online insulation fault detection of stator winding of induction motor based on a non-intrusive impedance extraction technique," in *Proc. IEEE Int. Conf. Intell. Rail Transp. (ICIRT)*, Singapore, 2018.

# References

[1]  L. Asiminoaei, R. Teodorescu, F. Blaabjerg, and U. Borup, "A new method of on-line grid impedance estimation for PV inverter," in *Proc. IEEE Appl. Power Electron. Conf. Expo. (APEC)*, Anaheim, CA, USA, 2004, pp. 1527-1533.

[2]  S. Cobreces, E. J. Bueno, D. Pizarro, F. J. Rodriguez, and F. Huerta, "Grid impedance monitoring system for distributed power generation electronic interfaces," *IEEE Trans. Instrum. Meas.,* vol. 58, no. 9, pp. 3112-3121, Sep. 2009.

[3]  Y. He, S.-H. Chung, C.-T. Lai, X. Zhang, and W. Wu, "Active cancelation of equivalent grid impedance for improving stability and injected power quality of grid-connected inverter under variable grid condition," *IEEE Trans. Power Electron.,* vol. 33, no. 11, pp. 9387-9398, Nov. 2018.

[4]  M. Liserre, R. Teodorescu, and F. Blaabjerg, "Stability of photovoltaic and wind turbine grid-connected inverters for a large set of grid impedance alues," *IEEE Trans. Power Electron.,* vol. 21, no. 1, pp. 263-272, Jan. 2006.

[5]  S. Cobreces, E. Bueno, F. J. Sanchez, F. Huerta, and P. Rodriguez, "Influence analysis of the effects of an inductive-resistive weak grid over L and LCL filter current hysteresis controllers," in *Proc. Eur. Power Electron. Conf.*, Aalborg, Denmark, 2007, pp. 1-10.

[6]  A. Tarkiainen, R. Pollanen, M. Niemela, and J. Pyrhonen, "Identification of grid impedance for purposes of voltage feedback active filtering," *IEEE Power Electron. Lett.,* vol. 2, no. 1, pp. 6-10, Mar. 2004.

[7]  M. Liserre, F. Blaabjerg, and R. Teodorescu, "Grid impedance estimation via excitation of LCL-filter resonance," *IEEE Trans. Ind. Appl.,* vol. 43, no. 5, pp. 1401-1407, Sep./Oct. 2007.

[8]  A. Cuadras and O. Kanoun, "SoC Li-ion battery monitoring with impedance spectroscopy," in *Proc. Int. Multi-Conf. Syst., Signals Devices*, Djerba, Tunisia, 2009, pp. 1-5.

[9]  D. A. Howey, P. D. Mitcheson, V. Yufit, G. J. Offer, and N. P. Brandon, "Online measurement of battery impedance using motor controller excitation," *IEEE Trans. Veh. Technol.,* vol. 63, no. 6, pp. 2557-2566, Jul. 2014.

[10] J. A. A. Qahouq, "Online battery impedance spectrum measurement method," in *Proc. IEEE Appl. Power Electron. Conf. Expo. (APEC)*, Long Beach, CA, USA, 2016, pp. 3611-3615.

[11] M. Coleman, C. K. Lee, C. Zhu, and W. G. Hurley, "State-of-charge determination from EMF voltage estimation: Using impedance, terminal voltage, and current for lead-acid and lithium-ion batteries," *IEEE Trans. Ind. Electron.,* vol. 54, no. 5, Oct. 2007.

[12] A. Zenati, P. Desprez, and H. Razik, "Estimation of the SOC and the SOH of Li-ion batteries, by combining impedance measurements with the fuzzy logic inference," in *Proc. Annu. Conf. IEEE Ind. Electron. Soc. (IECON)*, Glendale, AZ, USA, 2010, pp. 1773-1778.

[13] W. Waag, S. Kabitz, and D. U. Sauer, "Experimental investigation of the lithium-ion battery impedance characteristic at various conditions and aging states and its influence on the application," *Appl. Energy,* vol. 102, pp. 885-897, Feb. 2013.

[14] J. Zhu, Z. Sun, X. Wei, and H. Dai, "A new lithium-ion battery internal temperature on-line estimate method based on electrochemical impedance spectroscopy measurement," *J. Power Sources,* vol. 274, pp. 990-1004, 2015.

[15] M. J. Nave, *Power Line Filter Design for Switched-Mode Power Supplies*. New York: Van Nostrand Reinhold, 1991, pp. 43-48.

[16] C. R. Paul, *Introduction to Electromagnetic Compatibility*. New York: Wiley, 1992, pp. 477-487.

[17] J. A. Ferreira, P. R.Willcock, and S. R. Holm, "Sources, paths and traps of conducted EMI in switch mode circuits," in *Proc. IEEE Ind. Appl. Conf.*, New Orleans, LA, USA, 1997, pp. 1584–1591.

[18] K. Y. See and J. Deng, "Measurement of noise source impedance of SMPS using a two probes approach," *IEEE Trans. Power Electron.,* vol. 19, no. 3, pp. 862-868, May 2004.

[19] C. González, J. Pleite, V. Valdivia, and J. Sanz, "An overview of the on line application of frequency response analysis (FRA)," in *Proc. IEEE Int. Conf. Ind. Electron.*, Vigo, Spain, 2007, pp. 1294-1299.

[20] T. Funaki, N. Phankong, T. Kimoto, and T. Hikihara, "Measuring terminal capacitance and its voltage dependency for high-voltage power devices," *IEEE Trans. Power Electron.,* vol. 24, no. 6, pp. 1486-1493, Jun. 2009.

[21] X. Shang, D. Su, H. Xu, and Z. Peng, "A noise source impedance extraction method for operating SMPS using modified LISN and simplified calibration procedure," *IEEE Trans. Power Electron.,* vol. 32, no. 6, pp. 4132-4139, Jun. 2017.

[22] M. Sumner, B. Palethorpe, and D. Thomas, "Impedance measurement for improved power quality-part 1: The measurement technique," *IEEE Trans. Power Del.,* vol. 19, no. 3, pp. 1442-1448, Jul. 2004.

[23] A. V. Timbus, R. Teodorescu, F. Blaabjerg, and U. Borup, "Online grid measurement and ENS detection for PV inverter running on highly inductive grid," *IEEE Power Electron. Lett.,* vol. 2, no. 3, pp. 77-82, Sep. 2004.

[24] L. Asiminoaei, R. Teodorescu, F. Blaabjerg, and U. Borup, "Implementation and test of an online embedded grid impedance estimation technique for PV inverters," *IEEE Trans. Ind. Electron.,* vol. 52, no. 4, pp. 1136-1144, Aug. 2005.

[25] N. Hoffmann and F. W. Fuchs, "Minimal invasive equivalent grid impedance estimation in inductive resistive power networks using extended Kalman filter," *IEEE Trans. Power Electron.,* vol. 29, no. 2, pp. 631-641, Feb. 2014.

[26] W. Huang and J. A. A. Qahouq, "An online battery impedance measurement method using DC–DC power converter control," *IEEE Trans. Power Electron.,* vol. 61, no. 11, pp. 5987-5995, Nov. 2014.

[27] D. K. Alves, R. Ribeiro, F. B. Costa, and T. Rocha, "Real-time wavelet-based grid impedance estimation method," *IEEE Trans. Ind. Electron.,* vol. 66, no. 10,

pp. 8263-8265, Oct. 2019.

[28] R. A. Southwick and W. C. Dolle, "Line impedance measuring instrumentation utilizing current probe coupling," *IEEE Trans. Electromagn. Compat.,* vol. EMC-13, no. 4, pp. 31-36, Nov. 1971.

[29] V. Tarateeraseth, K. Y. See, F. G. Canavero, and R. W. Chang, "Systematic electromagnetic interference filter design based on information from in-circuit impedance measurements," *IEEE Trans. Electromagn. Compat.,* vol. 52, no. 3, pp. 588-598, Aug. 2010.

[30] K. Li, K. Y. See, and X. Li, "Inductive coupled in-circuit impedance monitoring of electrical system using two-port ABCD network approach," *IEEE Trans. Instrum. Meas.,* vol. 64, no. 9, pp. 2489-2495, Sep. 2015.

[31] S. B. Rathnayaka, K. Y. See, and K. Li, "Online impedance monitoring of transformer based on inductive coupling approach," *IEEE Trans. Dielectr. Electr. Insul.,* vol. 24, no. 2, pp. 1273-1279, Apr. 2017.

[32] D. M. Pozar, *Microwave engineering*. Hoboken, NJ: Wiley, 2012, pp. 165-227.

[33] C. J. Kikkert, "An on-line plc frequency impedance analyzer," in *Proc. IEEE Int. Conf. Smart Gird Comm.*, Vancouver, BC, Canada, 2013, pp. 606-611.

[34] R. L. Steigerwald, A. Ferraro, and F. G. Turnbull, "Application of power transistors to residential and intermediate rating photovoltaic array power conditioners," *IEEE Trans. Ind. Appl.,* vol. 19, no. 2, pp. 254-267, Mar. 1983.

[35] L. J. Feiner, "Power electronics for transport aircraft applications," in *Proc. Annu. Conf. IEEE Ind. Electron. Soc.(IECON)*, Maui, HI, USA, 1993, pp. 719-724.

[36] T. M. Jahns and V. Blasko, "Recent advances in power electronics technology for industrial and traction machine drives," *Proc. IEEE,* vol. 89, no. 6, pp. 963-975, Jun. 2001.

[37] F. Blaabjerg, M. Liserre, and K. Ma, "Power electronics converters for wind turbine systems," *IEEE Trans. Ind. Appl.,* vol. 48, no. 2, pp. 708-719, Mar./Apr. 2012.

[38] B. K. Bose, "Global energy scenario and impact of power electronics in 21st century," *IEEE Trans. Ind. Electron.,* vol. 60, no. 7, pp. 2638-2651, Jul. 2013.

[39] "PXI-5922 Oscilloscope," Nat. Instrum., Austin, TX, USA, 2018.

[40] "PXIe-5186 Digitizer," Nat. Instrum., Austin, TX, USA, 2015.

[41] J. A. Rosendo Macias and A. Gomez Exposito, "Efficient moving-window DFT algorithms," *IEEE Trans. Circuits Syst. II. Analog Digit. Signal Process. (1993-2003),* vol. 45, no. 2, pp. 256-260, Feb. 1998.

[42] Y. Liu, Z. Zhao, W. Wang, and J.-S. Lai, "Characterization and extraction of power loop stray inductance with SiC half-bridge power module," *IEEE Trans. Electron Devices,* vol. 67, no. 10, pp. 4040-4045, Oct. 2020.

[43] J. Wang, H. S. Chung, and R. T. Li, "Characterization and experimental assessment of the effects of parasitic elements on the MOSFET switching performance," *IEEE Trans. Power Electron.,* vol. 28, no. 1, pp. 573-590, Jan. 2013.

[44] H. Karkkainen, L. Aarniovuori, M. Niemela, and J. Pyrhonen, "Converter-fed induction motor efficiency: Practical applicability of IEC methods," *IEEE Ind. Electron. Mag.*, vol. 11, no. 2, pp. 45-57, Jun. 2017.

[45] P. F. Albrecht, J. C. Appiarius, R. M. McCoy, E. L. Owen, and D. K. Sharma, "Assessment of the reliability of motors in utility applications—Updated," *IEEE Trans. Energy Convers.*, vol. EC-1, no. 1, pp. 39-46, Mar. 1986.

[46] P. Zhang, Y. Du, T. G. Habetler, and B. Lu, "A survey of condition monitoring and protection methods for medium-voltage induction motors," *IEEE Trans. Ind. Appl.*, vol. 47, no. 1, Jan./Feb. 2011.

[47] C. Rojas, M. Melero, M. Cabanas, J. Cano, G. Orcajo, and F. Pedrayes, "Finite element model for the study of inter-turn short circuits in induction motors," in *Proc. IEEE Int. Symp. Diagnostics Electr. Mach., Power Electron. Drives*, Cracow, Poland, 2007, pp. 415–419.

[48] D. Diaz, M. Amaya, and A. Paz, "Inter-turn short-circuit analysis in an induction machine by finite elements method and field tests," in *Proc. IEEE Int. Conf. Electr. Mach (ICEM)*, Marseille, France, 2012, pp. 1757–1763.

[49] K. Prasob, N. P. Kumar, and T. Isha, "Inter-turn short circuit fault analysis of PWM inverter fed three-phase induction motor using finite element method," in *Proc. Int. Conf. Circuit, Power Comput. Technol. (ICCPCT)*, Kollam, India, 2017, pp. 1-6.

[50] W. T. Thomson, "On-line MCSA to diagnose shorted turns in low voltage stator windings of 3-phase induction motors prior to failure," in *Proc. Int. Electric. Machines and Drives Conf. (IEMDC)*, Cambridge, MA, USA, 2001, pp. 891-898.

[51] J. Sottile, F. C. Trutt, and J. L. Kohler, "Condition monitoring of stator windings in induction motors. II. Experimental investigation of voltage mismatch detectors," *IEEE Trans. Ind. Appl.*, vol. 38, no. 5, pp. 1454–1459, Sep./Oct. 2002.

[52] G. N. Surya, Z. J. Khan, M. S. Ballal, and H. M. Suryawanshi, "A simplified frequency-domain detection of stator turn fault in squirrel-cage induction motors using an observer coil technique," *IEEE Trans. Ind. Electron.*, vol. 64, no. 2, pp. 1495-1506, Feb. 2017.

[53] J. S. Hsu, "Monitoring of defects in induction motors through air-gap torque observation," *IEEE Trans. Ind. Appl.*, vol. 31, no. 5, pp. 1016-1021, Sep./Oct. 1995.

[54] F. B. Batista, P. C. Filho, R. Pederiva, and V. A. Silva, "An empirical demodulation for electrical fault detection in induction motors," *IEEE Trans. Instrum. Meas.*, vol. 65, no. 3, pp. 559-569, Mar. 2016.

[55] A. Mohammed, J. I. Melecio, and S. Djurović, "Stator winding fault thermal signature monitoring and analysis by in situ FBG sensors," *IEEE Trans. Ind. Electron.*, vol. 66, no. 10, pp. 8082-8092, Oct. 2019.

[56] S. F. Legowski, A. H. M. S. Ula, and A. M. Trzynadlowsh, "Instantaneous power as a medium for the signature analysis of induction motors," *IEEE Trans. Ind. Appl.*, vol. 32, no. 4, pp. 904-909, Jul./Aug. 1996.

[57] S. B. Lee, R. M. Tallam, and T. G. Habetler, "A robust, on-line turn-fault detection technique for induction machines based on monitoring the sequence component impedance matrix," *IEEE Trans. Power Electron.,* vol. 18, no. 3, pp. 865-872, May 2003.

[58] G. Betta, C. Liguori, A. Paolillo, and A. Pietrosanto, "A DSP-based FFT-analyzer for the fault diagnosis of rotating machine based on vibration analysis," *IEEE Trans. Instrum. Meas.,* vol. 51, no. 6, pp. 1316-1322, Dec. 2002.

[59] R. Romary, R. Pusca, J. P. Lecointe, and J. F. Brudny, "Electrical machines fault diagnosis by stray flux analysis," in *Proc. IEEE Workshop Electr. Mach. Des. Control Diagn. (WEMDCD)*, Paris, France, 2013, pp. 245-256.

[60] T. Yang, H. Pen, Z. Wang, and C. S. Chang, "Feature knowledge based fault detection of induction motors through the analysis of stator current data," *IEEE Trans. Instrum. Meas.,* vol. 65, no. 3, pp. 549-557, Mar. 2016.

[61] A. Glowacz and Z. Glowacz, "Diagnosis of the three-phase induction motor using thermal imaging," *Infrared Phys. Technol.,* vol. 81, pp. 7-16, 2017.

[62] R. M. Tallam, T. G. Habetler, and R. G. Harley, "Self-commissioning training algorithms for neural networks with applications to electric machine fault diagnostics," *IEEE Trans. Power Electron.,* vol. 17, no. 6, pp. 1089-1095, Nov. 2002.

[63] J. Seshadrinath, B. Singh, and B. K. Panigrahi, "Vibration analysis based interturn fault diagnosis in induction machines," *IEEE Trans. Ind. Informat.,* vol. 10, no. 1, pp. 340-350, Feb. 2014.

[64] G. H. Bazan, P. R. Scalassara, W. Endo, A. Goedtel, R. H. Palácios, and W. F. Godoy, "Stator short-circuit diagnosis in induction motors using mutual information and intelligent systems," *IEEE Trans. Ind. Electron.,* vol. 66, no. 4, pp. 3237-3246, Apr. 2019.

[65] A. Stief, J. R. Ottewill, J. Baranowski, and M. Orkisz, "A PCA and two-stage bayesian sensor fusion approach for diagnosing electrical and mechanical faults in induction motors," *IEEE Trans. Ind. Electron.,* vol. 66, no. 12, pp. 9510-9520, Dec. 2019.

[66] M. Singh and A. G. Shaik, "Incipient fault detection in stator windings of an induction motor using Stockwell Transform and SVM," *IEEE Trans. Instrum. Meas.,* vol. 69, no. 12, pp. 9496-9504, Dec 2020.

[67] M. S. Toulabi, L. Wang, L. Bieber, S. Filizadeh, and J. Jatskevich, "A universal high-frequency induction machine model and characterization method for arbitrary stator winding connections," *IEEE Trans. Energy Convers.,* vol. 34, no. 3, pp. 1164-1177, Sep. 2019.

[68] X. Song, Z. Wang, S. Li, and J. Hu, "Sensorless speed estimation of an inverter-fed induction motor using the supply-side current," *IEEE Trans. Energy Convers.,* vol. 34, no. 3, pp. 1432-1441, Sep. 2019.

[69] P. Baudesson, J. Lafontaine, and A. Nourrisson, "EMC filtering device in a variable speed drive," U.S. Patent 7738268, Jun. 15, 2010.

[70] M. Moreau, N. Idir, and P. L. Moigne, "Modeling of conducted EMI in adjustable speed drives," *IEEE Trans. Electromagn. Compat.,* vol. 51, no. 3, pp.

665–672, Aug. 2009.

[71] E. Mazzola, F. Grassi, and A. Amaducci, "Novel measurement procedure for switched-mode power supply modal impedances," *IEEE Trans. Electromagn. Compat.,* vol. 62, no. 4, pp. 1349-1357, Aug. 2020.

[72] D. Zhao, J. A. Ferreira, A. Roc'h, and F. Leferink, "Common-mode DC-bus filter fesign for variable-speed drive system via transfer ratio measurements," *IEEE Trans. Power Electron.,* vol. 24, no. 2, pp. 518-524, Feb. 2009.

[73] X. Liang, M. Z. Ali, and H. Zhang, "Induction motors fault diagnosis using finite element method: a review," *IEEE Trans. Ind. Appl.,* vol. 56, no. 2, pp. 1205-1217, Mar./Apr. 2020.

[74] J. Biela, M. Schweizer, S. Waffler, and J. W. Kolar, "SiC versus Si—Evaluation of potentials for performance improvement of inverter and DC–DC converter systems by SiC power semiconductors," *IEEE Trans. Ind. Electron.,* vol. 58, no. 7, pp. 2872-2882, Jul. 2011.

[75] X. Gong and J. A. Ferreira, "Investigation of conducted EMI in SiC JFET inverters using separated heat sinks," *IEEE Trans. Ind. Electron.,* vol. 61, no. 1, pp. 115-125, Jan. 2014.

[76] Y. Liu *et al.*, "LCL Filter Design of 50 kW 60 kHz SiC Inverter with Size and Thermal Considerations for Aerospace Applications," *IEEE Trans. Ind. Electron.,* vol. 64, no. 10, pp. 8321-8333, Oct. 2017.

[77] S. Yin, K. J. Tseng, R. Simanjorang, Y. Liu, and J. Pou, "A 50-kW High-Frequency and High-Efficiency SiC Voltage Source Inverter for More Electric Aircraft," *IEEE Trans. Ind. Electron.,* vol. 64, no. 11, pp. 9124-9134, Nov. 2017.

[78] R. Khanna, A. Amrhein, W. Stanchina, G. Reed, and Z.-H. Mao, "An analytical model for evaluating the influence of device parasitics on Cdv/dt induced false turn-on in SiC MOSFETs," in *Proc. IEEE Appl. Power Electron. Conf. Expo. (APEC)*, Long Beach, CA, USA, 2013, pp. 518-525.

[79] A. Lemmon, M. Mazzola, J. Gafford, and C. Parker, "Stability considerations for silicon carbide field-effect transistors," *IEEE Trans. Power Electron.,* vol. 28, no. 10, pp. 4453-4459, Oct. 2013.

[80] N. Oswald, P. Anthony, N. McNeill, and B. H. Stark, "An experimental investigation of the tradeoff between switching losses and EMI generation with hard-switched all-Si, Si-SiC, and all-SiC device combinations," *IEEE Trans. Power Electron.,* vol. 29, no. 5, pp. 2393-2407, May 2014.

[81] Z. Zhao *et al.*, "Extraction of Loop Inductances of SiC Half-Bridge Power Module Using An Improved Two-port Network Method," presented at the *Proc. Annu. Conf. IEEE Ind. Electron. Soc. (IECON)*, Washington, DC, USA, 2018.

[82] Z. Chen, D. Boroyevich, R. Burgos, and F. Wang, "Characterization and modeling of 1.2 kv, 20 A SiC MOSFETs," in *Proc. IEEE Energy Convers. Congr. Expo. (ECCE)*, San Jose, CA, USA, 2009, pp. 1480-1487.

[83] P. Anthony, N. McNeill, and D. Holliday, "A first approach to a design method for resonant gate driver architectures," *IEEE Trans. Power Electron.,* vol. 27, no. 8, pp. 3855-3868, Aug. 2012.

[84] Y. Liu, K. Y. See, R. Simanjorang, Z. Lim, and Z. Zhao, "Modeling and

simulation of switching characteristics of half-bridge SiC power module in single leg T-type converter for EMI prediction," in *Joint Proc. IEEE Int. Symp. Electromagn. Compat. and Proc. Asia-Pacific Int. Symp. Electromagn. Compat. (EMC/APEMC)*, Singapore, 2018, pp. 1314-1318.

[85] Z.-N. Ariga, K. Wada, and T. Shimizu, "TDR measurement method for voltage-dependent capacitance of power devices and components," *IEEE Trans. Power Electron.,* vol. 27, no. 7, pp. 3444-3451, Jul. 2012.

[86] R. Elferich, T. López, and N. Koper, "Accurate behavioural modelling of power MOSFETs based on device measurements and FE-simulations," in *Proc. Eur. Conf. Power Electron. Appl.*, Dresden, Germany, 2005, pp. 1-9.

[87] K. Li, A. Videt, and N. Idir, "Multiprobe measurement method for voltage-dependent capacitances of power semiconductor devices in high voltage," *IEEE Transactions on Power Electronics,* vol. 28, no. 11, pp. 5414-5422, 2013.

[88] Z. Chen, "Characterization and modeling of high-switching-speed behavior of SiC active devices," Master of Science, Department of Electrical Engineering, Virginia Tech., 2009.

[89] M. Ando and K. Wada, "Design of acceptable stray inductance based on scaling method for power electronics circuits," *IEEE J. Emerg. Sel. Topics Power Electron.,* vol. 5, no. 1, pp. 568-575, Mar. 2017.

[90] Y. Mukunoki *et al.*, "Modeling of a silicon-carbide MOSFET with focus on internal stray capacitances and inductances, and its verification," *IEEE Trans. Ind. Appl.,* vol. 54, no. 3, pp. 2588-2597, May/Jun. 2018.

[91] "Silicon carbide power MOSFET," C2M0080120D Cree datasheet, 2015.

[92] "Silicon caribde power MOSFET," SCT2H12NZ ROHM datasheet, 2017.

[93] "Silicon carbide power MOSFET," UJ3C120150K3S UnitedSiC datasheet, 2018.

[94] "Silicon carbide power MOSFET," LSIC1MO120E0160 Littelfuse datasheet, 2019.